\pdfoutput=1
\documentclass[11pt]{article}
\usepackage[preprint]{acl}
\usepackage{times}
\usepackage{latexsym}
\usepackage[T1]{fontenc}
\usepackage[utf8]{inputenc}
\usepackage{graphicx,booktabs,amsmath,amssymb,mathtools}
\usepackage{amsmath}
\usepackage{algorithm}
\usepackage{mathrsfs}
\usepackage[noend]{algpseudocode}
\usepackage{float,microtype,enumitem,tabularx,multirow,placeins,array,pifont,comment}
\usepackage{bbm}
\hypersetup{hidelinks,hypertexnames=false,pdftitle={Q-BWE: Quantum Chaos-Informed Speech Bandwidth Extension},pdfauthor={Tarikul Islam Tamiti, Sajid Fardin Dipto, David Vergano, Luke Baja-Ricketts, Anomadarshi Barua}}
\graphicspath{{./}}
\newcommand{\tablefontsize}{\fontsize{8}{9.5}\selectfont}
\newcommand{\method}{\textsc{Q-BWE}}
\newcommand{\cis}{\textsc{CIS-BWE}}
\newcommand{\qcd}{\textsc{QCD}}
\newcommand{\mrld}{\textsc{MRLD}}
\newcommand{\msdfa}{\textsc{MSDFA}}
\newcommand{\mrad}{\textsc{MRAD}}
\newcommand{\mrpd}{\textsc{MRPD}}
\newcommand{\cmark}{\ding{51}}
\newcommand{\xmark}{\ding{55}}
\newcommand{\R}{\mathbb{R}}

\title{Q-BWE: Quantum Chaos-Informed Speech Bandwidth Extension}
\author{Tarikul Islam Tamiti \and Sajid Fardin Dipto \and David Vergano \\
\textbf{Luke Baja-Ricketts \and Anomadarshi Barua}}
\begin{document}
\maketitle

\begin{abstract}

High-frequency reconstruction in bandwidth extension (BWE) depends on joined interactions among nonlinear spectral, temporal, and perceptual features; assigning a separate discriminator to each such feature may be redundant and may fail to model their joint interactions. We introduce quantum-informed generative model \method{}, where we design a five-qubit Quantum Chaos Discriminator (QCD) instead of using separate discriminator for every features, for modeling nonlinear joined-feature interactions efficiently. Our QCD explores a compact five-qubit quantum-inspired inductive bias based on data re-uploading circuit, trainable controlled NOT (CNOT) rotations, quantum entangling gates, and statevector measurement map. Five features for five qubits are selected from a 26-dimensional speech-dynamics bank and are encoded by a two-layer statevector circuit, producing 31 measurements map of local behavior, correlations, mixedness, and Born entropy. QCD guides a Lattice-ConformerNeXt generator with FiLM conditioning and a cosine-masked residual pathway, while delayed-reward Thompson sampling adaptively controls its contribution. The circuit is simulated on classical hardware. Overall, \method{} achieves competitive performance on VCTK English, and MLS French datasets in multilingual and multi speaker setups under clean and noisy conditions across a broad set of evaluation metrics with $\sim$ 
1.72x fewer parameters (33.5M vs. 19.4M) compared to recent CIS-BWE baseline published in ACL.

\end{abstract}

\vspace{-0.6em}
\section{Introduction and Related Work}
\vspace{-0.3em}
\label{sec:intro}
\label{sec:related}


Bandwidth extension (BWE) estimates missing high-frequency (HF) content, 
improving perceptual quality and preserving acoustic cues important for downstream text-to-speech (TTS) and automatic speech recognition (ASR) systems \cite{9413575}. 

\textbf{Research Gap:} 
In BWE, missing multiple high-frequencies may depend on same low-frequency (LF) content known as one-to-many mapping. Hence, a BWE model must reconstruct HF contents with subtle non-linear and uncertain artifacts similar to ground truth rather than just recover plausible HF continuation. Existing magnitude--phase models, including AP-BWE \cite{lu2024towards}, and CTFT-Net \cite{tamiti2025ctft} improve joint spectral reconstruction, while AERO \cite{mandel2023aero}, EBEN \cite{hauret2023eben}, and MS-BWE \cite{lu2024multi} address spectral super-resolution, 
and flexible sampling-rate issues. However, objectives dominated by averaged spectral error can still favor \textbf{over-smoothed} outputs that weaken transients, harmonics, and phase-dependent structure. 

\textbf{Multi-scale Chaos Feedback:} 
To discourage oversmoothing that pointwise losses capture, CIS-BWE \cite{tamiti2026cis} and NLDSI-BWE \cite{tamiti2026nldsi} incorporate nonlinear-dynamics and chaos-informed supervision into BWE. 
They show that chaos-derived discriminators offer complementary fine-grained cues by using finite-time divergence and detrended fluctuation analysis (DFA). 
However, these chaos-informed models assign a separate discriminator to each chaotic feature. 
Although these critics provide complementary feedback, this independent one-feature-per-discriminator design increases size and computational overhead. 
Moreover, they do not explicitly characterize the \textit{joint dispersion of the chaotic and other features}, which is critical to solve 
the one-to-many mapping problem. 
Hence, we investigate that, instead of using separate discriminator for every features, whether a single compact discriminator  (i.e., solving overhead) can be used to automatically select most relevant features for \textit{modeling joined-feature interactions efficiently}. 

\textbf{Contribution:} 
For the first time, we show that joined interactions among chaotic and other nonlinear features can be modeled with a single Quantum--Chaos Discriminator (QCD) with minimum computational overhead. Our QCD explores a compact five-qubit quantum-inspired inductive bias based on data re-uploading circuit, trainable  controlled-NOT (CNOT) rotations, quantum entangling gates, and statevector measurements \cite{havlicek2019supervised,schuld2019quantum,cerezo2021variational,perezsalinas2020data}. 
In particular, entangling gates allow each feature to influence the representation of the others, while local and joint measurements expose both feature-specific behavior and cross-feature dependence. 

We design QCD as a training-only critic 
that automatically (i.e., no human selection) selects five most relevant features 
from a 26-feature bank, which contains nonlinear, fractal, spectral, and temporal features. After normalization and bounded-angle encoding, the features/descriptors are processed by a two-layer data re-uploading circuit containing trainable $R_y$ and $R_z$ rotations and alternating forward and reverse CNOT rings. The exact state vector produces 31 measurements: 15 local Pauli-$X/Y/Z$ expectations, 10 nearest-neighbor $XX/ZZ$ correlations, 5 reduced-state mixedness values, and one normalized Born-entropy value. 

As improvements differ across spectral, and perceptual metrics, we report 
our experiments at different 
conditioning and robustness settings to justify our design choice. 
Most notably, Q-BWE achieves better results in
most over-smoothing (i.e., LSD) and perceptual (i.e., NISQA, PESQ, STOI) metrics with slightly
lower SI-SDR/SI-SNR with $\sim$1.72x fewer parameters compared to the most recent baseline CIS-BWE on VCTK English and MLS French datasets under multispeaker and multilingual settings. 
QCD achieves this improvement with only 4,393 parameters, including 40 quantum-circuit parameters. 
Therefore, for the first time, this paper shows that quantum--chaos supervision can jointly model interacting speech-dynamics and reduce reliance on multiple feature-specific critics, while preserving parameter efficiency in BWE tasks.
 
\vspace{-0.24em}
\section{\method{} Architecture}
\vspace{-0.3em}
\label{sec:method}


The complete end-to-end \method{} pipeline
is shown in Fig.~\ref{fig:overall_qbwe_architecture} and detailed in Appendix~\ref{app:overall_architecture}.
 
\vspace{-0.5em}
\subsection{Generator Design}
\vspace{-0.2em}
\label{sec:generator}

\begin{figure*}[t]
    \centering
    \includegraphics[width=0.99\textwidth]{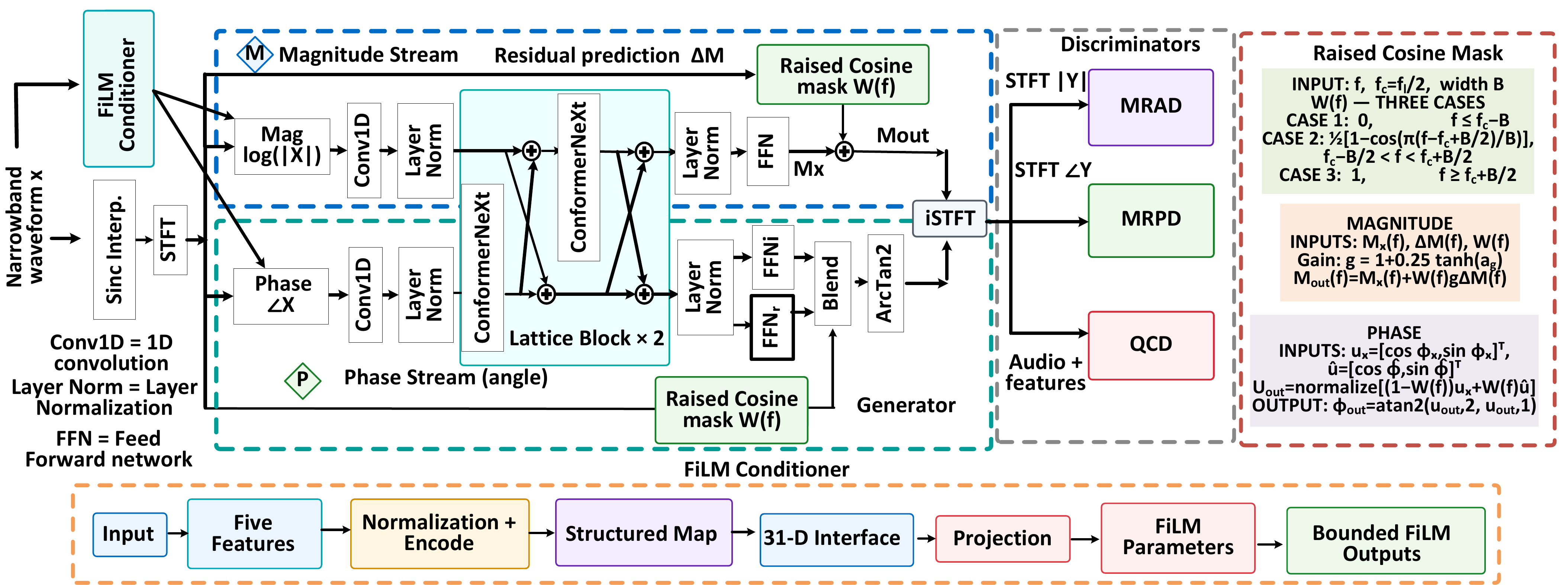}
    \vspace{-0.4em}
    \caption{Overall \method{} architecture with FiLM conditioning and raised cosine mask and QCD.} 
    \label{fig:overall_qbwe_architecture}
    \vspace{-01.5em}
\end{figure*}
\vspace{-0.0em}

Motivated by prior joint magnitude--phase modeling and complex-spectral reconstruction methods \citep{yin2020phasen,mandel2023aero,lu2024towards,lu2025explicit}, the generator processes log magnitude 
and wrapped phase 
in coupled Lattice-ConformerNeXt streams  \cite{tamiti2026cis}. 

\textbf{Feature-wise Linear Modulation (FiLM) \cite{perez2018film} Conditioner:}  
The missing HF spectrum 
depends strongly on the input signal type.  
A shared unconditioned generator applies the same transformation to all such inputs. 
This can be inefficient 
because inputs
may require substantially different HF extensions. Therefore, we provide the generator with an explicit summary of the input 
with five features: log-energy, spectral centroid, spectral flatness, spectral flux, and phase to capture signal level, spectral balance, tonality, temporal variation, and phase organization. 

After normalization and bounded angle encoding, the features are processed by a two-layer five-qubit map {\color{black}(see Section \ref{sec:circuit})}, 
having 40 trainable parameters, and produces 31-dimensional representation. 
A shared linear projection converts the 31-dimensional representation into five FiLM parameter groups $(\gamma_m,\beta_m,\gamma_p,\beta_p,a_g)$. Here, $m$ is magnitude stream, $p$ is phase stream;
$\gamma_m,\beta_m,\gamma_p,\beta_p\in\mathbb{R}^{C}$ are channel-wise FiLM scale and bias vectors, $a_g\in\mathbb{R}$ is a scalar controlling the magnitude-residual gain. 
Before joint magnitude--phase processing, each stream is modulated  as $H'_s=[1+0.1\tanh(\gamma_s)]\odot H_s+0.1\tanh(\beta_s)$, 
where $ s\in\{m,p\}$; $\gamma_s$ rescales feature channels, $\beta_s$ shifts them, and $\odot$ denotes element-wise multiplication broadcast across time. Moreover, the magnitude-residual strength is controlled separately by $g=1+0.25\tanh(a_g)\in[0.75,1.25]$.
This near-identity parameterization allows the FiLM conditioner to emphasize, suppress, or shift feature channels according to the input while preventing large, unstable deviations from the base generator.

\textbf{FiLM size:} FiLM module maps the narrowband input to 513 magnitude and phase bins by a 1,024-point STFT and generates $C=512$ latent channels. 
Therefore, the shared linear projection maps 31 measurements to $4C+1=2049$ conditioning coefficients and contains
$31\times2049+2049=65{,}568$ parameters, including its bias. Together with the 40-parameter map, the FiLM adds 65,608 parameters, approximately $0.36\%$ of the generator parameter. 


\textbf{Raised-Cosine Mask:}
BWE should generate information above the observed Nyquist frequency while preserving the reliable spectrum already present below it. Unrestricted full-spectrum prediction does not enforce this separation: prediction errors can alter valid low-frequency magnitude and phase, while a hard replacement at the band boundary can introduce an abrupt spectral transition, distorting build-up harmonics. To solve this, for the first time, we design \textit{raised-cosine mask}, which is combined with residual prediction 
to control where the generator may modify the input. 

It is termed as raised-cosine mask because the cosine is shifted and scaled from $[-1,1]$ to $[0,1]$, producing a monotonic half-cosine transition. Moreover, its slope is zero at both transition endpoints, so it connects the preserved and predicted frequency regions smoothly and reduces boundary discontinuities and associated ringing artifacts.

\vspace{-0.1em}
Let $f_c$ 
is the observed Nyquist frequency and $B$ is the transition bandwidth. Defining a clipping function, $t(f)=\operatorname{clip}([f-(f_c-B/2)]/B,0,1)$, gives the raised cosine mask, $W(f)=\tfrac{1}{2}[1-\cos(\pi t(f))]$. Hence, $W(f)=0$ in the confidently observed LF band, $W(f)=1$ in the missing high-frequency band, and $0<W(f)<1$ within a smooth transition centered at $f_c$. 

\vspace{-0.1em}
{\color{black} In magnitude stream, the residual prediction head predicts a log-magnitude correction $\Delta M$, and the masked output is $M_{\mathrm{out}}=M_x+W(f)g\Delta M$. This expression separates three
functions: $M_x$ supplies the observed spectrum, $\Delta M$ represents the learned correction, $g$ controls its input-dependent strength, and $W(f)$ localizes it in frequency.}

\vspace{-0.1em}
In phase stream, phase is blended in the unit-vector domain because direct interpolation of
wrapped angles is ill-defined near $\pm\pi$. With
$\mathbf{u}_x=[\cos\phi_x,\sin\phi_x]^\top$ and
$\widehat{\mathbf{u}}=[\cos\widehat{\phi},\sin\widehat{\phi}]^\top$, the
output direction is
$\mathbf{u}_{\mathrm{out}}=\operatorname{norm}[(1-W)\mathbf{u}_x+
W\widehat{\mathbf{u}}]$, and the phase is recovered using
$\phi_{\mathrm{out}}=\operatorname{atan2}(u_{\mathrm{out},2},
u_{\mathrm{out},1})$. Thus, the raised cosine mask is a parameter-free integration, which ensures that both log-magnitude prediction and phase-vector interpolation are reliably copied from the
input, 
gradually blended near the boundary, and replaced by the prediction only in the missing band.

Next, the reconstructed wideband magnitude and phase are combied by inverse-STFT and supplied to discriminators for adversarial feedback. 

\begin{figure*}[t]
\centering
\includegraphics[width=0.88\textwidth,trim=0 45 0 0,clip]{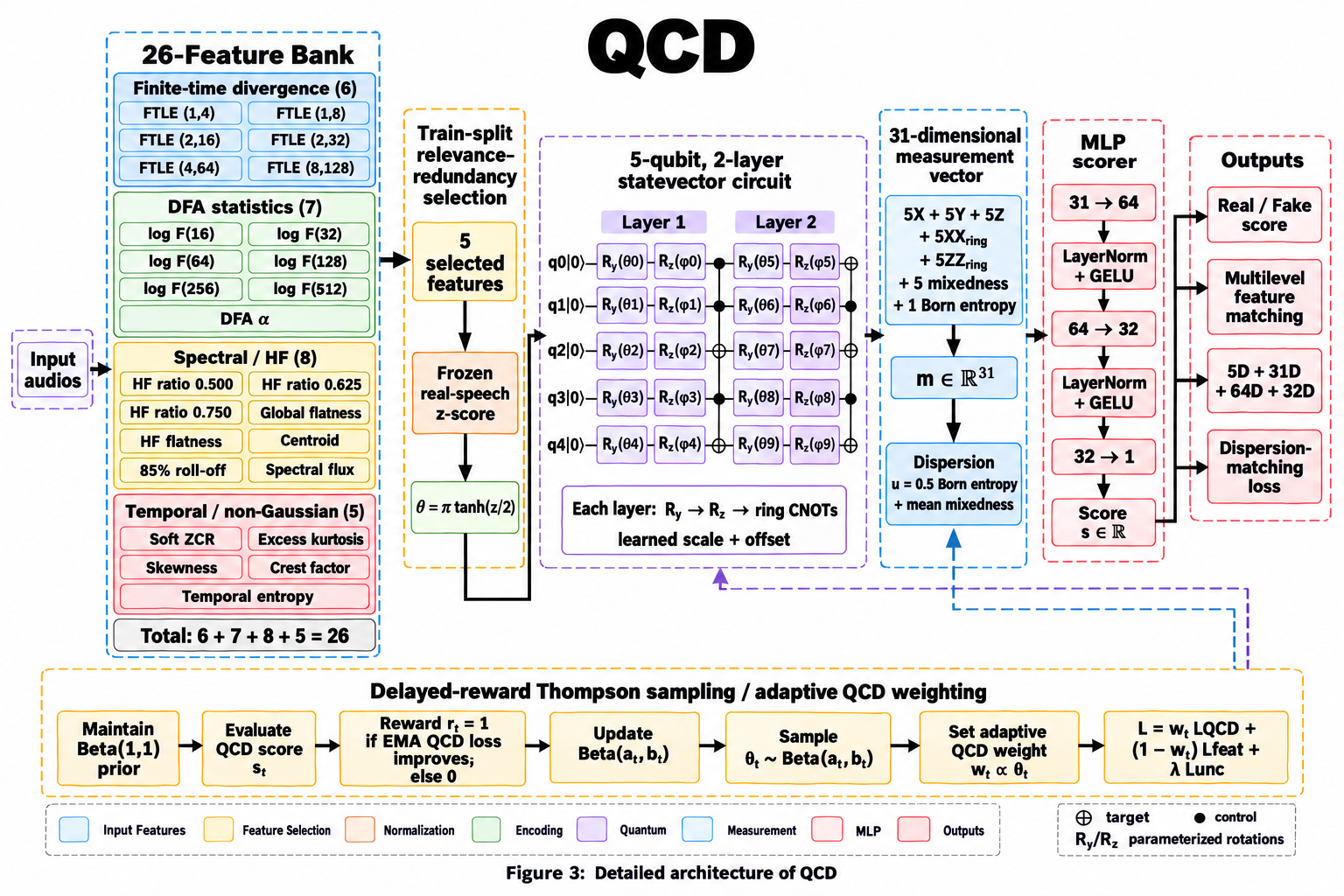}
\vspace{-0.7em}
\caption{Overall QCD architecture: feature-bank selection, normalization, bounded angle encoding, simulated statevector map, and the scorer and matching heads.}
\vspace{-0.5em}
\label{fig:qcd_architecture}
\vspace{-0.9em}
\end{figure*}

\vspace{-0.4em}
\subsection{QCD Architecture (Appendix ~\ref{app:featurebank}, ~\ref{app:math}, ~\ref{app:qcd_architecture})}
\label{sec:qcd}

Alg.~\ref{alg:qcd_forward} \& Fig.~\ref{fig:qcd_architecture} summarize the QCD architecture. 

\vspace{-0.4em}
\subsubsection{Training-Only Feature Selection} 
\vspace{-0.2em}
\label{sec:featurebank}

Although recent Multi-Resolution Amplitude Discriminator (\mrad{}) and Multi-Resolution Phase Discriminator (\mrpd{}) \cite{lu2024towards} evaluate magnitude and phase at multiple resolutions, they do not explicitly characterize  nonlinear dynamical properties \cite{tamiti2026cis}. 
To solve this problem, QCD is designed to provide a small set of automatically selected, complementary nonlinear features for additional training cues, which are weakly represented by spectral MRAD and MRPD.


First, we construct a differentiable 26-feature bank
$c(z)\in\mathbb{R}^{26}$ with wide range of chaos, 
spectral, and non-Gaussian features (Alg.\ref{alg:qcd_forward}, \textbf{Row \textcircled{\scriptsize 2}}). 
The bank has: (i) 6 finite-time Lyapunov-divergence 
to describe short- to long-horizon sensitivity in waveform; (ii) 7 DFA statistics to describe multiscale temporal persistence and long-range fractal correlation; (iii) 8 spectral and
HF statistics to measure missing-band energy, spectral shape, centroid, rolloff, flatness, and frame-to-frame change; and (iv) 5 temporal or non-Gaussian statistics to describe oscillation rate, impulsiveness, asymmetry, peak structure, and amplitude entropy. 

\vspace{-0.1em}
From the 26 features, only 5 features are selected as the default QCD contains five qubits, with one selected feature encoded per qubit. The selection is performed automatically on the training split for each declared bandwidth, noise, or codec regime. 
Different settings may produce different five-feature subsets. Once selected, however, the features and calibration statistics are frozen and reused across all experimental settings.

\vspace{-0.1em}
\textbf{5-feature Subset Selection:}
Each candidate is first scored by how strongly it distinguishes real wideband speech from the interpolated narrowband speech. For candidate
$j$, the relevance score is 
$ r_j=
\frac{|\mu_j^r-\mu_j^b|}
{\sqrt{\bigl[(\sigma_j^r)^2+(\sigma_j^b)^2\bigr]/2}+\epsilon},
$
where $(\mu_j^r,\sigma_j^r)$ and $(\mu_j^b,\sigma_j^b)$ are training-set moments for real 
speech and the corresponding degraded baseline. A large $r_j$ indicates that degradation removes or distorts information captured by that feature and that matching may provide a useful BWE training cues. However, relevance score alone can select several features that measure nearly the same property. Therefore, we use a redundancy penalty $\lambda_{\mathrm{red}}=0.35$. Starting from a void selection set, $S=\varnothing$, the selector adds candidate $j^\ast$ to set $S$ where
$
j^\ast=
\arg\max_{j\notin S}
[
r_j-\lambda_{\mathrm{red}}r_{\max}
\max_{k\in S}|\rho_{jk}|
].
$
Here $r_{\max}=\max_j r_j$, $\rho_{jk}$ is the absolute training-set correlation between candidates $j$ and $k$. For the first selection, the redundancy penalty is zero. 
The selector balances separation against similarity to features already exist in $S$, and the procedure stops when $|S|=5$. This may be viewed as a deterministic \emph{exploitation--exploration trade-off}. The relevance term $r_j$ exploits the strongest relevance score, while the redundancy penalty encourages coverage of complementary feature categories. It is not the Thompson-sampling policy used later to adapt the QCD loss weight.

\vspace{-0.1em}
\textbf{An Example:} In noisy 8-to-16~kHz settings, the automatic procedure selects high-frequency spectral flatness, global spectral flatness, short-horizon finite-time Lyapunov log-divergence, DFA log fluctuation at scale 32, and waveform-amplitude skewness. This mixture spans spectral texture, nonlinear short-term evolution, multiscale temporal structure, and non-Gaussian waveform shape rather than relying on five variants of one descriptor.

\vspace{-0.1em}
\textbf{Feature Normalization:} The selected features can have substantially different physical scales, so their raw values cannot be passed directly to bounded angle encoding as large-range features would dominate the rotations. 
For each selected feature $i$, original training speech defines the fixed reference statistics $(\mu_i,\sigma_i)$, and both real and generated samples are normalized as $\widetilde c_i=(c_i-\mu_i)/(\sigma_i+\epsilon)$; (\textbf{Rows \textcircled{\scriptsize 4}--\textcircled{\scriptsize 7}}). The normalized value is then bounded 
(\textbf{Row \textcircled{\scriptsize 7}}), preventing outliers 
from producing uncontrolled rotations (see Appendix ~\ref{app:featurebank}). 

\vspace{-0.32em}
\subsubsection{Quantum-inspired 5-Qubit Feature Map}
\vspace{-0.1em}
\label{sec:circuit}

The 5 selected QCD features describe complementary spectral, temporal, and nonlinear-dynamical properties, but treating them independently would ignore interactions between these properties. For example, a spectral flatness change may have different significance in a periodic signal than in a transient or highly irregular signal. A conventional large neural network could learn such interactions, but it is 
difficult to determine whether an improvement arises from the representation or simply 
from additional parameters. We 
hypothesize that a small quantum-inspired data-reuploading map can transform the selected five features into a richer representation containing individual-feature, cross-feature, and state-dispersion information using only 40 trainable parameters. This hypothesis is tested against a parameter-matched classical map. 

Each normalized feature $\widetilde c_i$ is first encoded by bounded rotation as
$\theta_i=\pi\tanh(\widetilde c_i/2)$ (\textbf{Row \textcircled{\scriptsize 7}}). 
This smooth mapping preserves ordering around the original-speech reference, bounds every angle to $[-\pi,\pi]$, and suppresses extreme standardized values without the discontinuous gradients introduced by hard clipping. Each of the five resulting angles is assigned to one qubit, maintaining a direct correspondence between the automatically selected features and the five-qubit input.

{
\setlength{\textfloatsep}{0pt plus 0pt minus 0pt}   
\begin{algorithm}[t]
\caption{QCD forward pass}
\label{alg:qcd_forward}
\small
\begin{algorithmic}[1]
\Require Real batch $y$, generated batch $\hat y$, selected set $S$ with $|S|=5$, real-training moments $(\mu,\sigma)$
\Ensure Scores $s^{r/g}$, feature maps $\mathcal H_q^{r/g}$, dispersion statistics $u^{r/g}$ (Here, $r$ = real and $g$ = generated)
\State $Z\gets\operatorname{concat}(y,\hat y)$
\State $C\gets\operatorname{FeatureBank}_{26}(Z)$; $c\gets C[:,S]$
\State Split $c$ into $(c^r,c^g)$ at the real/generated batch boundary
\If{statistics calibration is active}
    \State Update $(\mu,\sigma)$ from $\operatorname{sg}(c^r)$ only
    \State Synchronize sufficient statistics across workers
\EndIf
\State $\tilde c\gets(c-\mu)/(\sigma+\epsilon)$; $\theta\gets\pi\tanh(\tilde c/2)$
\State Initialize five qubits $|\psi\rangle\gets|0\rangle^{\otimes5}$ for every row of $\theta$
\For{$\ell\in\{1,2\}$} (i.e., two layer data re-uploading)
    \For{$i\in\{1,\ldots,5\}$}
        \State Apply $R_y(a^y_{\ell i}\theta_i+b^y_{\ell i})$
        \State Apply $R_z(a^z_{\ell i}\theta_i+b^z_{\ell i})$
    \EndFor
    \State Apply the forward CNOT ring if $\ell=1$; otherwise apply the reverse ring
\EndFor
\State Generate measurement map, $m=[\{X_i,Y_i,Z_i,X_iX_{i+1},Z_iZ_{i+1},\xi_i\}_{i=1}^{5},H_B]$
\State $u\gets\tfrac12H_B+\tfrac15\sum_{i=1}^{5}\xi_i$
\State $h_1\gets\operatorname{GELU}(\operatorname{LN}(W_1m+b_1))$
\State $h_2\gets\operatorname{GELU}(\operatorname{LN}(W_2h_1+b_2))$
\State $s\gets W_oh_2+b_o$; $\mathcal H_q\gets\{\tilde c,m,h_1,h_2\}$
\State Split $(s,\mathcal H_q,u)$ into real and generated outputs
\State \Return $(s^r,s^g)$, $(\mathcal H_q^r,\mathcal H_q^g)$, and $(u^r,u^g)$
\end{algorithmic}
\end{algorithm}
}

\enlargethispage{-\baselineskip}


The quantum-inspired feature map is implemented by exact differentiable statevector simulation and comprises state initialization, two-layer data re-uploading, entanglement, and measurement
(\textbf{Rows \textcircled{\scriptsize 8}--\textcircled{\scriptsize 14}}). Starting from $|0\rangle^{\otimes5}$
(\textbf{Row \textcircled{\scriptsize 8}}), two data re-uploading circuit layers iterate over the five qubits
(\textbf{Rows \textcircled{\scriptsize 9}--\textcircled{\scriptsize 10}})
and re-upload the feature angles through learned $R_y$ and $R_z$ rotations through two rotation axes, Y and Z
(\textbf{Rows \textcircled{\scriptsize 11}--\textcircled{\scriptsize 12}}).


For feature angle $\theta_i$, qubit $i$, and layer $\ell$, the local transformation is $U_{\ell i}(\theta_i)=R_z(a^z_{\ell i}\theta_i+b^z_{\ell i})
R_y(a^y_{\ell i}\theta_i+b^y_{\ell i})$. The coefficients
$a^y_{\ell i}$ and $a^z_{\ell i}$ control feature sensitivity, while
$b^y_{\ell i}$ and $b^z_{\ell i}$ provide learned offsets. With two layers, five qubits, two rotation axes, and a scale and offset for each rotation, the feature map contains $2\times5\times2\times2=40$ trainable parameters.

After the local rotations, the first circuit layer applies the forward CNOT ring
$\{(0,1),(1,2),(2,3),(3,4),(4,0)\}$, and the second circuit layer applies the reverse ring
$\{(4,3),(3,2),(2,1),(1,0),(0,4)\}$ (\textbf{Row \textcircled{\scriptsize 13}}).
Alternating the ring direction makes every qubit act as both a control and a target, allowing the representation of one selected feature to depend on the others. This provides a compact mechanism for modeling cross-feature relationships rather than processing the five features independently.
For five qubits, the final state
$|\psi\rangle=\sum_{s=0}^{31}\alpha_s|s\rangle$
contains 32 complex magnitudes. Here, $\alpha_s\in\mathbb{C}$ is the probability magnitude associated with computational-basis state $|s\rangle$, while
$p_s=|\alpha_s|^2$ is its corresponding measurement probability, satisfying
$\sum_{s=0}^{31}p_s=1$. From this state, QCD computes the measurement map
$m=[X,Y,Z,XX_{\mathrm{ring}},ZZ_{\mathrm{ring}},\xi,H_B]
\in\mathbb{R}^{31}$
(\textbf{Row \textcircled{\scriptsize 14}}). The map contains 15 single-qubit expectations
$\{\langle X_i\rangle,\langle Y_i\rangle,\langle Z_i\rangle\}_{i=1}^{5}$,
which describe the transformed response of each feature; ten ring
correlations
$\{\langle X_iX_{i+1}\rangle,\langle Z_iZ_{i+1}\rangle\}_{i=1}^{5}$,
which describe learned pairwise interactions; and five reduced-state
mixedness values
$\xi_i=1-\langle X_i\rangle^2-\langle Y_i\rangle^2-\langle Z_i\rangle^2$,
which summarize how strongly each qubit is coupled to the remaining state.
The final measurement is the normalized computational-basis Born entropy
$H_B=-[\log 32]^{-1}\sum_{s=0}^{31}p_s\log(p_s+\epsilon)$. 
The total dimension is 
$3(5)+2(5)+5+1=31$. The 31 measurements expose three complementary representation levels: single-feature responses, learned inter-feature correlations, and global or local state dispersion. 
The mixedness and entropy terms are combined into
$u=\tfrac{1}{2}H_B+\tfrac{1}{5}\sum_{i=1}^{5}\xi_i$
(\textbf{Row \textcircled{\scriptsize 15}}). Here, $u$ is a \emph{dispersion statistic}: it summarizes the spread of the simulated state across basis configurations and 
qubit states. 
QCD's state and computational costs scale exponentially with the number of qubits, although the five-qubit state used in \method{}  contains only 32 magnitudes (i.e., $2^{5}$) and remains inexpensive relative to the generator and feature bank. See Appendix~\ref{app:math} for details. 

\vspace{-0.32em}
\subsubsection{Scoring and Auxiliary Supervision}
\vspace{-0.1em}
\label{sec:qcd_supervision}

The 31-dimensional measurement map feeds a $31\!\rightarrow\!64\!\rightarrow\!32\!\rightarrow\!1$ scorer with LayerNorm and GELU in its hidden layers (\textbf{Rows \textcircled{\scriptsize 16}-\textcircled{\scriptsize 18}}). Hinge adversarial supervision is complemented by feature matching at four levels: the five normalized features, quantum circuit measurements, and two scorer hidden states. A separate
term matches the deterministic state-dispersion statistic $u$; it does not represent calibrated predictive uncertainty. Together, these objectives supervise both  final decision and  intermediate speech structure. The resulting scores, feature maps, and dispersion statistics are separated into real and generated outputs and returned to the training objectives (\textbf{Rows \textcircled{\scriptsize 19}--\textcircled{\scriptsize 20}}). During generator updates, critic parameters and original-speech targets are frozen, while gradients propagate through the generated branch. QCD has 4,393 trainable parameters and is removed at inference. Appendix~\ref{app:qcd_supervision} gives details. 

\vspace{-0.2em}
\subsubsection{Delayed-reward Thompson Weighting}
\vspace{-0.0em}
\label{sec:thompson_main}

A fixed QCD weight may underuse an informative critic later in training or place excessive emphasis on immature features early in training. Gradual activation and online adaptation address this nonstationarity.  To accommodate changes in critic usefulness during training, the QCD influence is gradually activated and original-speech statistics are frozen after 500 calibration steps. Delayed-reward Thompson sampling \cite{thompson1933likelihood} then selects the discriminator-side QCD weight from $\{0.05,0.10,0.25,0.50,1.00\}$ using Beta posteriors. An applied arm is credited only at the next raw QCD-loss observation, according to whether that loss falls below its preceding exponential-moving-average baseline. This ordering avoids rewarding an action with a loss measured before its critic update; skipped optimizer steps create no pending action. Generator-side QCD coefficients remain fixed, and the policy is discarded at inference. The reward is a training heuristic, so benefits must be assessed on held-out BWE metrics. Appendix~\ref{app:thompson} details the policy and credit assignment, Appendix~\ref{app:warmup} gives the schedules, and Algorithm~\ref{alg:qcis_training} specifies the alternating updates.

\vspace{-0.4em}
\subsection{Loss Functions and Training Objectives}
\label{sec:losses}
\vspace{-0.2em}

The total generator loss is
$
\mathcal{L}_{G} =
45\mathcal{L}_{\mathrm{mag}}
+30\mathcal{L}_{\mathrm{pha}}
+90\mathcal{L}_{\mathrm{com}}
+10\mathcal{L}_{\mathrm{stft}}
+20\mathcal{L}_{\mathrm{wav}}
+10\mathcal{L}_{\mathrm{sc}}
+5\mathcal{L}_{\mathrm{flux}}
+\mathcal{L}_{\mathrm{FM}}
+\mathcal{L}_{\mathrm{adv}}
+0.25\mathcal{L}_{\mathrm{QCD}}^{\mathrm{adv}}
+0.10\mathcal{L}_{\mathrm{QCD}}^{\mathrm{FM}}
+0.05\mathcal{L}_{\mathrm{QCD}}^{\mathrm{disp}},
$
where $\mathcal{L}_{\mathrm{mag}}$, $\mathcal{L}_{\mathrm{pha}}$, $\mathcal{L}_{\mathrm{com}}$, $\mathcal{L}_{\mathrm{stft}}$, $\mathcal{L}_{\mathrm{wav}}$, $\mathcal{L}_{\mathrm{sc}}$, and $\mathcal{L}_{\mathrm{flux}}$ denote the masked log-magnitude, phase, complex-spectrum, STFT-consistency, waveform, high-band spectral-convergence, and spectral-flux reconstruction losses, respectively. $\mathcal{L}_{\mathrm{FM}}$ and $\mathcal{L}_{\mathrm{adv}}$ are the feature-matching and hinge adversarial losses from the classical critics, while $\mathcal{L}_{\mathrm{QCD}}^{\mathrm{adv}}$, $\mathcal{L}_{\mathrm{QCD}}^{\mathrm{FM}}$, and $\mathcal{L}_{\mathrm{QCD}}^{\mathrm{disp}}$ are the QCD hinge, multilevel feature-matching, and dispersion-matching objectives. The generator-side QCD coefficients are fixed at $(0.25,0.10,0.05)$ throughout training.

The total discriminator loss is
$
\mathcal{L}_{D}
=
\sum_{d\in\mathcal{D}_{\mathrm{classical}}}
\mathcal{L}_{D}^{(d)}
+
w_{d}^{\mathrm{QCD}}\mathcal{L}_{\mathrm{QCD}}^{D},
$
where $\mathcal{L}_{D}^{(d)}$ is the hinge loss of classical critic $d$=\{MRAD,MRPD\}, $\mathcal{L}_{\mathrm{QCD}}^{D}$ is the QCD discriminator-side hinge objective, and $w_{d}^{\mathrm{QCD}}$ is the coefficient adjusted by the adaptive weighting policy. Appendix~\ref{app:repro} provides the complete loss definitions 
and training details.

\vspace{-0.42em}
\section{Comprehensive Analysis}
\label{sec:experiments}
\label{sec:results}
\vspace{-0.42em}

\subsection{Metrics, Hyperparameters, and Datasets}
\label{subsec:EvaluationMetrics}
\vspace{-0.20em}

We evaluate reconstruction quality using LSD, STOI, PESQ, SI-SDR, SI-SNR, NISQA-MOS, and WER. We aslo devise HF and ITU-inspired metrics. See Appendix~\ref{app:data_protocol}, ~\ref{app:comparative}, and~\ref{app:completed_experiments} for details.


\textbf{Hyperparameters:} We train for 50 epochs using AdamW with an effective batch size of 16. The generator and discriminator learning rates are initialized to $2\times10^{-4}$ and $3.5\times10^{-4}$, respectively. 
Training is performed using an NVIDIA RTX 4090 GPU and an AMD Ryzen 9 7950X3D CPU with seed 1234. 
Details are provided in Appendix~\ref{app:repro}.

\textbf{Datasets:} We use the English VCTK v0.92 corpus~\cite{yamagishi2019cstr}, containing 110 speakers, and a 35-hour subset of the MLS French corpus~\cite{Pratap2020MLSAL}, containing 114 speakers, to evaluate multilingual and cross-speaker generalization. Together, the two corpora 
contain  $\sim$ 75 hours of speech, with source sampling rates of 16 and 48~kHz. 
Dataset construction, preprocessing, 
split, and evaluation 
are described in Appendix~\ref{app:data_protocol}.

\vspace{-0.5em}
\subsection{Comparison with Baseline Methods}
\vspace{-0.2em}
\label{subsec:baseline_comparison}

Table~\ref{tab:Comparative_analysis} compares \method{} with EBEN, AP-BWE, and \cis{} on clean English VCTK and French MLS speech. 
Since PESQ supports only 8 and 16~kHz sampling rates, for the 16$\rightarrow$48~kHz setting, we downsample both reconstructed and reference signals to 16~kHz before computing PESQ.

\begin{table*}[t]
\centering
\setlength{\tabcolsep}{0.85pt}
\renewcommand{\arraystretch}{0.99}
\resizebox{\textwidth}{!}{%
\begin{tabular}{ l l l ccc ccc ccc ccc ccc ccc ccc}
\toprule

\multirow{2}{*}{Method}
& \multirow{2}{*}{Size}
& \multirow{2}{*}{Data}
& \multicolumn{3}{c}{NISQA-MOS$\uparrow$}
& \multicolumn{3}{c}{STOI$\uparrow$}
& \multicolumn{3}{c}{PESQ$\uparrow$}
& \multicolumn{3}{c}{SI-SDR$\uparrow$}
& \multicolumn{3}{c}{SI-SNR$\uparrow$}
& \multicolumn{3}{c}{LSD$\downarrow$}
& \multicolumn{3}{c}{WER \%$\downarrow$}\\
\cmidrule(lr){4-6}
\cmidrule(lr){7-9}
\cmidrule(lr){10-12}
\cmidrule(lr){13-15}
\cmidrule(lr){16-18}
\cmidrule(lr){19-21}
\cmidrule(lr){22-24}
 & &
& 4--16 & 8--16 & 16--48
& 4--16 & 8--16 & 16--48
& 4--16 & 8--16 & 16--48
& 4--16 & 8--16 & 16--48
& 4--16 & 8--16 & 16--48
& 4--16 & 8--16 & 16--48
& 4--16 & 8--16 & 16--48\\
\midrule

\multirow{2}{*}{Unprocessed}
& \multirow{2}{*}{--}
& VCTK
& 2.79 & 3.67 & 4.43
& 0.55 & 0.61 & 0.61
& 1.15 & 1.51 & 1.41
& -11.03 & -8.07 & -6.07
& -10.53 & -7.62 & -5.63
& 3.27 & 2.27 & 2.85
& 90.0 & 6.1 & 2.1\\
 & & MLS
& 2.29 & 3.12 & --
& 0.51 & 0.57 & --
& 1.05 & 1.34 & --
& -14.15 & -18.89 & --
& -13.67 & -17.71 & --
& 3.48 & 2.48 & --
& 91.1 & 6.8 & --\\
\hline

 \multirow{2}{*}{EBEN, 2023}
& \multirow{2}{*}{29.7M}
& VCTK
& 2.59 & 2.69 & 2.53
& 0.89 & 0.98 & 0.98
& 2.64 & 3.69 & 3.71
& 11.94 & \textbf{19.94} & 20.82
& 11.94 & \textbf{19.94} & 20.83
& 1.03 & 0.78 & 0.92
& 19.1 & 12.4 & 8.5\\
 & & MLS
& 2.14 & 2.28 & --
& 0.87 & 0.96 & --
& 2.47 & 3.55 & --
& 11.81 & \textbf{18.24} & --
& 11.74 & \textbf{18.38} & --
& 1.17 & 0.88 & --
& 19.4 & 12.7 & --\\
\hline


 \multirow{2}{*}{AP-BWE, 2024}
& \multirow{2}{*}{72M}
& VCTK
& 3.86 & 3.97 & 4.49
& 0.94 & 0.99 & 0.99
& 2.55 & 3.69 & 3.72
& \textbf{13.42} & 18.26 & 20.86
& \textbf{13.35} & 18.07 & 20.74
& 0.96 & 0.74 & 0.75
& 13.7 & 4.2 & 1.7\\
 & & MLS
& 3.18 & 3.42 & --
& 0.92 & \textbf{0.99} & --
& 2.51 & 3.62 & --
& \textbf{13.57} & 17.39 & --
& \textbf{13.17} & 16.33 & --
& 1.11 & 0.84 & --
& 13.9 & 4.2 & --\\
\hline

\multirow{2}{*}{CIS-BWE, 2026}
& \multirow{2}{*}{33.5M}
& VCTK
& \textbf{4.24} & 4.26 & 4.53
& \textbf{0.95} & 0.99 & 0.99
& 2.64 & 3.72 & 3.75
& 13.24 & 18.13 & 19.53
& 13.15 & 17.98 & 19.44
& 0.95 & 0.72 & 0.71
& 13.5 & 3.9 & 1.5\\
 & & MLS
& \textbf{3.47} & \textbf{3.91} & --
& 0.92 & 0.99 & --
& 2.58 & \textbf{3.71} & --
& 13.29 & 17.68 & --
& 12.86 & 16.81 & --
& 1.10 & 0.82 & --
& 13.8 & \textbf{4.1} & --\\
\hline

\hline

 \multirow{2}{*}{\method{} (ours)}
& \multirow{2}{*}{19.4M}
& VCTK
& 4.21 & \textbf{4.36} & \textbf{4.53}
& \textbf{0.95} & \textbf{0.99} & \textbf{1.00}
& \textbf{2.76} & \textbf{3.93} & \textbf{4.62}
& 13.28 & 19.50 & \textbf{21.84}
& 13.28 & 19.50 & \textbf{21.84}
& \textbf{0.95} & \textbf{0.72} & \textbf{0.71}
& \textbf{13.47} & \textbf{3.81} & \textbf{1.13}\\
& & MLS
& 3.41 & 3.58 & --
& \textbf{0.93} & \textbf{0.99} & --
& \textbf{2.58} & 3.67 & --
& 12.01 & 15.47 & --
& 11.79 & 15.27 & --
& \textbf{1.09} & \textbf{0.81} & --
& \textbf{13.78} & 5.5 & --\\
\bottomrule
\vspace{-2.2em}
\end{tabular}%
\vspace{-1.2em}
}
\caption{Comparison with baseline on clean English VCTK and French MLS. 
MLS does not
contain 48~kHz speech.}
\label{tab:Comparative_analysis}
\vspace{-1.2em}
\end{table*}

Please note that LSD is a measure for over-smoothing, and NISQA-MOS, PESQ, and STOI are measures of perceptual quality. Table~\ref{tab:Comparative_analysis} shows that Q-BWE has the highest NISQA-MOS for 8-16 kHz and similar to CIS-BWE at 16-48 kHz. The STOI is better/similar compared to the most recent baseline CIS-BWE. The PESQ and WER are best in most ranges. The SI-SDR \& SI-SNR is best in 16-48 kHz but slightly lower in other ranges. The LSD is better/similar to CIS-BWE. 

Most notably, Q-BWE achieves better results in most over-smoothing (i.e., LSD) and perceptual results (i.e., NISQA, PESQ, STOI) with slightly lower SI-SDR/SI-SNR with \textbf{$\sim$1.72x fewer parameters} compared to CIS-BWE. This also holds for the MLS French dataset, indicating CIS-BWE outperforms most baselines in \textbf{multilingual and multi-speaker settings} with the lowest parameters.

\begin{figure}[t]
  \centering
  \includegraphics[width=0.49\textwidth,height=0.16\textheight]{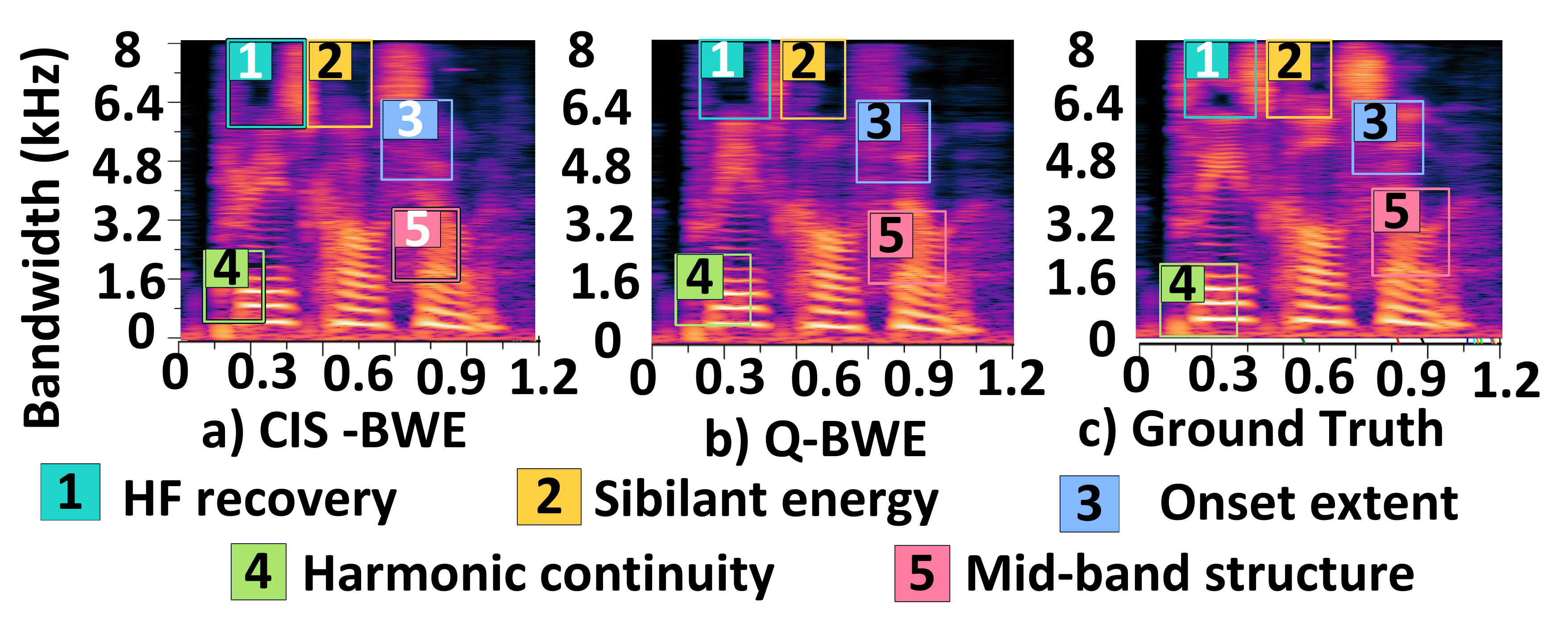}
  \vspace{-2.0em}
  \caption{Qualitative comparison at 4--16~kHz.}
  \label{fig:figcomparison}
  \vspace{-1.5em}
\end{figure}


\noindent\textbf{Qualitative observation:}
Fig.~\ref{fig:figcomparison} shows that, compared to CIS-BWE, Q-BWE more closely reproduces the highlighted acoustic structures, including \textbf{HF recovery} and \textbf{sibilant energy} in the 6.4--8~kHz region, the \textbf{extent of transient onsets} around 4.8--6.4~kHz, as well as \textbf{harmonic continuity} and \textbf{mid-band spectral structure}. 
These observations are consistent with the role of the QCD, which jointly models nonlinear, fractal, spectral, and temporal descriptors through its quantum-inspired measurement representation. 

\vspace{-0.42em}
\subsection{Ablation on Generator and Discriminator}
\label{subsec:controlled_protocol}
\vspace{-0.20em}

We conduct ablations to separate the effects of QCD supervision,
FiLM conditioning, raised cosine mask, and quantum circuit structure in Table~\ref{tab:one_seed_results}. 

In the generator side, if we have no FiLM conditioning (Row 1) and raised cosine mask (Row 2), the overall performance degrades across NISQA-MOS, STOI, PESQ, SI-SNR, and LSD compared to Q-BWE (Row 9) where we have both FiLM and mask (except NISQA-MOS for no mask). These findings support the use of mask and FiLM conditioning for the task at hand.

In the discriminator side, if we remove QCD and keep only MRAD \& MRPD (Row 3), the overall performance degrades compared to Q-BWE (i.e., Row 9 with QCD), proving the importance of our newly designed quantum discriminator QCD. Row 4 also indicates that removing FiLM from the generator and QCD from the discriminator degrades performance compared to Q-BWE (Row 9).

Row 5 removes entanglement, Row 6 removes $XX_{\mathrm{ring}}$, $ZZ_{\mathrm{ring}}$, $\xi$, and $H_B$ from the measurement map, Row 7 removes dispersion $u$, and Row 8 only uses one layer data re-uploading circuit. All these combinations provide degrading results in most of the metrics whereas our full QCD in Q-BWE (Row 9) provides overall better performance.

\begin{table}[t]
\centering
\setlength{\tabcolsep}{1pt}
\renewcommand{\arraystretch}{0.91}
\resizebox{\columnwidth}{!}{%
\begin{tabular}{@{}clccccc@{}}
\toprule
Row & System  & N-MOS$\uparrow$ & STOI$\uparrow$ & PESQ$\uparrow$
& SI-SNR$\uparrow$ & LSD$\downarrow$ \\
\midrule
\textbf{\textcircled{\scriptsize 1}}  & No FiLM         & 4.15 & 0.95 & 2.83 & 13.09 & 1.03 \\

\textbf{\textcircled{\scriptsize 2}} & No-Mask    & 4.39 & 0.95 & 2.72 & 12.85 & 0.97 \\
\midrule

\textbf{\textcircled{\scriptsize 3}}  & No QCD (only MRAD+MRPD)        & 4.15 & 0.95 & 2.85 & 13.15 & 1.01 \\

\textbf{\textcircled{\scriptsize 4}}  & No FiLM \& QCD          & 4.20 & 0.95 & 2.81 & 13.05 & 0.98 \\

\midrule

\textbf{\textcircled{\scriptsize 5}}  & No Entanglement.     & 4.27 & 0.95 & 2.76 & 13.05 & 0.99 \\
\textbf{\textcircled{\scriptsize 6}} & XYZ-only   & 4.2 & 0.95 & 2.83 & 13.14 & 0.98 \\

\textbf{\textcircled{\scriptsize 7}} & No Dispersion    & 4.28 & 0.95 & 2.76 & 12.79 & 0.98 \\
\textbf{\textcircled{\scriptsize 8}} & One-Layer re-uploading   & 4.12 & 0.95 & 2.86 & 13.41 & 1.01 \\



\midrule
\textbf{\textcircled{\scriptsize 9}}  & Q-BWE      & 4.21 & 0.95 & 2.84 & 13.28 & 0.95 \\

\bottomrule
\end{tabular}%
}

\vspace{-0.5em}
\caption{Ablation study on VCTK at 4$\rightarrow$16\,kHz.} 
\label{tab:one_seed_results}
\vspace{-0.20em}
\end{table}

\vspace{-0.42em}
\subsection{Study on Qubit States}
\label{subsec:Study on Seeds and Qubit States}
\vspace{-0.20em}

Table \ref{tab:seedandqubit} indicates that 5 qubit gives overall better performance compared to 7 qubit states.

\begin{table}[!tbp]
\centering
\small
\setlength{\tabcolsep}{2.3pt}
\begin{tabular}{clrrrr}
\toprule
SL & QCD inputs  & STOI$\uparrow$ & PESQ$\uparrow$ &
SI-SNR$\uparrow$ & N-MOS$\uparrow$\\
\midrule

\textcircled{\textbf{\scriptsize 1}} & 5 features  & $\boldsymbol{0.9478}$ & $\boldsymbol{2.735}$ &
$\boldsymbol{13.27}$ & $3.99$ \\

\textcircled{\textbf{\scriptsize 2}} & 7 features  & $0.9462$ & $2.731$ &
$13.11$ & $\boldsymbol{4.06}$ \\
\bottomrule
\end{tabular}
\vspace{-0.70em}
\caption{5 and 7 qubit ablation for 4$\rightarrow$16~kHz BWE.}
\label{tab:seedandqubit}
\vspace{-01.50em}
\end{table}

\vspace{-0.60em}
\subsection{Evaluation Under Noisy Conditions}
\label{subsec:noisy_conditions}
\vspace{-0.30em}

We use eight noise sources from the AURORA dataset \cite{pearce00_icslp} and
five SNR levels ($-10$, $-5$, $0$, $5$, and $10$~dB). The selected noises cover both stationary and non-stationary conditions, including airport, babble, car, exhibition, station, street, speech-shaped noise, and white Gaussian noise. Table~\ref{tab:noiseComparative_analysis} reports the average performance over all eight noise sources and all five SNR levels. For Q-BWE, we consider two settings: \textit{zero-shot} (ZS),
where the clean-trained model is directly evaluated on noisy speech, and
\textit{matched}, where a separate model is trained using the declared noise
mixture. Other baselines use matched settings. Q-BWE  remains competitive under noisy conditions while using
roughly half the model parameters. 
See
Appendix~\ref{app:robustness} for detailed  results.

\begin{table}[t]
\centering
\scriptsize
\setlength{\tabcolsep}{3.0pt}
\renewcommand{\arraystretch}{0.95}
\resizebox{\columnwidth}{!}{%
\begin{tabular}{clccccc}
\toprule
SL & Metric
& EBEN
& AERO
& AP-BWE
& Q-BWE ZS
& Q-BWE Matched \\
\midrule

\textbf{\textcircled{\scriptsize 1}}
& NISQA-MOS$\uparrow$
& 1.01
& 1.52
& \textbf{2.74}
& 1.466
& 1.854 \\

\textbf{\textcircled{\scriptsize 2}}
& SI-SNR$\uparrow$
& 4.23
& 4.39
& 4.36
& 5.459
& \textbf{5.557} \\

\textbf{\textcircled{\scriptsize 3}}
& LSD$\downarrow$
& 1.41
& 1.43
& 1.37
& 1.592
& \textbf{1.367} \\

\bottomrule
\end{tabular}%
}
\vspace{-0.82em}
\caption{Performance on noisy VCTK for 4$\rightarrow$16~kHz.}
\label{tab:noiseComparative_analysis}
\vspace{-0.80em}
\end{table}


\vspace{-0.4em}
\subsection{Robustness to Channel Degradation}
\label{subsec:channel_robustness}
\vspace{-0.1em}

Table \ref{tab:channel_main} shows that Q-BWE remains robust under different telephony codecs and packet loss. Codec tests use
G.711 $\mu$/A-law and AMR-NB at 4.75/12.2~kb/s, comparing zero-shot (ZS)
and matched (M) training. Packet-loss tests erase independent 20-ms frames at
3\%/5\% and apply repeat (R) or zero (Z) concealment. Full codec, packet-loss, and SNR results are provided in Appendix~\ref{app:robustness}.

\begin{table}[t]
\centering
\scriptsize
\setlength{\tabcolsep}{2.2pt}
\renewcommand{\arraystretch}{0.94}
\resizebox{\columnwidth}{!}{%
\begin{tabular}{cllrrrrr}
\toprule
SL & Condition & Pair & LSD$\downarrow$ & STOI$\uparrow$ &
PESQ$\uparrow$ & SI-SNR$\uparrow$ & N-MOS$\uparrow$ \\
\midrule
\textbf{\textcircled{\scriptsize 1}}
& G.711 A-law & ZS/M
& 0.98/0.99 & 0.94/0.94 & 2.62/2.72 & 13.05/13.04 & 4.12/4.24 \\

\textbf{\textcircled{\scriptsize 2}}
& G.711 $\mu$-law & ZS/M
& 0.98/0.99 & 0.94/0.95 & 2.68/2.75 & 13.03/13.024 & 4.14/4.24 \\

\textbf{\textcircled{\scriptsize 3}}
& AMR-NB 4.75 & ZS/M
& 1.36/1.13 & 0.82/0.83 & 1.85/1.80 & -16.88/-16.95 & 2.88/3.46 \\

\textbf{\textcircled{\scriptsize 4}}
& AMR-NB 12.2 & ZS/M
& 1.17/1.08 & 0.87/0.88 & 2.41/2.33 & -16.88/-16.89 & 3.88/4.13 \\
\midrule
\textbf{\textcircled{\scriptsize 5}}
& Loss 3\% & R/Z
& 0.99/0.99 & 0.93/0.92 & 2.26/2.14 & 9.91/10.97 & 3.85/2.75 \\

\textbf{\textcircled{\scriptsize 6}}
& Loss 5\% & R/Z
& 1.01/1.01 & 0.92/0.91 & 2.05/1.89 & 8.45/9.94 & 3.59/2.26 \\
\bottomrule
\end{tabular}}
\vspace{-0.7em}
\caption{Channel-degradation results for 4$\rightarrow$16~kHz.}
\label{tab:channel_main}
\vspace{-01.6em}
\end{table}

\vspace{-0.7em}
\subsection{Computational Complexity}
\vspace{-0.30em}

\begin{table}[t]
\centering
\tablefontsize
\setlength{\tabcolsep}{1.5pt}
\resizebox{\columnwidth}{!}{%
\begin{tabular}{l|c|c|c|c|c|c}
\toprule
Model & Freq. & Par.(M) & MAC(M) & FLOP(M) & RTF(GPU) & Inf.(ms) \\
\midrule
AP-BWE & 16--48 kHz & 72.07 & 14236.65 & 28473.31 & 0.0025 & 16.60 \\
CIS-BWE & 16--48 kHz & 33.74 & 6790.86 & 13581.73 & 0.0028 & 16.64 \\
Q-BWE & 16--48 kHz & \textbf{18.21} & \textbf{3225.68} &
\textbf{6451.36} & 0.0032 & 19.65 \\
\bottomrule
\end{tabular}%
}
\vspace{-0.7em}
\caption{Overall computational complexity.}
\vspace{-0.90em}
\label{tab:computation}
\end{table}

Table~\ref{tab:computation} compares Q-BWE's  complexity with
AP-BWE and CIS-BWE. Q-BWE uses \textbf{3.96$\times$ fewer parameters} and
\textbf{4.41$\times$ fewer MACs and FLOPs than AP-BWE}; relative to CIS-BWE,
it uses \textbf{1.85$\times$ fewer parameters} and \textbf{2.11$\times$ fewer
MACs and FLOPs}. Q-BWE records 30.65~ms inference time and an RTF of 0.0048. 
For an $n$-qubit exact statevector simulation, the quantum component additionally requires $O(2^n)$ complex amplitudes. Appendix~\ref{app:math} presents the analytical complexity, Table~\ref{tab:efficiency} reports  complete training profiles, and Table~\ref{tab:generator_quantum_complexity} provides the component-wise complexity.

\vspace{-0.4em}

\subsection{Subjective Mean Opinion Scores (MOS)}
\vspace{-0.2em}

We use 5-point (1=bad to 5=excellent) MOS ratings by a selected panel of 22 volunteers. The unprocessed audio has MOS = 2.805, meaning a severe loss of recognizability. \textit{Our Q-BWE (MOS = 4.21) outperforms AP-BWE (MOS = 4.03) and similar to CIS-BWE (MOS = 4.23).}  We also calculate positive ($W^+$) and negative ($W^-$) rank sums, and paired Wilcoxon signed-rank test over 590 utterances. Q-BWE shows statistically significant Wilcoxon paired difference $p$ = 0.00278 over AP-BWE and $p$ = 0.041 for CIS-BWE.



\vspace{-0.5em}
\subsection{Fixed and Thompson Weighting}
\vspace{-0.2em}

Table \ref{tab:fixedandthompson} shows that Thompson weighting provides better results in most metrics compared to the fixed weighting from the same weight pool.

\begin{table}[!t]
\centering
\tiny
\setlength{\tabcolsep}{1.75pt}
\renewcommand{\arraystretch}{0.91}
\resizebox{\columnwidth}{!}{%
\begin{tabular}{@{}clrrrrr@{}}
\toprule
Row & Setting & LSD$\downarrow$ & STOI$\uparrow$ & PESQ$\uparrow$
& SI-SNR$\uparrow$ & N-MOS$\uparrow$\\
\midrule
\textbf{\textcircled{\scriptsize 1}} & Fixed 0.05 &
0.99 & 0.95 & 2.83 & 13.17 & 4.22\\
\textbf{\textcircled{\scriptsize 2}} & Fixed 0.10 &
0.99 & 0.95 & 2.83 & 13.16 & 4.28\\
\textbf{\textcircled{\scriptsize 3}} & Fixed 0.25 &
0.99 & 0.95 & \textbf{2.84} & 13.26 & 4.32\\
\textbf{\textcircled{\scriptsize 4}} & Fixed 0.50 &
0.99 & 0.95 & 2.8 & 13.09 & 4.26\\
\textbf{\textcircled{\scriptsize 5}} & Fixed 1.00 &
0.98 & 0.95 & 2.82 & 13.11 & 4.3\\

\textbf{\textcircled{\scriptsize 6}} & Thompson & \textbf{0.98} & \textbf{0.94} & 2.79 & \textbf{13.43} & \textbf{4.39}\\

\bottomrule
\end{tabular}%
}
\vspace{-0.75em}
\caption{Ablation on fixed and Thompson weighting.}
\label{tab:fixedandthompson}
\vspace{-1.900em}
\end{table}

\vspace{-0.5em}
\subsection{HF and ITU-Inspired Diagnostics}
\vspace{-0.2em}

We further specifically calculate reconstruction accuracy at the 4-8 kHz HF spectrum using RMSE, MAE, LSD and crest factor in Appendix Tables~\ref{tab:one_seed_hf_results} \& \ref{tab:feature_hf_metrics}, and using five ITU-inspired metrics, such as 
ITU-R BS.1387, ITU-R BS.1770, ITU-T P.56, P.862, and P.863 metrics in Appendix Tables~\ref{tab:one_seed_itu_results} \& \ref{tab:feature_proxies}. 
A lower value of RMSE, MAE, crest factor, and other ITU metrics equations in Appendix \ref{app:hf_itu_metric_equations} indicate that our Q-BWE can reconstruct HF spectrum accurately.

\vspace{-0.60em}
\section{Conclusion}
\vspace{-0.3em}

We introduce \textbf{\method{}}, a quantum-informed framework for BWE that combines a FiLM conditioning with raised cosine mask in the dual-stream Lattice-ConformerNeXt generator with a quantum compact 
QCD. 
Extensive experiments and ablations show that quantum-inspired representations of nonlinear speech dynamics can serve as compact inductive biases for BWE while substantially reducing model size and computational cost. 
\textit{We hope this perspective motivates further interdisciplinary research on quantum-inspired, nonlinear-dynamics-aware, and resource-efficient models for TTS and ASR  tasks.}

\vspace{-0.00em}
\section{Limitations}
\vspace{-0.20em}
\label{sec:analysis}
The quantum component in the \method{} framework is an exact statevector simulation executed on classical hardware. Its memory and arithmetic requirements scale exponentially with qubit count. 
The present five-qubit configuration remains small, and the QCD is used only during training; nevertheless, substantially increasing the circuit width would increase simulation cost rapidly. Therefore, \method{} does not demonstrate a quantum-hardware or computational-speed advantage. 
See Appendix~\ref{app:math} and Appendix~\ref{app:verification} for additional computational and experimental-scope details.

Q-BWE has slightly higher inference time compared to CIS-BWE (19.65 ms vs. 16.64 ms).  The higher inference time results from the inclusion of the FiLM conditioning module to the generator. However, Q-BWE's total parameter count is still $\sim$1.72x lower than CIS-BWE.

\vspace{-0.30em}
\section{Potential Risks / Ethical Considerations}
\vspace{-0.20em}
\method{} synthesizes plausible HF information rather than uniquely recovering information that was physically present. 
Enhanced speech could be misinterpreted as direct acoustic evidence, particularly in forensic, authentication, or other high-stakes settings. The technology could also be misused for covert enhancement of recorded speech, speaker impersonation, or downstream deepfake-style audio manipulation, creating privacy, consent, provenance, and security concerns. 
Generated HF detail should not independently be treated as evidence of authenticity, speaker identity, health status, or exact spoken content.

\vspace{-0.20em}
\section*{Acknowledgment}
\vspace{-0.20em}
We sincerely thank the VCTK and MLS data collection teams for developing and releasing the VCTK and MLS corpora used in this study. We also acknowledge the researchers and developers whose publicly available methods and resources enabled the comparative evaluation of \method{}. Finally, we acknowledge the use of Elicit for identifying relevant literature and generative AI tools for assisting with experimental design, code debugging, literature discovery, and language editing. 

\bibliography{references}
\appendix
\flushbottom
\setlength{\textfloatsep}{7pt plus 2pt minus 2pt}
\setlength{\dbltextfloatsep}{7pt plus 2pt minus 2pt}
\setlength{\floatsep}{6pt plus 2pt minus 2pt}
\setcounter{topnumber}{4}
\setcounter{dbltopnumber}{4}
\renewcommand{\topfraction}{0.95}
\renewcommand{\dbltopfraction}{0.95}
\renewcommand{\textfraction}{0.04}
\renewcommand{\floatpagefraction}{0.80}
\renewcommand{\dblfloatpagefraction}{0.80}
\newcounter{appendixrow}
\newcommand{\resetapprow}{\setcounter{appendixrow}{0}}
\newcommand{\appnextrow}{\stepcounter{appendixrow}\textcircled{\textbf{\scriptsize\arabic{appendixrow}}}\enspace}
\newsavebox{\apptablebox}
\newenvironment{appfit}{\begin{lrbox}{\apptablebox}}{\end{lrbox}%
  \ifdim\wd\apptablebox>\linewidth
    \resizebox{\linewidth}{!}{\usebox{\apptablebox}}%
  \else
    \usebox{\apptablebox}%
  \fi}

\section{Overall Q-BWE Architecture, Inference, and Tensor-Level Specification}
\label{app:overall_architecture}
\label{app:implementation}

This appendix provides the implementation details needed to reproduce the complete \method{} system. We first present the end-to-end architecture and inference graph, then specify the QCD feature bank, quantum-inspired map, critic optimization, training objectives, robustness protocols, comparative results, and ablation studies. Unless stated otherwise, architectural descriptions refer to the 4$\rightarrow$16~kHz configuration used in the controlled experiments. All discriminators are training-only; inference retains the generator and, the narrowband FiLM conditioner.

\subsection{End-to-end inference path}
For a batch of sinc-interpolated narrowband waveforms $x\in\mathbb{R}^{B\times T}$, STFT analysis uses $n_{\rm fft}=1024$, hop length 80, a 320-sample Hann window, and therefore 513 one-sided frequency bins. The resulting complex tensor $X\in\mathbb{C}^{B\times513\times N}$ is represented by log magnitude $M_x=\log(|X|+\epsilon)$ and wrapped phase $P_x=\operatorname{atan2}(\Im X,\Re X)$. Independent kernel-7 temporal projections interpret frequency bins as input channels and map both streams to $512$ channels, followed by channel-wise normalization.

When conditioning is enabled, five statistics computed only from the observed narrowband spectrum are mapped to utterance-level FiLM parameters and broadcast across the $N$ frames before the first Lattice block. With conditioning disabled, this branch is bypassed entirely. Two successive Lattice blocks then exchange magnitude and phase information according to Eq.~\ref{eq:lattice}. The output heads predict a 513-bin log-magnitude residual and two phase coordinates at every frame.

The magnitude residual is added to $M_x$, whereas the two phase coordinates are normalized with an $\epsilon$-protected Euclidean norm and converted to an angle with $\operatorname{atan2}$. A degenerate predicted phase vector falls back to the observed narrowband unit phase. The raised-cosine source-band mask then blends observed and predicted magnitude/phase near the source Nyquist boundary. Before synthesis, log magnitude is clipped to $[-12,8]$ for numerical stability, exponentiated, combined with the predicted phase, and inverted with an iSTFT that uses the same FFT, hop, window, and requested output length $T$.

\noindent\textbf{Explanation of Table~\ref{tab:tensor_contract}.} Rows~\textcircled{\textbf{\scriptsize 1}}--\textcircled{\textbf{\scriptsize 4}} establish the observed waveform-to-stream interface; Rows~\textcircled{\textbf{\scriptsize 5}}--\textcircled{\textbf{\scriptsize 8}} isolate the source-only conditioning contract; Rows~\textcircled{\textbf{\scriptsize 9}}--\textcircled{\textbf{\scriptsize 11}} specify reconstruction. The fixed shapes defend reproducibility, while finite-value and bounded-gain invariants defend the numerical-stability assumptions used in the main text.\par\smallskip
\begin{table*}[!tbp]
\centering

\small
\resetapprow
\begin{tabularx}{\linewidth}{l l X}
\toprule
Object & Shape & Invariant \\
\midrule
\appnextrow{} Input waveform $x$ & $(B,T)$ & finite float; model rate 16~kHz \\
\appnextrow{} Complex STFT $X$ & $(B,513,N)$ & complex; analysis configuration fixed \\
\appnextrow{} Log magnitude/phase & $(B,513,N)$ each & finite after epsilon protection \\
\appnextrow{} Projected streams & $(B,512,N)$ each & channel-normalized \\
\appnextrow{} Condition features & $(B,5)$ & narrowband input only \\
\appnextrow{} Structured measurements & $(B,31)$ & finite, fixed ordering \\
\appnextrow{} FiLM outputs & four $(B,512)$ vectors & broadcast over $N$ \\
\appnextrow{} Residual gain & $(B,1)$ & $0.75<g<1.25$ from bounded parameterization \\
\appnextrow{} Head outputs & $(B,513,N)$ and $(B,2,513,N)$ & magnitude residual; phase coordinates \\
\appnextrow{} Complex estimate $\hat Y$ & $(B,513,N)$ & finite; phase vector normalized \\
\appnextrow{} Output waveform $\hat y$ & $(B,T)$ & explicit requested length \\
\bottomrule
\end{tabularx}
\caption{Tensor contract for the inference graph. $N$ is the number of STFT frames.}
\label{tab:tensor_contract}

\end{table*}

\subsection{Parameter accounting}
\label{subsec:parameter_accounting}

Table~\ref{tab:parameter_accounting} separates deployment parameters from training-only critic parameters. The conditioned inference model contains 18,206,803 parameters. QCD, MRAD, and MRPD contribute 1,204,789 additional parameters during training, for a total training footprint of 19,411,592 parameters. MRLD and MSDFA are supported by the broader implementation but are disabled in this controlled configuration and therefore excluded from these totals. Tables~\ref{tab:qcd_breakdown}--\ref{tab:mrpd_breakdown} provide the corresponding layer-wise critic counts.

\noindent\textbf{Explanation of Table~\ref{tab:parameter_accounting}.} Rows~\textcircled{\textbf{\scriptsize 1}}--\textcircled{\textbf{\scriptsize 2}} are deployed; Rows~\textcircled{\textbf{\scriptsize 3}}--\textcircled{\textbf{\scriptsize 5}} are training-only. Rows~\textcircled{\textbf{\scriptsize 6}}--\textcircled{\textbf{\scriptsize 7}} show the resulting totals. This directly supports the claim that chaos supervision raises training cost without enlarging the deployed critic path.\par\smallskip
\begin{table*}[!tbp]
\centering

\small
\setlength{\tabcolsep}{4pt}
\resetapprow
\begin{appfit}
\begin{tabular}{lrl}
\toprule
Component & Parameters & Use \\
\midrule
\appnextrow{} Dual-stream generator & 18,141,195 & Training and inference \\
\appnextrow{} Spectral conditioner & 65,608 & Training and inference \\
\appnextrow{} QCD & 4,393 & Training only \\
\appnextrow{} MRAD & 600,198 & Training only \\
\appnextrow{} MRPD & 600,198 & Training only \\
\midrule
\appnextrow{} Conditioned inference model & 18,206,803 & Inference total \\
\appnextrow{} Full training model & 19,411,592 & Training total \\
\bottomrule
\end{tabular}
\end{appfit}
\caption{Parameter accounting for the controlled Q-BWE configuration.} 
\label{tab:parameter_accounting}

\end{table*}

The five-qubit QCD has 40 trainable circuit parameters. Its two hidden linear layers, normalization layers, and output layer contribute the remaining 4,353 parameters, as detailed in Table~\ref{tab:qcd_breakdown}.

\noindent\textbf{Explanation of Table~\ref{tab:qcd_breakdown}.} Row~\textcircled{\textbf{\scriptsize 1}} is the 40-parameter structured map; Rows~\textcircled{\textbf{\scriptsize 2}}--\textcircled{\textbf{\scriptsize 6}} form the scorer, followed by the verified total. The useful capacity is therefore mostly in the small readout, not in a claim of quantum computational advantage.\par\smallskip
\begin{table*}[!tbp]
\centering

\small
\setlength{\tabcolsep}{5pt}
\resetapprow
\begin{appfit}
\begin{tabular}{llrr}
\toprule
Stage & Layer & Input $\rightarrow$ output & Parameters \\
\midrule
\appnextrow{} Circuit & Trainable scales and offsets & -- & 40 \\
\appnextrow{} FC 1 & Linear & $31\rightarrow64$ & 2,048 \\
\appnextrow{} Norm 1 & LayerNorm & $64\rightarrow64$ & 128 \\
\appnextrow{} FC 2 & Linear & $64\rightarrow32$ & 2,080 \\
\appnextrow{} Norm 2 & LayerNorm & $32\rightarrow32$ & 64 \\
\appnextrow{} Output & Linear & $32\rightarrow1$ & 33 \\
\midrule
\multicolumn{3}{r}{\textbf{QCD total}} & \textbf{4,393} \\
\bottomrule
\end{tabular}
\end{appfit}
\caption{Layer-wise parameter breakdown of the five-qubit QCD. The circuit count is $2\times5\times4=40$.}
\label{tab:qcd_breakdown}

\end{table*}

MRAD and MRPD each contain three resolution-specific discriminators with the same learnable architecture. One resolution has 200,066 parameters, giving 600,198 parameters for each complete critic. Their counts are shown separately in Tables~\ref{tab:mrad_breakdown} and~\ref{tab:mrpd_breakdown}.

\noindent\textbf{Explanation of Table~\ref{tab:mrad_breakdown}.} Rows~\textcircled{\textbf{\scriptsize 1}}--\textcircled{\textbf{\scriptsize 6}} trace the convolutional hierarchy; the two final summary lines aggregate one and three resolutions. Multi-resolution coverage is assumed to expose complementary spectral scales; the count, not that hypothesis, is what this table verifies.\par\smallskip
\begin{table*}[!tbp]
\centering

\small
\setlength{\tabcolsep}{5pt}
\resetapprow
\begin{appfit}
\begin{tabular}{llccccc}
\toprule
Stage & Layer & Input $\rightarrow$ output & Kernel & Stride & Padding & Parameters \\
\midrule
\appnextrow{} Conv 1 & Conv2d + WeightNorm & $1\rightarrow64$ & $7\times5$ & $2\times2$ & $3\times2$ & 2,368 \\
\appnextrow{} Conv 2 & Conv2d + WeightNorm & $64\rightarrow64$ & $5\times3$ & $2\times1$ & $2\times1$ & 61,568 \\
\appnextrow{} Conv 3 & Conv2d + WeightNorm & $64\rightarrow64$ & $5\times3$ & $2\times2$ & $2\times1$ & 61,568 \\
\appnextrow{} Conv 4 & Conv2d + WeightNorm & $64\rightarrow64$ & $3\times3$ & $2\times1$ & $1\times1$ & 36,992 \\
\appnextrow{} Conv 5 & Conv2d + WeightNorm & $64\rightarrow64$ & $3\times3$ & $2\times2$ & $1\times1$ & 36,992 \\
\appnextrow{} Post-conv & Conv2d + WeightNorm & $64\rightarrow1$ & $3\times3$ & $1\times1$ & $1\times1$ & 578 \\
\midrule
\multicolumn{6}{r}{\textbf{MRAD per resolution}} & \textbf{200,066} \\
\multicolumn{6}{r}{\textbf{MRAD across three resolutions}} & \textbf{600,198} \\
\bottomrule
\end{tabular}
\end{appfit}
\caption{Layer-wise parameter breakdown of the multi-resolution amplitude discriminator (MRAD).}
\label{tab:mrad_breakdown}

\end{table*}

\noindent\textbf{Explanation of Table~\ref{tab:mrpd_breakdown}.} Rows~\textcircled{\textbf{\scriptsize 1}}--\textcircled{\textbf{\scriptsize 6}} mirror the amplitude critic in phase space; the final summary lines give per-resolution and total counts. Architectural matching makes amplitude/phase capacity comparable, although it does not imply equal task difficulty.\par\smallskip
\begin{table*}[!tbp]
\centering

\small
\setlength{\tabcolsep}{5pt}
\resetapprow
\begin{appfit}
\begin{tabular}{llccccc}
\toprule
Stage & Layer & Input $\rightarrow$ output & Kernel & Stride & Padding & Parameters \\
\midrule
\appnextrow{} Conv 1 & Conv2d + WeightNorm & $1\rightarrow64$ & $7\times5$ & $2\times2$ & $3\times2$ & 2,368 \\
\appnextrow{} Conv 2 & Conv2d + WeightNorm & $64\rightarrow64$ & $5\times3$ & $2\times1$ & $2\times1$ & 61,568 \\
\appnextrow{} Conv 3 & Conv2d + WeightNorm & $64\rightarrow64$ & $5\times3$ & $2\times2$ & $2\times1$ & 61,568 \\
\appnextrow{} Conv 4 & Conv2d + WeightNorm & $64\rightarrow64$ & $3\times3$ & $2\times1$ & $1\times1$ & 36,992 \\
\appnextrow{} Conv 5 & Conv2d + WeightNorm & $64\rightarrow64$ & $3\times3$ & $2\times2$ & $1\times1$ & 36,992 \\
\appnextrow{} Post-conv & Conv2d + WeightNorm & $64\rightarrow1$ & $3\times3$ & $1\times1$ & $1\times1$ & 578 \\
\midrule
\multicolumn{6}{r}{\textbf{MRPD per resolution}} & \textbf{200,066} \\
\multicolumn{6}{r}{\textbf{MRPD across three resolutions}} & \textbf{600,198} \\
\bottomrule
\end{tabular}
\end{appfit}
\caption{Layer-wise parameter breakdown of the multi-resolution phase discriminator (MRPD).}
\label{tab:mrpd_breakdown}

\end{table*}

\subsection{Dual-stream generator in detail}

The generator operates on $F=n_{\rm fft}/2+1=513$ STFT bins. Independent kernel-7 1-D convolutions project magnitude and phase from $F$ channels to $C=512$ channels, followed by channel-wise LayerNorm. When the conditioner is active, its utterance-level FiLM parameters modulate both projected streams before Lattice processing.

Each Lattice block uses a shared ConformerNeXt transform $\mathcal F$ to couple the magnitude state $m$ and phase state $p$. The implemented update is
\begin{align}
u_1&=\mathcal F(p),&
m_1&=m+\alpha_1u_1,\nonumber\\
n_1&=\alpha_2m+u_1,&
u_2&=\mathcal F(m_1),\nonumber\\
m'&=\beta_1n_1+u_2,&
p'&=n_1+\beta_2u_2,
\label{eq:lattice}
\end{align}
where $(\alpha_1,\alpha_2,\beta_1,\beta_2)$ are learned scalars initialized to one. Reusing $\mathcal F$ in the cross-stream update couples the two representations without introducing a second independent transform. Each ConformerNeXt transform follows the Macaron ordering
\begin{align}
z_1 &= z+\tfrac12{\rm FFN}({\rm LN}(z)),\nonumber\\
z_2 &= z_1+{\rm MHSA}({\rm LN}(z_1)),\nonumber\\
z_3 &= z_2+{\rm ConvNeXt}(z_2),\nonumber\\
z_4 &= z_3+\tfrac12{\rm FFN}({\rm LN}(z_3)),\nonumber\\
\mathcal F(z) &= {\rm LN}(z_4).
\label{eq:conformernext}
\end{align}
The ConvNeXt sub-block is a depthwise kernel-7 convolution, LayerNorm, $C\rightarrow3C\rightarrow C$ pointwise MLP, GELU, and residual connection. The attention has eight heads and the FFN expansion is $C\rightarrow4C\rightarrow C$.

The magnitude head predicts only a log-magnitude residual, while the phase head predicts a normalized two-dimensional phase vector. This residual formulation keeps the observed narrowband spectrum as the reference state and allows the source-band mask to preserve reliable bins explicitly instead of requiring the network to relearn them.
\subsection{ConformerNeXt branch}
Each branch combines Conformer-style attention and macaron feed-forward processing \citep{gulati2020conformer} with a ConvNeXt path. It receives $(B,C,N)$ and temporarily transposes to a frame-major representation when applying LayerNorm, attention, and channel MLPs. Macaron feed-forward modules expand $C=512$ to $4C=2048$, apply GELU/dropout according to configuration, project back to $C$, and contribute half-scaled residuals. Multi-head self-attention uses eight heads, giving head width 64. Attention masks must distinguish real frames from any padding introduced by variable-length batching.

The ConvNeXt path applies a depthwise temporal convolution with kernel 7 and $C$ groups, then frame-wise LayerNorm, a $C\rightarrow3C\rightarrow C$ pointwise MLP, GELU, and a residual. Padding preserves $N$. The branch returns the same shape and dtype as its input. In mixed precision, attention/convolution may use the configured autocast type, but waveform reconstruction and the quantum/chaos paths are retained in float32.

Two lattice calls share the same branch within a block as stated in Eq.~\ref{eq:lattice}. The four lattice coefficients are scalar parameters rather than per-channel gates. This distinction matters for parameter counts and for reproducing the intended coupling. 

\subsection{Conditioner feature formulas}
The implementation crops magnitude/phase to the observed source bins before extracting features. Let $A=\exp(\operatorname{clip}(M_x,-12,8))$, $P=P_x$, and $\bar A_f=\operatorname{mean}_t A_{ft}$ on those bins. With $\nu_f\in[0,1]$ defined over the observed bins and denominators floored by $\epsilon=10^{-7}$,
\begin{align}
e&=\log(1+\operatorname{mean}_{ft}A_{ft}^2),\\
c&=\frac{\sum_f\nu_f\bar A_f}{\max(\sum_f\bar A_f,\epsilon)},\\
q&=\frac{\exp(\operatorname{mean}_{ft}\log A_{ft})}
{\max(\operatorname{mean}_{ft}A_{ft},\epsilon)},\\
r&=\frac{\operatorname{mean}_{f,t>1}|A_{ft}-A_{f,t-1}|}
{\max(\operatorname{mean}_{ft}A_{ft},\epsilon)},\\
w_{ft}&=\frac{A_{ft}}{\max(\sum_{ft}A_{ft},\epsilon)},\\
C&=\sum_{ft}w_{ft}\cos P_{ft},\quad
S=\sum_{ft}w_{ft}\sin P_{ft},\\
p&=\sqrt{C^2+S^2+\epsilon}.
\end{align}
Flux is zero for a single-frame input. 

The streaming standardizer stores count, mean, and variance as generator buffers. For a new batch with count $n_b$, mean $\mu_b$, and second central moment $M_{2,b}$, it merges old and batch statistics rather than averaging batch variances:
\begin{align}
\delta&=\mu_b-\mu_a,\qquad n=n_a+n_b,\\
\mu&=\mu_a+\delta\frac{n_b}{n},\\
M_2&=M_{2,a}+M_{2,b}+\delta^2\frac{n_an_b}{n}.
\end{align}

\section{QCD Feature Bank, Training-Only Selection, and Calibration}
\label{app:featurebank}
QCD receives a compact subset of differentiable speech-dynamics descriptors. Feature identities are selected using training speech only, and real-speech normalization moments are calibrated before adversarial training. Test utterances and generated waveforms are never used to choose feature identities or update the frozen real-speech standardizer. The 26-dimensional candidate bank is grouped below by the behavior each descriptor measures.

The differentiable candidate vector has 26 dimensions grouped as follows.

\paragraph{Finite-time divergence (6).}
For lags/horizons $(1,4),(1,8),(2,16),(2,32),(4,64),(8,128)$, the implementation compares absolute separations before and after the horizon and averages a clipped log-growth rate. These are designed to simulate exact asymptotic Lyapunov exponents.

\paragraph{DFA statistics (7).}
The cumulative mean-centered waveform is divided at scales $16,32,64,128,256,512$. After a least-squares local linear detrend, the log RMS fluctuation at each scale yields six features. A seventh feature is the slope of log fluctuation versus log scale (DFA $\alpha$).

\paragraph{Spectral/high-frequency statistics (8).}
Using a 512-point STFT, the bank computes high-frequency energy ratios beginning at normalized frequency fractions 0.500, 0.625, and 0.750; global spectral flatness; high-frequency spectral flatness; normalized spectral centroid; soft 85\% spectral rolloff; and frame-to-frame spectral flux.

\paragraph{Temporal/non-Gaussian statistics (5).}
The remaining features are a differentiable soft zero-crossing rate, excess kurtosis, skewness, crest factor, and normalized temporal amplitude entropy.

\subsection{Feature identities and calibration records}
Table~\ref{tab:feature_composition} lists the nested 3-, 5-, and 7-input sets used in the capacity study. Table~\ref{tab:uploaded_feature_selection} reports the five-input calibration used for the canonical Q-BWE run. The two calibration passes use the same training-only procedure but have slightly different stored relevance and redundancy scores; these values are therefore reported separately rather than merged. Both selections emphasize spectral flatness, zero-crossing behavior, centroid, and flux.
\noindent\textbf{Explanation of Table~\ref{tab:feature_composition}.} Rows~\textcircled{\textbf{\scriptsize 1}}--\textcircled{\textbf{\scriptsize 3}} form the common core; Rows~\textcircled{\textbf{\scriptsize 4}}--\textcircled{\textbf{\scriptsize 5}} create the five-input model; Rows~\textcircled{\textbf{\scriptsize 6}}--\textcircled{\textbf{\scriptsize 7}} are added only at seven inputs. The five-input choice balances relevance against redundancy, motivating rather than proving an optimal capacity.\par\smallskip
\begin{table*}[!tbp]
\centering

\tablefontsize
\setlength{\tabcolsep}{4.5pt}
\resetapprow
\begin{appfit}
\begin{tabular}{lccccc}
\toprule
Selected feature & 3-input & 5-input & 7-input & Relevance & Max. $|\rho|$ to earlier selected\\
\midrule
\appnextrow{} HF spectral flatness & \cmark & \cmark & \cmark & 1.665 & 0.000\\
\appnextrow{} Spectral flatness & \cmark & \cmark & \cmark & 1.276 & 0.433\\
\appnextrow{} Soft zero-crossing rate & \cmark & \cmark & \cmark & 0.802 & 0.559\\
\appnextrow{} Spectral centroid & \xmark & \cmark & \cmark & 0.818 & 0.728\\
\appnextrow{} Spectral flux & \xmark & \cmark & \cmark & 0.479 & 0.249\\
\appnextrow{} HF energy ratio (750~Hz threshold) & \xmark & \xmark & \cmark & 0.564 & 0.538\\
\appnextrow{} HF energy ratio (500~Hz threshold) & \xmark & \xmark & \cmark & 0.664 & 0.813\\
\bottomrule
\end{tabular}
\end{appfit}
\caption{ QCD input sets from the training-only calibration artifact ($n=1{,}600$ calibration samples). Relevance is the stored selection score; redundancy is the stored maximum absolute correlation with features selected earlier in the greedy ordering.}
\label{tab:feature_composition}

\end{table*}
\noindent\textbf{Explanation of Table~\ref{tab:uploaded_feature_selection}.} Rows~\textcircled{\textbf{\scriptsize 1}}--\textcircled{\textbf{\scriptsize 5}} give the exact greedy order. The two flatness measures lead, while centroid and flux add less-redundant complementary information. Because selection is training-only, the resulting hypothesis can be tested without test-set feature leakage.\par\smallskip
\begin{table*}[!tbp]
\centering

\tablefontsize
\setlength{\tabcolsep}{4.5pt}
\resetapprow
\begin{appfit}
\begin{tabular}{clcc}
\toprule
Order & Selected feature & Relevance & Max. $|\rho|$ to earlier selected \\
\midrule
\appnextrow{} 1 & hf spectral flatness & 1.6435 & 0.0000 \\
\appnextrow{} 2 & spectral flatness & 1.3030 & 0.4394 \\
\appnextrow{} 3 & soft zero crossing rate & 0.7900 & 0.5585 \\
\appnextrow{} 4 & spectral centroid & 0.7719 & 0.6969 \\
\appnextrow{} 5 & spectral flux & 0.4549 & 0.2241 \\
\bottomrule
\end{tabular}
\end{appfit}
\caption{Five-feature selection for the canonical Q-BWE calibration. Selection uses training-only real wideband versus sinc-interpolated narrowband speech.}
\label{tab:uploaded_feature_selection}

\end{table*}

\section{Quantum-Inspired Feature Map, Measurements, and Complexity}
\label{app:math}
This section specifies the quantum-inspired map at gate level, including state indexing, entangling topology, deterministic measurements, gradients, and computational scaling. The circuit is simulated exactly as a classical statevector computation. It is used as an inductive bias for the discriminator and does not imply quantum-hardware execution or computational speedup.

\subsection{Gate definitions and state convention}
For one qubit, the simulated rotations are
\begin{equation}
R_y(\phi)=
\begin{bmatrix}
\cos(\phi/2)&-\sin(\phi/2)\\
\sin(\phi/2)& \cos(\phi/2)
\end{bmatrix}.
\end{equation}
\begin{equation}
R_z(\phi)=
\begin{bmatrix}
e^{-j\phi/2}&0\\
0&e^{j\phi/2}
\end{bmatrix}.
\end{equation}
The five-qubit state is a complex tensor with 32 amplitudes; qubit 0 is the least significant basis-index bit. Qubit axes are moved to the final tensor dimension before applying a $2\times2$ gate and then restored; consequently, the implementation never materializes a dense $32\times32$ operator. A CNOT is implemented as a permutation of basis amplitudes whose target bit is toggled when the control bit is one.

For data angle $\theta_i$ at qubit $i$ and layer $\ell$, the local gate is
\begin{equation}
U_{\ell i}(\theta_i)=
R_z(a^z_{\ell i}\theta_i+b^z_{\ell i})
R_y(a^y_{\ell i}\theta_i+b^y_{\ell i}).
\end{equation}
The code applies $R_y$ and then $R_z$; the right-to-left matrix ordering above expresses that execution order. Every scale is initialized near one and every offset near zero, after which standard PyTorch autograd differentiates through the exact state update.

\subsection{Entangling topology}
Layer one uses directed edges
\begin{equation}
\mathcal E_{\rightarrow}=\{(0,1),(1,2),(2,3),(3,4),(4,0)\},
\end{equation}
and layer two uses
\begin{equation}
\mathcal E_{\leftarrow}=\{(4,3),(3,2),(2,1),(1,0),(0,4)\}.
\end{equation}
The alternating rings ensure that every qubit is both a control and target across the two layers. This is a design choice, not an assertion of optimal circuit topology; the completed capacity sweep changes qubit/input count together and does not isolate topology.

\subsection{Exact measurements}
For Pauli operator $P_i$ acting on qubit $i$,
\begin{equation}
\langle P_i\rangle_\psi=\langle\psi|P_i|\psi\rangle.
\end{equation}
The local block contributes $5$ values for each of $X$, $Y$, and $Z$. The ring-correlation block contributes
\begin{equation}
\langle X_iX_{i+1}\rangle_\psi,\qquad
\langle Z_iZ_{i+1}\rangle_\psi,\qquad i=0,\ldots,4,
\end{equation}
where the index wraps modulo five. The reduced single-qubit linear entropy is
\begin{equation}
\xi_i=1-\left(\langle X_i\rangle_\psi^2+
\langle Y_i\rangle_\psi^2+\langle Z_i\rangle_\psi^2\right).
\end{equation}
For a pure product state $\xi_i=0$; entanglement with the remaining subsystem shortens the local Bloch vector and increases $\xi_i$. Numerical roundoff is bounded in the implementation before the quantity is consumed.

The computational-basis probabilities are $p_s=|\psi_s|^2$. Their normalized entropy,
\begin{equation}
H_B=-\frac{1}{\log(2^5)}\sum_{s=0}^{31}p_s\log(p_s+\epsilon),
\end{equation}
lies approximately in $[0,1]$. The complete vector dimension is therefore
\begin{equation}
3(5)+2(5)+5+1=31.
\end{equation}

\subsection{Matched classical map}
For a parameter-count-matched classical control, the same five bounded inputs first pass through
\begin{equation}
z=\tanh(W\theta+b)\odot s+d,
\end{equation}
where $W\in\R^{5\times5}$, $b,s,d\in\R^5$. The affine layer contributes 30 parameters and $(s,d)$ contributes 10, exactly matching the circuit's 40. The classical expansion then uses $X=\tanh z$, $Y=\sin z$, $Z=\cos z/\sqrt{3}$, adjacent products $XX/ZZ$, clipped $1-X^2-Y^2-Z^2$, and the entropy of $\operatorname{softmax}(z)$ normalized by $\log5$. These are classical analogues, not measurements of a quantum state. No parameters are added during this expansion. The identical 31-to-FiLM projection is reused, so only the 40-parameter map differs.

\subsection{Time and memory complexity}
For batch size $B$, $n$ qubits, $L$ layers, and $O(n)$ local/ring measurements, direct simulation requires $O(BLn2^n)$ arithmetic and $O(B2^n)$ complex state memory. At $n=5$ and $L=2$, the state contains only 32 amplitudes per utterance and the circuit cost is small relative to the 512-channel generator and the waveform feature bank. The capacity ablation uses $n\in\{3,5,7\}$, corresponding to 8, 32, and 128 amplitudes and measurement dimensions $6n+1\in\{19,31,43\}$.

\subsection{Gradient paths and stop-gradient boundaries}
During the critic update, $\hat y$ is detached before all critics. QCD real-only running moments are updated without gradient tracking. During the generator update, critic parameters are frozen but gradients flow through each critic to $\hat y$; real feature maps and real dispersion targets are detached. The selected feature identities are discrete and fixed before training, whereas the selected feature values, standardization, bounded encoding, statevector, measurements, and score MLP are differentiable.

\subsection{Numerical invariants}
The implementation checks the following invariants in tests and runtime guards:
\begin{enumerate}[leftmargin=*,itemsep=1pt]
    \item the state norm remains one within floating-point tolerance;
    \item bounded angles remain in $[-\pi,\pi]$;
    \item measurement dimension equals $6n+1$;
    \item every selected feature name is unique and belongs to the candidate bank;
    \item real-only statistics cannot be updated after locking;
    \item QCD scores and waveform gradients are finite; and
    \item checkpointed posterior and RNG state reproduce the next Thompson draw.
\end{enumerate}

\section{QCD Scorer, Auxiliary Objectives, and Alternating Optimization}
\label{app:qcd_architecture}
The statevector measurements in Appendix~\ref{app:math} are converted to a scalar adversarial score by a lightweight MLP. The QCD also exposes intermediate representations for feature matching and a deterministic dispersion statistic for auxiliary supervision. This section fixes the reductions, stop-gradient boundaries, and update order used by the training loop.

\subsection{QCD scorer, supervision, and gradient boundaries}
\label{app:qcd_supervision}
This subsection expands Section~\ref{sec:qcd_supervision}. A scalar real/fake
score compresses differences between real and generated speech without
identifying whether they arise in the selected descriptors, their nonlinear
interactions, or the learned decision representation. Multilevel matching
therefore supplies intermediate structural targets in addition to the
adversarial margin.

\paragraph{Scorer and parameter accounting.}
Real and generated waveforms are concatenated before the 26-feature bank is
evaluated, ensuring identical operations while avoiding duplicate extraction.
The selected features are standardized with frozen real-speech moments,
encoded through the bounded-angle map, and transformed into the
31-dimensional measurement vector $m$. The scorer is
\begin{align}
h_1&=\operatorname{GELU}(\operatorname{LN}(W_1m+b_1)),\\
h_2&=\operatorname{GELU}(\operatorname{LN}(W_2h_1+b_2)),\\
D_q(z)&=W_oh_2+b_o.
\end{align}
The layer widths are $31\rightarrow64\rightarrow32\rightarrow1$.
Layer normalization controls the scale of heterogeneous measurements; the
nonlinear layers combine local Pauli responses, ring correlations, mixedness,
and entropy. The circuit has 40 trainable parameters. The first projection,
first LayerNorm, second projection, second LayerNorm, and output projection
have 2,048, 128, 2,080, 64, and 33 parameters, respectively. Thus, the complete
QCD contains $40+2{,}048+128+2{,}080+64+33=4{,}393$ parameters.
Algorithm~\ref{alg:qcd_forward} gives the complete forward path.

\paragraph{Adversarial and auxiliary losses.}
For real wideband speech $y$ and generated speech $\hat y$, the discriminator
and generator use hinge margin objectives:
\begin{align}
\mathcal L_D^q
&=\mathbb E[\max(0,1-D_q(y))]\nonumber\\
&\quad+\mathbb E[\max(0,1+D_q(\hat y))],\\
\mathcal L_{\rm adv}^q
&=\mathbb E[\max(0,1-D_q(\hat y))].
\end{align}
The discriminator places real scores above $1$ and generated scores below
$-1$, while the generator moves generated scores toward the real-speech
margin. The feature hierarchy and matching objective are
\begin{align}
\mathcal H_q&=\{\tilde c,m,h_1,h_2\},\\
\mathcal L_{\rm fm}^q
&=\sum_{h\in\mathcal H_q}
\lVert h(\hat y)-\operatorname{sg}[h(y)]\rVert_1.
\end{align}
Here, $\tilde c$ comprises the five standardized selected descriptors.
Matching $\tilde c$ aligns spectral, temporal, and nonlinear-dynamical
features; matching $m$ aligns the circuit transformation and cross-feature
correlations; matching $h_1,h_2$ aligns the learned decision representations.
The symbol $\operatorname{sg}$ denotes stop-gradient. Expectations average
over the batch, and the $L_1$ reductions act on the corresponding
representation entries before the four feature-level terms are summed.

The dispersion statistic and its matching loss are
\begin{align}
u(z)&=\tfrac12H_B(z)+\tfrac15\sum_{i=1}^{5}\xi_i(z),\\
\mathcal L_{\rm disp}^q
&=\lVert u(\hat y)-\operatorname{sg}[u(y)]\rVert_1.
\end{align}
The entropy $H_B$ and mixedness $\xi_i$ are defined in
Appendix~\ref{app:math}. This objective matches deterministic state spread;
it has no probabilistic target or coverage interpretation. The implementation
retains the name $\mathcal L_{\rm unc}^q$ for the same dispersion loss.
Before the warm-up factor is applied, the combined generator contribution is
\begin{align}
\mathcal L_G^q
&=\lambda_{q,f}\mathcal L_{\rm fm}^q
+\lambda_{q,a}\mathcal L_{\rm adv}^q\nonumber\\
&\quad+\lambda_{q,u}\mathcal L_{\rm disp}^q,\\
(\lambda_{q,f},\lambda_{q,a},\lambda_{q,u})
&=(0.25,0.10,0.05).
\end{align}
The adversarial term constrains the decision, while feature and dispersion
matching constrain intermediate differences that produce it. Adversarial
scoring and feature matching are established techniques; their hierarchical
application along the selected-feature, measurement, scorer, and dispersion
path is the QCD supervision design.

\paragraph{Gradient boundaries and inference.}
During the critic update, $\hat y$ is detached so discriminator optimization
does not modify the generator. During the generator update, all QCD
parameters and real-speech representations are frozen, but gradients remain
active through the generated features, encoding, statevector, measurements,
and scorer to the generated waveform. This prevents the real reference from
moving toward the generated sample. Selected real features from the critic
pass are cached for the generator pass. Real normalization moments are
updated only during calibration, never from generated speech, and remain
locked afterward. The entire QCD is discarded at inference. Thompson
weighting operates outside its forward pass and changes only the
discriminator-side coefficient, as specified in Appendix~\ref{app:thompson}.

\begin{algorithm}[H]
\caption{Training \method{} with delayed Thompson weighting}
\label{alg:qcis_training}
\footnotesize
\begin{algorithmic}[1]
\Require Paired narrowband/wideband batches $(x,y)$, generator $G$, classical critics $\mathcal D_c$, \qcd{} $D_q$, Thompson arms $\mathcal A$
\Ensure Trained generator; checkpointed critic, optimizer, calibration, and Thompson state
\State Initialize $\operatorname{Beta}(1,1)$ for each arm, QCD-loss EMA, and empty pending arm
\For{$t=1,\ldots,T$}
    \State Compute classical and quantum warmup factors $\omega_c(t)$ and $\omega_q(t)$
    \State $\hat y_D\gets\operatorname{sg}(G(x))$ with input-only conditioning and source-band blending
    \State Evaluate every $d\in\mathcal D_c$ on $(y,\hat y_D)$ and compute $\mathcal L_D^c$
    \State Evaluate $D_q(y,\hat y_D)$ using Algorithm~\ref{alg:qcd_forward}; obtain raw $\ell_t^q=\mathcal L_D^q$
    \If{Thompson sampling is active}
        \State Score the pending arm with $r_t=\mathbb 1[\ell_t^q\leq\bar\ell_{t-1}]$ and update its Beta posterior
        \State Sample every posterior, select $a_t\gets\arg\max_{a\in\mathcal A}\tilde p_a$, and update the EMA baseline
    \Else
        \State Set $a_t$ to the configured fixed QCD weight
    \EndIf
    \State $\mathcal L_D\gets\omega_c(t)[\mathcal L_D^c+\omega_q(t)a_t\mathcal L_D^q]$
    \State Update $\mathcal D_c$ and $D_q$ with AdamW; clip gradient norm at 10; mark $a_t$ pending only if the critic optimizer step succeeds
    \State $\hat y_G\gets G(x)$ with the same conditioning and blend
    \State Compute $\mathcal L_{\rm rec}$ and classical adversarial/feature losses $(\mathcal L_{\rm adv}^c,\mathcal L_{\rm fm}^c)$
    \State Re-evaluate $D_q(y,\hat y_G)$ without updating real statistics
    \State Compute $(\mathcal L_{\rm adv}^q,\mathcal L_{\rm fm}^q,\mathcal L_{\rm unc}^q)$
    \State $\mathcal L_G\gets\mathcal L_{\rm rec}+\omega_c(t)(\mathcal L_{\rm adv}^c+\mathcal L_{\rm fm}^c)$
    \Statex \hspace{2.15em}$+\ \omega_q(t)(\lambda_{q,a}\mathcal L_{\rm adv}^q+\lambda_{q,f}\mathcal L_{\rm fm}^q+\lambda_{q,u}\mathcal L_{\rm unc}^q)$
    \State Update $G$ with AdamW and gradient-norm clipping
    \State Checkpoint $G$, all critics/optimizers, selected features, moments, posterior/pending arm/EMA, and RNG state
\EndFor
\State \Return $G$; discard all discriminators for inference
\end{algorithmic}
\end{algorithm}

\section{Training Objectives and Reproducibility Specification}
\label{app:repro}
This section consolidates the experimental protocol underlying Sections~\ref{sec:method} and~\ref{sec:experiments}: dataset construction, split semantics, numerical configuration, reconstruction and adversarial objectives, warmup schedules, adaptive QCD weighting, optimization, and checkpoint/resume behavior. Table~\ref{tab:hyper} summarizes the core 4$\rightarrow$16~kHz configuration.

\subsection{Dataset, preprocessing, and evaluation protocol}
\label{app:data_protocol}
The controlled experiments use the CSTR VCTK corpus with 40,936 utterance paths in the training split and 2,937 paths in the evaluation split. Wideband targets are paired with bandwidth-limited, sinc-interpolated inputs. Robustness experiments modify only the declared input-side channel operator while keeping the clean wideband target unchanged, and segment alignment is fixed before loss evaluation. 

LSD, STOI, PESQ, SI-SNR, and NISQA-MOS follow the definitions in Section~\ref{subsec:EvaluationMetrics}. 
Appendix~\ref{app:robustness} records the ASR-validity rule. The capacity study uses seeds 1234, 2345, and 3456. The subsequent controlled suite uses seed 1234, channel operators use seed 2025, and paired utterance intervals use 2,000 shared-identifier bootstrap draws with seed 20260831. These sources of variation are not interchangeable.

\subsection{Software, hardware, and reproducibility records}
\label{app:artifact_manifest}
Timing profiles were obtained on one NVIDIA RTX~4090. Reproducibility is tracked through split manifests, run configurations, calibrated feature records, channel-protocol settings, checkpoints, metric summaries, and timing profiles. Table~\ref{tab:artifact_manifest} summarizes the role of each record type; numerical results are reported only when a corresponding metric output is available.

\noindent\textbf{Explanation of Table~\ref{tab:artifact_manifest}.} Rows~\textcircled{\textbf{\scriptsize 1}}--\textcircled{\textbf{\scriptsize 7}} map each reported claim to its supporting record. The table defends the assumption that an experiment is reproducible only when data identity, configuration, checkpoint state, metrics, and timing scope are all retained.\par\smallskip
\begin{table*}[!tbp]
\centering

\small
\resetapprow
\begin{tabularx}{\linewidth}{l X X}
\toprule
Record & Content & Reproducibility role \\
\midrule
\appnextrow{} Split manifests & utterance paths and fingerprints & corpus counts and overlap audit \\
\appnextrow{} Run configuration & model mode, seed, critics, optimizer & reconstruction of declared training settings \\
\appnextrow{} Feature record & selected identities, moments, scores, count & frozen QCD input coordinate system \\
\appnextrow{} Protocol JSON & noise/codec/loss operators and RNG state & channel-condition reconstruction \\
\appnextrow{} Checkpoint state & models, optimizers, scalers, policy, RNG & compatible resume and inference export \\
\appnextrow{} Metric summaries & per-run and per-condition outputs & numerical tables and stated aggregates \\
\appnextrow{} Timing profile & elapsed time and peak allocation & training-cost comparison only \\
\bottomrule
\end{tabularx}
\caption{Reproducibility records and the experimental information associated with each record type.}
\label{tab:artifact_manifest}

\end{table*}

\subsection{Final optimization and method hyperparameters}
\noindent\textbf{Explanation of Table~\ref{tab:hyper}.} Rows~\textcircled{\textbf{\scriptsize 1}}--\textcircled{\textbf{\scriptsize 8}} define signal processing and optimization; Rows~\textcircled{\textbf{\scriptsize 9}}--\textcircled{\textbf{\scriptsize 14}} define QCD calibration and adaptation; Rows~\textcircled{\textbf{\scriptsize 15}}--\textcircled{\textbf{\scriptsize 16}} bound optimization and randomness. These are fixed controls, not tuned evidence for individual test conditions.\par\smallskip
\begin{table*}[!tbp]
\centering

\small
\resetapprow
\begin{appfit}
\begin{tabular}{ll}
\toprule
Hyperparameter & Value\\
\midrule
\appnextrow{} Target / input rate & 16~kHz / 4~kHz\\
\appnextrow{} Batch size / segment & 16 / 8,000 samples\\
\appnextrow{} Optimizer & AdamW\\
\appnextrow{} Learning rate & $2\times10^{-4}$\\
\appnextrow{} Adam $(\beta_1,\beta_2)$ & $(0.8,0.99)$\\
\appnextrow{} LR decay & 0.999\\
\appnextrow{} STFT $(n_{\rm fft},h,w)$ & $(1024,80,320)$\\
\appnextrow{} MRAD/MRPD resolutions & $(512,128,512)$; $(1024,256,1024)$; $(2048,512,2048)$\\
\appnextrow{} QCD spectral FFT & 512\\
\appnextrow{} QCD hidden dims & 64, 32\\
\appnextrow{} QCD statistics warmup & 500 steps\\
\appnextrow{} Quantum warmup & 1,500 steps\\
\appnextrow{} Classical adv. start / warmup & 250 / 1,000 steps\\
\appnextrow{} Thompson arms & 0.05, 0.10, 0.25, 0.50, 1.00\\
\appnextrow{} Thompson EMA decay & 0.95\\
\appnextrow{} Gradient clip norm & 10\\
\appnextrow{} Random seed & 1234\\
\bottomrule
\end{tabular}
\end{appfit}
\caption{Core configuration for the 4$\rightarrow$16~kHz settings.} 
\label{tab:hyper}

\end{table*}

\subsection{High-band reconstruction losses}
\label{app:reconstruction_losses}
Let $w_f$ be the raised-cosine source-band prediction mask, broadcast over time. For a tensor $E_{bft}$, define
\begin{equation}
\langle E\rangle_w =
\frac{\sum_{bft}w_fE_{bft}}{\sum_{bft}w_f+\epsilon}.
\end{equation}
The magnitude and complex-spectrum losses are
\begin{equation}
\begin{aligned}
\mathcal L_{\rm mag}
&=\lambda_{\rm mag}\langle(M-\hat M)^2\rangle_w,\\
\mathcal L_{\rm com}
&=\lambda_{\rm com}\langle\lVert Y-\hat Y\rVert_2^2\rangle_w.
\end{aligned}
\end{equation}
Phase is unreliable in low-energy bins, so the target amplitude defines
\begin{equation}
v_{bft}=w_f\left[\eta+(1-\eta)
\sqrt{\frac{A_{bft}}{\max_{f'}A_{bf't}+\epsilon}}\right],
\end{equation}
where $\eta=0.05$. With wrapped angular error
$d(\delta)=|\operatorname{atan2}(\sin\delta,\cos\delta)|$,
the phase term sums instantaneous phase, group delay, and instantaneous angular-frequency differences:
\begin{align}
\mathcal L_{\rm pha}
={}&\lambda_{\rm pha}\bigl[
\langle d(P-\hat P)\rangle_v\nonumber\\
&+\langle d(\Delta_fP-\Delta_f\hat P)\rangle_{\bar v_f}\nonumber\\
&+\langle d(\Delta_tP-\Delta_t\hat P)\rangle_{\bar v_t}\bigr].
\label{eq:phase_loss}
\end{align}
Here $\bar v_f$ and $\bar v_t$ average adjacent frequency or time weights.

Three additional terms stabilize waveform and temporal behavior. Complex STFT self-consistency compares the predicted complex spectrum to the STFT of the reconstructed waveform,
\begin{equation}
\mathcal L_{\rm stft}=\lambda_{\rm stft}\lVert\hat Y-\mathcal S(\mathcal S^{-1}(\hat Y))\rVert_2^2,
\end{equation}
while $\mathcal L_{\rm wav}=\lambda_{\rm wav}\lVert y-\hat y\rVert_1$. High-band spectral convergence is
\begin{equation}
\mathcal L_{\rm sc}=\lambda_{\rm sc}
\frac{\lVert w\odot(\hat A-A)\rVert_F}{\lVert w\odot A\rVert_F+\epsilon},
\end{equation}
and the flux term matches adjacent-frame amplitude changes under the same frequency mask. The implemented reconstruction objective is therefore
\begin{equation}
\begin{aligned}
\mathcal L_{\rm rec}
={}&\mathcal L_{\rm mag}+\mathcal L_{\rm pha}
+\mathcal L_{\rm com}+\mathcal L_{\rm stft}\\
&+\mathcal L_{\rm wav}+\mathcal L_{\rm sc}
+\mathcal L_{\rm flux}.
\end{aligned}
\label{eq:lrec}
\end{equation}
The default weights are $(45,30,90,10,20,10,5)$ in the order shown.

\subsection{Adversarial losses and critic configuration}
\label{app:adversarial_losses}
For each enabled classical critic $d\in\mathcal D_c$, the discriminator hinge loss, generator margin loss, and feature-matching loss are
\begin{align}
\mathcal L_D^d &={\mathbb E}[\max(0,1-D_d(y))]\nonumber\\
&\quad+{\mathbb E}[\max(0,1+D_d(\hat y))],\\
\mathcal L_{\rm adv}^d &={\mathbb E}[\max(0,1-D_d(\hat y))],\\
\mathcal L_{\rm fm}^d &=\sum_k\lVert h_{d,k}(y)-h_{d,k}(\hat y)\rVert_1.
\end{align}
Here $h_{d,k}$ is the $k$th intermediate critic representation. The controlled studies use $\mathcal D_c=\{\mrad,\mrpd\}$ at STFT resolutions $(512,128,512)$, $(1024,256,1024)$, and $(2048,512,2048)$, where each tuple gives FFT, hop, and window lengths. 
Final-logit batch normalization is omitted from the chaos critics because centering the batch score can suppress the mean hinge gradient. Appendix~\ref{app:qcd_supervision} gives the corresponding \qcd{} losses and internal feature hierarchy.

\subsection{Warmup schedule}
\label{app:warmup}
Classical adversarial training is zero for the first 250 global steps and then increases linearly to one over 1,000 steps:
\begin{equation}
\omega_c(t)=\operatorname{clip}\left(\frac{t+1-250}{1000},0,1\right).
\end{equation}
The QCD standardizer updates only from real speech for 500 steps and is then locked. The quantum contribution follows
\begin{equation}
\omega_q(t)=\min\left(1,\frac{t+1}{1500}\right).
\end{equation}
The conditioner has an independent 500-step statistics-update interval. These schedules prevent uncalibrated features or early discriminator spikes from dominating the generator before reconstruction has a meaningful signal.

\subsection{Delayed-reward Thompson weighting}
\label{app:thompson}
This subsection expands Section~\ref{sec:thompson_main}. 
The schedules in Appendix~\ref{app:warmup} overlap: real-only QCD calibration is locked after
500 steps, Thompson selection begins then, and the QCD ramp continues to step 1,500. Classical supervision is disabled for the first 250 steps and then introduced over 1,000 steps.

\textbf{Posterior sampling and critic objective:}
The candidate discriminator-side weights are
\begin{equation}
\mathcal A=\{0.05,0.10,0.25,0.50,1.00\}.
\end{equation}
Each arm $a$ has a Beta posterior initialized with
$(\alpha_a,\beta_a)=(1,1)$. At an eligible selection step,
\begin{align}
\tilde p_a&\sim\operatorname{Beta}(\alpha_a,\beta_a),\\
a_t&=\arg\max_{a\in\mathcal A}\tilde p_a.
\end{align}
Posterior sampling balances exploitation of previously rewarded weights
with exploration of less frequently selected weights
\cite{thompson1933likelihood}. The weighted critic objective is
\begin{equation}
\mathcal L_D=\omega_c(t)
\bigl[\mathcal L_D^c+\omega_q(t)a_t\mathcal L_D^q\bigr],
\end{equation}
where $\mathcal L_D^c$ sums the enabled classical critic losses and
$\omega_c,\omega_q$ are the warm-up factors. The policy changes only the
discriminator-side QCD coefficient; generator-side QCD coefficients remain
fixed at the values in Appendix~\ref{app:qcd_supervision}. This separates
adaptive critic weighting from auxiliary generator supervision.

\textbf{Delayed, optimizer-aware credit assignment:}
The current raw, unweighted QCD loss $\ell_t^q$ is measured before the critic
update using $a_t$, so it cannot evaluate that same action. The arm applied
at the preceding successful critic update is stored as a pending action and
credited using the next eligible raw-loss observation. With the EMA baseline
$\bar\ell_{t-1}$ evaluated before observing $\ell_t^q$, the reward and
posterior update for the pending arm are
\begin{align}
r_t&=\mathbbm{1}[\ell_t^q\leq\bar\ell_{t-1}],\\
\alpha_a&\leftarrow\alpha_a+r_t,\\
\beta_a&\leftarrow\beta_a+1-r_t.
\end{align}
Only afterward is the baseline updated:
\begin{equation}
\bar\ell_t=0.95\bar\ell_{t-1}+0.05\ell_t^q.
\end{equation}
The first eligible observation initializes the EMA without rewarding an arm,
because no earlier action is pending. A newly selected arm becomes pending
only if its critic optimizer step is successfully applied. A mixed-precision
step skipped for nonfinite gradients therefore creates no pending action
and receives no later credit. In distributed training, rank zero evaluates
the delayed reward, samples the posteriors, and broadcasts the selected arm.
This ordering prevents an action from receiving credit for a loss measured
before it affected the critic.

\textbf{Alternating updates and checkpointing:}
Each iteration first evaluates real speech and a detached generated waveform
with all enabled critics, combines their losses with the schedules and
selected arm, and applies the critic update. It then freezes critic
parameters and optimizes the generator using reconstruction, classical
adversarial and feature-matching losses, and fixed-weight QCD adversarial,
multilevel-feature, and dispersion losses. Generated-branch gradients remain
active, while real targets are detached. Real selected features are cached
from the critic pass to avoid redundant extraction and a second statistics
update. Algorithm~\ref{alg:qcis_training} and
Appendix~\ref{app:optimization} specify the complete update order.
All policy state---Beta parameters, EMA, pending arm, sampled values, and
random-number-generator state---is checkpointed for deterministic
continuation; Appendix~\ref{app:checkpointing} gives the full training-state
contract. The QCD, classical critics, and adaptive policy are discarded after
training and add no inference-time cost.

\textbf{Validation and interpretation:}
Adaptive weighting is compared with every fixed arm in $\mathcal A$.
Evaluation reports seed-level reconstruction and perceptual metrics alongside
arm selection frequencies, posterior trajectories, delayed rewards, raw QCD
losses, warm-up factors, and QCD gradient norms. Policy replay checks whether
saved posterior, EMA, pending-arm, optimizer-success, and RNG state reproduce
subsequent decisions. A reduction in raw QCD loss remains a training
heuristic: it may reflect critic learning, batch difficulty, or generator
progress rather than improved perceptual quality. Evidence for a benefit
therefore requires held-out BWE improvements over fixed-arm controls across
seeds.

\subsection{Optimization, precision, and updates}
\label{app:optimization}
The default configuration uses 50 epochs, batch size 16, AdamW with learning rate $2\times10^{-4}$, $(\beta_1,\beta_2)=(0.8,0.99)$, per-epoch multiplicative decay 0.999, 8,000-sample segments, and a 1,024-point STFT with hop 80 and window 320. Critic and generator gradients are clipped independently to norm 10.

Classical critics may run under BF16 autocast (FP16 fallback), while the generator forward pass defaults to float32. STFT/ISTFT, phase normalization, exponential magnitude reconstruction, chaos statistics, and statevector operations remain float32. TF32, foreach AdamW, pinned memory, persistent workers, and cached Hann windows are enabled when supported. Strict finite checks cover input audio, generated spectra/waveforms, losses, gradients, optimizer state, model parameters, and checkpoint tensors.

Every iteration first generates a detached waveform for the critic step, enables critic gradients, and updates all enabled critics together. It then freezes critic parameters, regenerates the non-detached waveform, computes reconstruction and adversarial/feature terms, and updates the generator. The real selected QCD features from the critic pass are cached for the generator pass, but real normalization moments are not updated twice.

\subsection{Checkpoint and resume semantics}
\label{app:checkpointing}
The generator file contains only generator weights and is sufficient for inference. The training-state file contains every classical critic, \qcd{}, both optimizers, AMP scalers, epoch/global step, active-discriminator list, conditioner mode, fixed-weight override, selected feature metadata, QCD/conditioner moments, Thompson posterior, pending arm, EMA, and NumPy RNG state. Resume rejects incompatible critic sets, conditioner modes, or fixed-weight settings. This contract prevents an apparently successful restart from silently changing the experiment.

\section{Robustness Protocols and Channel Results}
\label{app:robustness}
This section expands Sections~\ref{subsec:noisy_conditions} and~\ref{subsec:channel_robustness} with the exact corruption operators and every completed noise, codec, and packet-loss condition. The zero-shot and matched-mixture blocks remain separate because they answer different questions; neither is pooled with the earlier partitioned cross-band tables in Appendix~\ref{app:comparative}.
\subsection{Noise construction and full condition matrix}
The robustness protocol uses channel seed 2025 and eight fixed noise recordings. Additive noise is cropped/aligned and scaled to the requested SNR after bandwidth limitation; the clean wideband waveform remains the target. For a zero-mean noise segment $n$, the scaling is
\begin{equation}
\tilde n=n\sqrt{\frac{\|x\|_2^2}{\|n\|_2^2\,10^{s/10}+\epsilon}},\qquad x_s=x+\tilde n.
\end{equation}
Each condition is evaluated on the complete 2,937-utterance evaluation list, unlike the smaller partitioned comparisons reproduced later for historical context.

\subsection{Telecommunication Codec transfer}
The experiment settings includes G.711 $\mu$-law/A-law and actual AMR-NB encode/decode at 4.75/12.2~kb/s. The two training regimes are clean-trained zero-shot and codec-mixture matched training. Codec processing is performed at 8~kHz before resampling back to the 16~kHz model rate; the 4~kHz source-band limitation is applied separately. 
Results are shown in Table ~\ref{tab:pcm_codec_results_seed1234}. 

\noindent\textbf{Explanation of Table~\ref{tab:pcm_codec_results_seed1234}.} Rows~\textcircled{\textbf{\scriptsize 1}}--\textcircled{\textbf{\scriptsize 4}} are clean-trained zero-shot transfer; Rows~\textcircled{\textbf{\scriptsize 5}}--\textcircled{\textbf{\scriptsize 8}} use matched training. G.711 remains comparatively stable, whereas AMR-NB dominates the degradation. Matched training helps some AMR metrics but does not remove the severe waveform mismatch.\par\smallskip
\begin{table*}[!tbp]
\centering

\tablefontsize
\setlength{\tabcolsep}{3.5pt}
\resetapprow
\begin{appfit}
\begin{tabular}{lrrrrrr}
\toprule
Condition & LSD & STOI & PESQ & SI-SDR & SI-SNR & N-MOS \\
\midrule
\appnextrow{} Zero-shot G.711 A-law & 0.98540 & 0.94616 & 2.6279 & 13.0536 & 13.0536 & 4.1251 \\
\appnextrow{} Zero-shot G.711 $\mu$-law & 0.98240 & 0.94642 & 2.6779 & 13.0310 & 13.0312 & 4.1388 \\
\appnextrow{} Zero-shot AMR-NB 4.75 kb/s & 1.3650 & 0.82551 & 1.8543 & -16.8879 & -16.8879 & 2.8870 \\
\appnextrow{} Zero-shot AMR-NB 12.2 kb/s & 1.1718 & 0.87424 & 2.4096 & -16.8838 & -16.8838 & 3.8867 \\
\appnextrow{} Matched G.711 A-law & 0.99753 & 0.94616 & 2.7188 & 13.0431 & 13.0432 & 4.2361 \\
\appnextrow{} Matched G.711 $\mu$-law & 0.99340 & 0.94650 & 2.7510 & 13.0239 & 13.0241 & 4.2423 \\
\appnextrow{} Matched AMR-NB 4.75 kb/s & 1.1287 & 0.83834 & 1.8007 & -16.9524 & -16.9524 & 3.4658 \\
\appnextrow{} Matched AMR-NB 12.2 kb/s & 1.0869 & 0.88204 & 2.3321 & -16.8930 & -16.8930 & 4.1298 \\
\bottomrule
\end{tabular}
\end{appfit}
\caption{Telecommunication channel degradation results for 4$\rightarrow$16~kHz, seed 1234; 2,937 utterances per condition. ZS uses clean-trained weights. Matched uses noise-mixture training.}
\label{tab:pcm_codec_results_seed1234}

\end{table*}

\subsection{Packet loss and concealment}
Packet-loss experiments use 20-ms frames. Zero-fill and repeat concealment are tested at independent loss probabilities 0.03/0.05. The first erased frame has no prior received frame to repeat and is zero-filled. A fixed operator seed ensures that compared systems use identical loss masks. Results are shown in Table ~\ref{tab:telephony_loss_results_seed1234}.

\noindent\textbf{Explanation of Table~\ref{tab:telephony_loss_results_seed1234}.} Rows~\textcircled{\textbf{\scriptsize 1}}--\textcircled{\textbf{\scriptsize 2}} use 3\% loss and Rows~\textcircled{\textbf{\scriptsize 3}}--\textcircled{\textbf{\scriptsize 4}} use 5\%. Repeat concealment preserves perceptual scores better, while zero fill can yield higher SI-SNR. This trade-off cautions against treating one aggregate score as channel robustness.\par\smallskip
\begin{table*}[!tbp]
\centering

\tablefontsize
\setlength{\tabcolsep}{3.5pt}
\resetapprow
\begin{appfit}
\begin{tabular}{lrrrrrr}
\toprule
Loss / concealment & LSD & STOI & PESQ & SI-SDR & SI-SNR & N-MOS \\
\midrule
\appnextrow{} 3\% repeat & 0.99700 & 0.93199 & 2.2684 & 9.9102 & 9.9106 & 3.8518 \\
\appnextrow{} 3\% zero & 0.99744 & 0.92676 & 2.1483 & 10.9766 & 10.9768 & 2.7544 \\
\appnextrow{} 5\% repeat & 1.0082 & 0.92139 & 2.0510 & 8.4508 & 8.4513 & 3.5946 \\
\appnextrow{} 5\% zero & 1.0128 & 0.91223 & 1.8910 & 9.9367 & 9.9370 & 2.2671 \\
\bottomrule
\end{tabular}
\end{appfit}
\caption{Packet loss experiments for 4$\rightarrow$16~kHz, seed 1234; 2,937 utterances per condition. Clean-trained full model; independent 20-ms erasures at 3 or 5 percent, with repeat or zero concealment.}
\label{tab:telephony_loss_results_seed1234}

\end{table*}

\subsection{ASR Evaluation and Metric Validity}
\label{sec:ASREvaluation}

Q-BWE is also evaluated on a downstream automatic speech recognition (ASR) task. Following prior work using the DeepSpeech framework \cite{gaddy-klein-2021-improved,hannun2014deep}, we transcribe the band-limited inputs, Q-BWE-enhanced outputs, and corresponding clean speech using the open-source \texttt{deepspeech-0.9.3-models} checkpoint. Word error rate (WER) and character error rate (CER) are computed as
\[
\mathrm{WER}=\frac{S_w+D_w+I_w}{N_w},
\mathrm{CER}=\frac{S_c+D_c+I_c}{N_c},
\]
where $S$, $D$, and $I$ denote substitutions, deletions, and insertions, respectively, and $N$ is the number of words or characters in the human reference.

Numerical human-reference WER/CER results from the capacity study are reported in Appendix~\ref{app:completed_experiments}. For PESQ, nearly all noisy evaluation pairs yield finite scores. Across 293,700 pairwise evaluations, only 18 values from 14 noisy conditions are non-finite. These observations are excluded rather than imputed, and the corresponding aggregate PESQ statistics are computed using the available finite evaluations.

\section{Comparative Results, Cross-Band Evaluation, and Computational Analysis}
\label{app:comparative}
\begin{sloppypar}
This section collects baseline comparisons, cross-band results, computational measurements, high-frequency diagnostics, and perceptual-model analyses that complement the controlled 4$\rightarrow$16~kHz suite. These tables originate from task-specific evaluations with their preprocessing/evaluator settings; they are therefore kept separate from the common-evaluator MRAD+MRPD experiments. Cross-band rows represent models evaluated for the stated source/target bandwidth and should not be interpreted as zero-shot transfer from a single source rate.
\end{sloppypar}

\subsection{Generator Quantum-Conditioning Complexity and Latency}
\label{subsec:generator_quantum_complexity}

Table~\ref{tab:generator_quantum_complexity} reports the arithmetic complexity and measured latency of the classically simulated quantum-conditioning components. The component benchmark uses batch size 1 and an 8,000-sample input at 48~kHz, corresponding to 101 frames, on an NVIDIA GeForce RTX 4090. Measurements use FP32/complex64 precision and 500 repetitions following 50 warm-up iterations. FLOPs are analytic
estimates, and MACs are reported as FLOP/2 equivalents.

\noindent\textbf{Explanation of Table~\ref{tab:generator_quantum_complexity}.} Rows~\textcircled{\textbf{\scriptsize 1}}--\textcircled{\textbf{\scriptsize 4}} cover preprocessing and initialization; Rows~\textcircled{\textbf{\scriptsize 5}}--\textcircled{\textbf{\scriptsize 10}} isolate gates and measurements; Rows~\textcircled{\textbf{\scriptsize 11}}--\textcircled{\textbf{\scriptsize 13}} are composite paths. Zero FLOPs in Rows~\textcircled{\textbf{\scriptsize 4}} and \textcircled{\textbf{\scriptsize 7}} do not mean zero latency, so arithmetic and timing claims are deliberately separated.\par\smallskip
\begin{table*}[!tbp]
\centering

\tablefontsize
\setlength{\tabcolsep}{4pt}
\resetapprow
\begin{appfit}
\begin{tabular}{clccc}
\toprule
Row & Generator quantum component
& MACs (M) & FLOPs (M) & Latency (ms) \\
\midrule
\textcircled{\textbf{\scriptsize 1}}
& Generator\_stats\_FE
& 0.155612 & 0.311225 & 0.523 \\
\textcircled{\textbf{\scriptsize 2}}
& Standardization
& 0.000010 & 0.000020 & 0.055 \\
\textcircled{\textbf{\scriptsize 3}}
& Angle\_encoding
& 0.000008 & 0.000015 & 0.042 \\
\textcircled{\textbf{\scriptsize 4}}
& State\_preparation
& 0.000000 & 0.000000 & 0.036 \\
\textcircled{\textbf{\scriptsize 5}}
& Ry\_single\_gate
& 0.000098 & 0.000196 & 0.372 \\
\textcircled{\textbf{\scriptsize 6}}
& Rz\_single\_gate
& 0.000241 & 0.000482 & 0.136 \\
\textcircled{\textbf{\scriptsize 7}}
& CNOT\_single\_gate
& 0.000000 & 0.000000 & 0.267 \\
\textcircled{\textbf{\scriptsize 8}}
& Bloch\_XYZ\_FE
& 0.002468 & 0.004935 & 2.179 \\
\textcircled{\textbf{\scriptsize 9}}
& Pauli\_XX\_ZZ\_FE
& 0.001592 & 0.003185 & 1.796 \\
\textcircled{\textbf{\scriptsize 10}}
& Quantum\_measurement\_FE
& 0.004207 & 0.008414 & 4.186 \\
\textcircled{\textbf{\scriptsize 11}}
& Full\_quantum\_map
& 0.007617 & 0.015234 & 10.261 \\
\textcircled{\textbf{\scriptsize 12}}
& FiLM\_projection
& 0.064544 & 0.129087 & 0.040 \\
\textcircled{\textbf{\scriptsize 13}}
& Full\_conditioner
& 0.333776 & 0.667552 & 11.008 \\
\bottomrule
\end{tabular}
\end{appfit}
\caption{Generator quantum-conditioning component complexity.
Batch 1, 8,000 samples at 48~kHz (101 frames), NVIDIA GeForce RTX 4090, FP32/complex64, and 500 repetitions after 50 warm-ups. FLOPs are analytic estimates, and MACs are FLOP/2 equivalents. Single gates operate on the full statevector. CNOT is a permutation and has zero arithmetic FLOPs in this simulator. STFT and generator stems are excluded from timing. }
\label{tab:generator_quantum_complexity}

\end{table*}

The preprocessing components in Table~\ref{tab:generator_quantum_complexity}, rows
\textcircled{\textbf{\scriptsize 1}}--\textcircled{\textbf{\scriptsize 3}}, comprise
generator-statistic extraction, standardization, and angle encoding, with measured latencies of 0.523, 0.055, and 0.042~ms, respectively. State preparation in row \textcircled{\textbf{\scriptsize 4}} has zero reported arithmetic FLOPs but takes 0.036~ms. The single-gate benchmarks in
rows \textcircled{\textbf{\scriptsize 5}}--\textcircled{\textbf{\scriptsize 7}} operate on the full statevector. In particular, the CNOT permutation in row \textcircled{\textbf{\scriptsize 7}} takes 0.267~ms despite its zero arithmetic FLOP count, illustrating that arithmetic estimates alone do not capture execution latency. Measurement-feature extraction also contributes measurable overhead. Bloch-coordinate and Pauli-correlation extraction in rows \textcircled{\textbf{\scriptsize 8}} and \textcircled{\textbf{\scriptsize 9}} take 2.179 and 1.796~ms, respectively, while the combined quantum-measurement feature extractor in row \textcircled{\textbf{\scriptsize 10}} takes 4.186~ms. The full quantum map in row \textcircled{\textbf{\scriptsize 11}} takes 10.261~ms, whereas the FiLM projection in row \textcircled{\textbf{\scriptsize 12}} takes 0.040~ms. The complete conditioner in row \textcircled{\textbf{\scriptsize 13}} requires 0.333776~M MAC equivalents,
0.667552~M estimated FLOPs, and 11.008~ms. Because the composite benchmarks include constituent operations, their costs must not beadded to the corresponding component rows.

A separate $16\rightarrow48$~kHz timing record measures Q-BWE generator-plus-ISTFT latency at 30.65~ms for 6.370~s of audio, corresponding to an RTF of 0.0048. This record uses a different
input duration and timing scope from the component benchmark. Consequently, the 11.008-ms conditioner latency in Table~\ref{tab:generator_quantum_complexity}, row
\textcircled{\textbf{\scriptsize 13}}, cannot be added to the generator-plus-ISTFT measurement. Baseline latency rankings require measurements under a matched protocol, and these existing 48~kHz
records do not establish deployment latency for the later $4\rightarrow16$~kHz suite.

\subsection{Task-specific cross-band quality and prior compute measurements}
\label{subsec:crossband_quality}

Table~\ref{tab:qcis-bwe-frequency-range} summarizes task-specific
Q-BWE results over 2,937 VCTK utterances. The upper panel reports
spectral distortion, intelligibility, perceptual quality, and
waveform reconstruction metrics; the lower panel reports NISQA
quality dimensions and ASR word error rate. Matching circled row
numbers identify the same task across both panels. The principal
$4\rightarrow16$~kHz results are reported separately in
Tables~\ref{tab:quick4k_metrics} and~\ref{tab:quick4k_nisqa_dims}.

\noindent\textbf{Explanation of Table~\ref{tab:qcis-bwe-frequency-range}.} Matching Rows~\textcircled{\textbf{\scriptsize 1}}--\textcircled{\textbf{\scriptsize 9}} identify the same bandwidth task in both panels. Higher source rates generally improve reconstruction and WER, but the NISQA dimensions are not uniformly monotone. We hypothesize that reduced extrapolation distance drives the trend; the table does not isolate that mechanism.\par\smallskip
\begin{table*}[!tbp]
\centering

\tablefontsize
\setlength{\tabcolsep}{4pt}
\resetapprow
\begin{appfit}
\begin{tabular}{clrrrrrr}
\toprule
Row & Task (kHz) & LSD & STOI & PESQ & SI-SDR & SI-SNR & N-MOS \\
\midrule
\textcircled{\textbf{\scriptsize 1}}
& 2--16  & 1.12 & 0.860 & 1.69 & 7.25  & 7.25  & 4.12 \\
\textcircled{\textbf{\scriptsize 2}}
& 8--16  & 0.84 & 0.998 & 3.93 & 19.50 & 19.50 & 4.36 \\
\textcircled{\textbf{\scriptsize 3}}
& 2--48  & 1.20 & 0.828 & 1.52 & 7.29  & 7.29  & 3.80 \\
\textcircled{\textbf{\scriptsize 4}}
& 4--48  & 1.21 & 0.927 & 2.42 & 13.44 & 13.44 & 4.00 \\
\textcircled{\textbf{\scriptsize 5}}
& 8--48  & 1.02 & 0.998 & 3.63 & 17.24 & 17.24 & 4.47 \\
\textcircled{\textbf{\scriptsize 6}}
& 12--48 & 0.87 & 1.000 & 4.24 & 19.33 & 19.33 & 4.50 \\
\textcircled{\textbf{\scriptsize 7}}
& 16--48 & 0.81 & 1.000 & 4.62 & 21.84 & 21.84 & 4.51 \\
\textcircled{\textbf{\scriptsize 8}}
& 24--48 & 0.69 & 1.000 & 4.64 & 27.51 & 27.51 & 4.55 \\
\textcircled{\textbf{\scriptsize 9}}
& 32--48 & 0.59 & 1.000 & 4.64 & 32.71 & 32.71 & 4.56 \\
\bottomrule
\end{tabular}
\end{appfit}

\par\medskip

\resetapprow
\begin{appfit}
\begin{tabular}{clrrrrr}
\toprule
Row & Task (kHz)
& Noisiness & Discontinuity & Coloration & Loudness & WER \\
\midrule
\textcircled{\textbf{\scriptsize 1}}
& 2--16  & 4.194 & 4.269 & 3.819 & 4.178 & 0.620 \\
\textcircled{\textbf{\scriptsize 2}}
& 8--16  & 4.330 & 4.410 & 4.061 & 4.304 & 0.050 \\
\textcircled{\textbf{\scriptsize 3}}
& 2--48  & 3.999 & 4.101 & 3.621 & 3.931 & 0.815 \\
\textcircled{\textbf{\scriptsize 4}}
& 4--48  & 4.114 & 4.036 & 3.799 & 4.075 & 0.323 \\
\textcircled{\textbf{\scriptsize 5}}
& 8--48  & 4.364 & 4.390 & 4.225 & 4.385 & 0.073 \\
\textcircled{\textbf{\scriptsize 6}}
& 12--48 & 4.353 & 4.449 & 4.281 & 4.421 & 0.024 \\
\textcircled{\textbf{\scriptsize 7}}
& 16--48 & 4.361 & 4.456 & 4.279 & 4.421 & 0.011 \\
\textcircled{\textbf{\scriptsize 8}}
& 24--48 & 4.359 & 4.484 & 4.322 & 4.442 & 0.003 \\
\textcircled{\textbf{\scriptsize 9}}
& 32--48 & 4.344 & 4.467 & 4.329 & 4.442 & 0.003 \\
\bottomrule
\end{tabular}
\end{appfit}

\caption{Task-specific Q-BWE cross-band results over 2,937 VCTK
utterances. Input--output labels denote sampling rates in kHz.
Circled row numbers identify corresponding tasks across both
panels. WER is a fraction.
Lower LSD and WER indicate better performance; higher values
indicate better performance for the remaining metrics.
Values retain the supplied numeric precision.}
\label{tab:qcis-bwe-frequency-range}

\end{table*}

For the 16~kHz target, Table~\ref{tab:qcis-bwe-frequency-range},
rows \textcircled{\textbf{\scriptsize 1}} and \textcircled{\textbf{\scriptsize 2}},
shows stronger reconstruction scores for the higher source rate.
Moving from $2\rightarrow16$ to $8\rightarrow16$~kHz, LSD decreases
from 1.12 to 0.84, STOI increases from 0.860 to 0.998, and PESQ
increases from 1.69 to 3.93. SI-SDR and SI-SNR increase from
7.25 to 19.50~dB, while WER decreases from 0.620 to 0.050,
equivalent to 62.0\% and 5.0\%, respectively.

The 48~kHz-target results in
Table~\ref{tab:qcis-bwe-frequency-range}, rows
\textcircled{\textbf{\scriptsize 3}}--\textcircled{\textbf{\scriptsize 9}},
show a generally favorable quality trend as the source rate
increases. The $2\rightarrow48$~kHz task in row
\textcircled{\textbf{\scriptsize 3}} obtains STOI of 0.828, PESQ of 1.52,
and WER of 0.815. At $16\rightarrow48$~kHz, row
\textcircled{\textbf{\scriptsize 7}}, these values reach 1.000, 4.62,
and 0.011, respectively. Rows \textcircled{\textbf{\scriptsize 8}} and
\textcircled{\textbf{\scriptsize 9}} report further reductions in LSD to
0.69 and 0.59, with WER of 0.003 for both tasks. STOI values
displayed as 1.000 reflect the reported precision and should
not be interpreted as proof of perfect intelligibility.

The lower panel of Table~\ref{tab:qcis-bwe-frequency-range}
provides complementary predicted-quality measurements.
Comparing rows \textcircled{\textbf{\scriptsize 3}} and
\textcircled{\textbf{\scriptsize 9}}, coloration increases from 3.621 to
4.329 and loudness from 3.931 to 4.442, alongside an increase
in overall N-MOS from 3.80 to 4.56 in the upper panel.
Individual dimensions are not uniformly monotonic: noisiness,
for example, is 4.364 in row \textcircled{\textbf{\scriptsize 5}} and
4.344 in row \textcircled{\textbf{\scriptsize 9}}. These task-specific
results describe quality across source--target rate pairs;
their deployment costs must be assessed using the separately
reported compute and timing protocols.

\subsection{Equations for HF and ITU-Inspired Metrics}
\label{app:hf_itu_metric_equations}

This subsection defines the high-frequency (HF) reconstruction
metrics and the ITU-inspired diagnostic measures used in our
evaluation. The latter borrow front-end or diagnostic concepts from
ITU-R BS.1387, ITU-R BS.1770, ITU-T P.56, P.862, and P.863
\cite{bs1387_2023,itu_bs1770_2023,itu_p862_2001}, but are not ITU-certified standards-like PEAQ, PESQ, or POLQA implementations.

\paragraph{Notation and Evaluation Bands.}
Let $y[n]$ and $\hat{y}[n]$ denote the time-aligned reference and
reconstructed waveforms, respectively, sampled at $f_s$ Hz, and let
$f_{\mathrm N}=f_s/2$ be the Nyquist frequency. For the waveform-domain
HF metrics, the evaluator uses the upper half of the available
one-sided spectrum:
\begin{equation}
f_{\mathrm{HF}}
=
f_{\mathrm N}(1-\eta);
\eta=0.5;
\mathcal F_{\mathrm{HF}}
=
[f_{\mathrm{HF}},f_{\mathrm N}].
\label{eq:appendix_hf_cutoff}
\end{equation}
Thus, $\mathcal F_{\mathrm{HF}}=4$--$8$~kHz for 16-kHz evaluation and
$12$--$24$~kHz for 48-kHz evaluation. Let
\begin{equation}
y_{\mathrm{HF}}=\mathcal H_{f_{\mathrm{HF}}}\{y\},
\qquad
\hat y_{\mathrm{HF}}
=\mathcal H_{f_{\mathrm{HF}}}\{\hat y\},
\label{eq:appendix_hf_filter}
\end{equation}
where $\mathcal H_{f_{\mathrm{HF}}}$ is the implemented fifth-order
Butterworth high-pass filter, applied in zero phase whenever the
signal length permits.

The ITU-inspired analysis uses fixed 4-kHz sub-bands
$
\mathcal B_i\in
\{
[0,4),[4,8),[8,12),[12,16),[16,20),[20,24]
\},
\label{eq:appendix_itu_bands}
$
truncated at Nyquist. The final band includes the Nyquist bin.
At 16~kHz, the reported generated-band diagnostic is therefore the
4--8~kHz sub-band. At 48~kHz, the evaluator additionally forms total-HF
diagnostics over 8--24~kHz.

\paragraph{HF-RMSE and HF-MAE.}
With $T$ aligned samples, the waveform-domain HF errors are
\begin{align}
\mathrm{HF\text{-}RMSE}
&=
\left[
\frac{1}{T}
\sum_{n=1}^{T}
\left(
\hat y_{\mathrm{HF}}[n]-y_{\mathrm{HF}}[n]
\right)^2
\right]^{1/2},
\label{eq:appendix_hf_rmse}\\
\mathrm{HF\text{-}MAE}
&=
\frac{1}{T}
\sum_{n=1}^{T}
\left|
\hat y_{\mathrm{HF}}[n]-y_{\mathrm{HF}}[n]
\right|.
\label{eq:appendix_hf_mae}
\end{align}
Lower values indicate smaller reconstruction errors in the evaluated
HF region.

\paragraph{HF Log-Spectral Distance.}
Let $Y(k,m)$ and $\hat Y(k,m)$ denote the complex STFT coefficients of
the reference and reconstructed signals. The evaluator uses a Hann
window, a nominal 0.64-s analysis interval, a 0.16-s hop, and an FFT
size rounded upward to the next power of two. Define the stabilized
log-power spectra
$
P(k,m)
=
\log_{10}\!\left(
\max\{|Y(k,m)|^2,10^{-8}\}
\right),
\hat P(k,m)
=
\log_{10}\!\left(
\max\{|\hat Y(k,m)|^2,10^{-8}\}
\right).
\label{eq:appendix_hf_log_power}
$
If $\Omega_{\mathrm{HF}}$ is the set of STFT bins in
$\mathcal F_{\mathrm{HF}}$, then
\begin{equation*}
\begin{aligned}
\mathrm{HF\text{-}LSD}
&=
\frac{1}{M}
\sum_{m=1}^{M}
\Biggl[
\frac{1}{|\Omega_{\mathrm{HF}}|}
\\
&\qquad
\sum_{k\in\Omega_{\mathrm{HF}}}
\left(
\hat P(k,m)-P(k,m)
\right)^2
\Biggr]^{1/2}.
\end{aligned}
\label{eq:appendix_hf_lsd}
\end{equation*}

\paragraph{HF Scale-Invariant SDR.}
The reported HF SI-SDR does not perform mean removal. Its projection
coefficient, target component, and residual are

$\alpha_{\mathrm{HF}} =
\frac{\langle \hat{y}_{\mathrm{HF}}, y_{\mathrm{HF}} \rangle + \epsilon}
{\|y_{\mathrm{HF}}\|_2^2 + \epsilon},$ $s_{\mathrm{target}} = \alpha_{\mathrm{HF}} y_{\mathrm{HF}},$ $e_{\mathrm{noise}} = \hat{y}_{\mathrm{HF}} - s_{\mathrm{target}}.$

The metric is
\begin{equation}
\mathrm{HF\text{-}SI\text{-}SDR}
=
10\log_{10}
\frac{
\|s_{\mathrm{target}}\|_2^2+\epsilon
}{
\|e_{\mathrm{noise}}\|_2^2+\epsilon
}.
\label{eq:appendix_hf_sisdr}
\end{equation}
Higher HF SI-SDR is preferred.

\paragraph{HF Crest-Factor Error.}
For a nonzero waveform $z$, the evaluator uses the linear, rather
than decibel, crest factor
\begin{equation}
\mathrm{CF}(z)
=
\frac{
\max_n |z[n]|
}{
\sqrt{T^{-1}\sum_{n=1}^{T}z[n]^2}
}.
\label{eq:appendix_hf_crest_factor}
\end{equation}
The reported discrepancy is
\begin{equation}
D_{\mathrm{CF}}
=
\left|
\mathrm{CF}(\hat y_{\mathrm{HF}})
-
\mathrm{CF}(y_{\mathrm{HF}})
\right|.
\label{eq:appendix_hf_crest_error}
\end{equation}
The implementation assigns a crest factor of zero to an all-zero
signal.

\paragraph{BS.1387-Inspired Spectral:}
The BS.1387-inspired diagnostics use $N=2048$ samples and a
50\%-overlapped scaled Hann window
\begin{equation}
h[r]
=
\frac{1}{2}\sqrt{\frac{8}{3}}
\left[
1-\cos\left(\frac{2\pi r}{N-1}\right)
\right].
\label{eq:appendix_bs1387_window}
\end{equation}
For frame $m$,
\begin{equation}
X(k,m)
=
\frac{1}{N}
\sum_{r=0}^{N-1}
x_m[r]h[r]e^{-j2\pi kr/N}.
\label{eq:appendix_bs1387_stft}
\end{equation}
Let $A(k,m)=|X(k,m)|$ and $Q(k,m)=A(k,m)^2$. The
outer/middle-ear weighting, with $f_k$ expressed in kHz, is

\begin{equation}
\begin{aligned}
W(f_k)
&= -2.184f_k^{-0.8}
+ 6.5e^{-0.6(f_k-3.3)^2}\\
&\quad - 10^{-3}f_k^{3.6}
\quad \mathrm{dB}.
\end{aligned}
\label{eq:appendix_bs1387_ear_weight}
\end{equation}

Including the evaluator's constant playback-calibration gain $g$, the
weighted magnitude and power are

\begin{equation}
\begin{aligned}
\widetilde A(k,m)
&= g\,10^{W(f_k)/20}A(k,m),\\
\widetilde Q(k,m)
&= \widetilde A(k,m)^2.
\end{aligned}
\label{eq:appendix_bs1387_weighted_power}
\end{equation}

Because the same calibration is applied to the reference and
candidate, it does not alter their relative spectral differences.

A BS.1387 diagnostic frame is retained when its reference-band energy
is no more than 60~dB below the maximum reference-band frame energy:

\begin{equation}
\begin{aligned}
\mathcal A_i
=
\Bigg\{
m:\;
&\sum_{k\in\Omega_i}\widetilde Q_y(k,m)\\
&\ge
10^{-6}\max_{\ell}
\sum_{k\in\Omega_i}\widetilde Q_y(k,\ell)
\Bigg\}.
\end{aligned}
\label{eq:appendix_bs1387_activity}
\end{equation}

\paragraph{BS.1387-Inspired Noise-to-Mask Ratio.}
Let $\mathcal G_{250}$ denote Gaussian smoothing along frequency with
a 250-Hz standard deviation. The reference-derived masker is
\begin{equation}
M(k,m)
=
\max\left\{
10^{-2}
\mathcal G_{250}
\left(
\widetilde Q_y(k,m)
\right),
10^{-12}
\right\}.
\label{eq:appendix_masking_estimate}
\end{equation}
The binwise noise-to-mask ratio is
\begin{equation}
R_i(k,m)
=
\frac{
\left|
\widetilde A_y(k,m)
-
\widetilde A_{\hat y}(k,m)
\right|^2
}{
M(k,m)
}.
\label{eq:appendix_nmr_ratio}
\end{equation}
The reported total and segmental NMR values are
\begin{align}
\mathrm{NMR}_i
&=
10\log_{10}
\Biggl[
\frac{1}{
|\mathcal A_i||\Omega_i|
}
\notag\\[-0.2em]
&\qquad
\sum_{m\in\mathcal A_i}
\sum_{k\in\Omega_i}
R_i(k,m)
\Biggr],
\label{eq:appendix_total_nmr}\\
\mathrm{NMR}^{\mathrm{seg}}_i
&=
\frac{1}{|\mathcal A_i|}
\sum_{m\in\mathcal A_i}
10\log_{10}
\Biggl[
\notag\\[-0.2em]
&\qquad
\frac{1}{|\Omega_i|}
\sum_{k\in\Omega_i}R_i(k,m)
\Biggr].
\label{eq:appendix_segmental_nmr}
\end{align}
The relative disturbed-frame rate is

\begin{equation}
\begin{aligned}
D_{\mathrm{RDF},i}
&=
\frac{1}{|\mathcal A_i|}
\sum_{m\in\mathcal A_i}
\mathbbm{1}
\Bigg[
10\log_{10}
\left(
\max_{k\in\Omega_i}R_i(k,m)
\right)\\
&\qquad\ge 1.5
\Bigg].
\end{aligned}
\label{eq:appendix_disturbed_frames}
\end{equation}

\paragraph{BS.1387-Inspired Magnitude and Bandwidth Errors.}
Define the weighted spectral levels

\begin{equation}
\begin{aligned}
L_y(k,m)
&=20\log_{10}
\max\{\widetilde A_y(k,m),10^{-30}\},\\
L_{\hat y}(k,m)
&=20\log_{10}
\max\{\widetilde A_{\hat y}(k,m),10^{-30}\}.
\end{aligned}
\label{eq:appendix_weighted_levels}
\end{equation}

The weighted magnitude error is
\begin{equation}
\begin{aligned}
D_{\mathrm{WM},i}
&=
\frac{1}{
|\mathcal A_i||\Omega_i|
}
\sum_{m\in\mathcal A_i}
\\[-0.2em]
&\qquad
\sum_{k\in\Omega_i}
\left|
L_{\hat y}(k,m)-L_y(k,m)
\right|.
\end{aligned}
\label{eq:appendix_weighted_mag_error}
\end{equation}
The unweighted error $D_{\mathrm{UM},i}$ is defined identically after
replacing $\widetilde A$ by $A$.

For the bandwidth diagnostic, let $\mathcal T_i$ contain the highest
10\% of the frequency bins in $\Omega_i$, and define the local
candidate noise threshold
\begin{equation}
q_i(m)
=
\max_{k\in\mathcal T_i}L_{\hat y}(k,m).
\label{eq:appendix_bandwidth_threshold}
\end{equation}
The occupied reference and candidate edges are

\begin{align}
b_{y,i}(m)
&=
\max\Bigl\{
f_k:
\notag\\
&\qquad
L_y(k,m)\ge q_i(m)+10
\Bigr\},
\label{eq:appendix_ref_bandwidth_edge}\\
b_{\hat y,i}(m)
&=
\max\Bigl\{
f_k\le b_{y,i}(m):
\notag\\
&\qquad
L_{\hat y}(k,m)\ge q_i(m)+5
\Bigr\}.
\label{eq:appendix_test_bandwidth_edge}
\end{align}

If a qualifying bin does not exist, the implementation uses the
lowest frequency of the sub-band. The bandwidth error is
\begin{equation}
\begin{aligned}
D_{\mathrm{BW},i}
&=
\Biggl|
\frac{1}{|\mathcal A_i|}
\sum_{m\in\mathcal A_i}b_{y,i}(m)
\\
&\qquad-
\frac{1}{|\mathcal A_i|}
\sum_{m\in\mathcal A_i}b_{\hat y,i}(m)
\Biggr|.
\end{aligned}
\label{eq:appendix_bandwidth_error}
\end{equation}

\paragraph{BS.1387-Inspired Energy Ratio and Spectral Correlation.}
The active-frame sub-band energy ratios are

\begin{equation}
\begin{aligned}
r_i(z)
&=
\frac{
\sum_{m\in\mathcal A_i}
\sum_{k\in\Omega_i}Q_z(k,m)
}{
\sum_{m\in\mathcal A_i}
\sum_{k\in\Omega}Q_z(k,m)+\epsilon
},\\
D_{\mathrm{ER},i}
&=
|r_i(\hat y)-r_i(y)|.
\end{aligned}
\label{eq:appendix_bs1387_energy_ratio}
\end{equation}

Let $\operatorname{vec}(\cdot)$ flatten all active frames and bins.
The spectral correlation is
\begin{equation}
\rho_{\mathrm{spec},i}
=
\operatorname{corr}
\left(
\operatorname{vec}
\left[
\log_{10}\widetilde Q_y
\right],
\operatorname{vec}
\left[
\log_{10}\widetilde Q_{\hat y}
\right]
\right),
\label{eq:appendix_spectral_correlation}
\end{equation}
where both arguments are restricted to
$\mathcal A_i\times\Omega_i$. Higher correlation is preferred.

\paragraph{BS.1387-Inspired Spectral-Shape Errors.}
For each frame, let

\begin{equation}
\begin{aligned}
p_z(k,m)
&= \widetilde A_z(k,m)^2,\\
E_z(m)
&= \sum_{k\in\Omega_i}p_z(k,m)+\epsilon.
\end{aligned}
\label{eq:appendix_shape_power}
\end{equation}

The framewise flatness, centroid, and spread are
\begin{align}
\phi_z(m)
&=
\frac{
\exp\!\left[
|\Omega_i|^{-1}
\sum_{k\in\Omega_i}\log p_z(k,m)
\right]
}{
|\Omega_i|^{-1}
\sum_{k\in\Omega_i}p_z(k,m)+\epsilon
},
\label{eq:appendix_spectral_flatness}\\
c_z(m)
&=
\frac{
\sum_{k\in\Omega_i}f_kp_z(k,m)
}{
E_z(m)
},
\label{eq:appendix_spectral_centroid}\\
s_z(m)
&=
\left[
\frac{
\sum_{k\in\Omega_i}
p_z(k,m)(f_k-c_z(m))^2
}{
E_z(m)
}
\right]^{1/2}.
\label{eq:appendix_spectral_spread}
\end{align}
The corresponding errors are

\begin{equation}
\begin{aligned}
D_{\phi,i}
&=
\left|
\overline{\phi}_{\hat y}-\overline{\phi}_y
\right|,\\
D_{c,i}
&=
\left|
\overline{c}_{\hat y}-\overline{c}_y
\right|,\\
D_{s,i}
&=
\left|
\overline{s}_{\hat y}-\overline{s}_y
\right|.
\end{aligned}
\label{eq:appendix_spectral_shape_errors}
\end{equation}

where the overbar denotes the mean over frames.

For spectral flux, define the normalized magnitude vector
\begin{equation}
\mathbf v_z(m)
=
\frac{
[\widetilde A_z(k,m)]_{k\in\Omega_i}
}{
\|[\widetilde A_z(k,m)]_{k\in\Omega_i}\|_2+\epsilon
}.
\label{eq:appendix_normalized_spectral_shape}
\end{equation}
The implemented flux and its error are

\begin{align}
F_i(z)
&=
\frac{1}{M-1}
\sum_{m=2}^{M}
\Bigg[
\frac{1}{|\Omega_i|}
\sum_{k\in\Omega_i}
\notag\\
&\qquad
\left(
v_z(k,m)-v_z(k,m-1)
\right)^2
\Bigg]^{1/2},
\label{eq:appendix_spectral_flux_value}\\
D_{\mathrm{SF},i}
&=
|F_i(\hat y)-F_i(y)|.
\label{eq:appendix_spectral_flux_error}
\end{align}

\paragraph{BS.1387-Inspired Diagnostic Quality Score.}
The evaluator also exports an ODG-like score, denoted $Q_i^\star$
rather than ODG because it is not the normative BS.1387 mapping:
\begin{equation}
Q_i^\star
=
-4\,
\frac{
u_{\mathrm{WM},i}
+
u_{\mathrm{NMR},i}
+
u_{\mathrm{BW},i}
+
u_{\mathrm{ER},i}
}{4},
\label{eq:appendix_qstar}
\end{equation}
where
\begin{align}
u_{\mathrm{WM},i}
&=
\operatorname{clip}_{[0,1]}
\left(
\frac{D_{\mathrm{WM},i}}{10}
\right),\\
u_{\mathrm{NMR},i}
&=
\operatorname{clip}_{[0,1]}
\left(
\frac{\mathrm{NMR}_i+10}{20}
\right),\\
u_{\mathrm{BW},i}
&=
\operatorname{clip}_{[0,1]}
\left(
\frac{D_{\mathrm{BW},i}}
{f_{i,\mathrm{high}}-f_{i,\mathrm{low}}}
\right),\\
u_{\mathrm{ER},i}
&=
\operatorname{clip}_{[0,1]}
\left(D_{\mathrm{ER},i}\right).
\end{align}
Consequently, $Q_i^\star\in[-4,0]$, with values closer to zero
indicating smaller diagnostic error.

\paragraph{Reference-Derived Active-Speech Mask.}
The remaining ITU-inspired diagnostics first divide the reference into
20-ms frames with a 10-ms hop. If
\begin{equation}
E_y(m)
=
\frac{1}{N_m}
\sum_{n\in\mathcal W_m}y[n]^2,
\label{eq:appendix_activity_energy}
\end{equation}
the reference-derived activity indicator is
\begin{equation}
a(m)
=
\mathbbm{1}
\left[
E_y(m)
\ge
10^{-4.5}\max_{\ell}E_y(\ell)
\right].
\label{eq:appendix_p56_activity}
\end{equation}
The same mask is applied to both signals. Let $y^a$ and $\hat y^a$
denote the retained active samples, and let
\begin{equation}
y_i^a=\mathcal B_i\{y^a\},
\qquad
\hat y_i^a=\mathcal B_i\{\hat y^a\},
\label{eq:appendix_active_band_signals}
\end{equation}
where $\mathcal B_i$ is the implemented sixth-order Butterworth
low-pass or band-pass filter.

\paragraph{BS.1770-Inspired K-Weighted Loudness.}
Let $\mathcal K(\cdot)$ denote the evaluator's K-weighting cascade.
At 48~kHz it uses the BS.1770 shelving and RLB biquad coefficients.
The sub-band loudness estimate is

\begin{equation}
\begin{aligned}
L_{K,i}(z)
&= -0.691 \\
&\quad + 10\log_{10}
\left[
\frac{1}{T_a}
\sum_{n=1}^{T_a}
\mathcal K(z_i^a)[n]^2
+\epsilon
\right].
\end{aligned}
\label{eq:appendix_k_loudness}
\end{equation}

The signed loudness difference and absolute loudness-distance error are
\begin{equation}
\Delta L_{K,i}
=
L_{K,i}(\hat y)-L_{K,i}(y),
\qquad
D_{K,i}
=
|\Delta L_{K,i}|.
\label{eq:appendix_k_loudness_error}
\end{equation}

For 20-ms K-weighted frames, define

\begin{equation}
\begin{aligned}
L_i(m;z)
&= -0.691 \\
&\quad + 10\log_{10}
\Biggl[
\frac{1}{N_m}
\sum_{n\in\mathcal W_m}
\mathcal K(z_i^a)[n]^2
\\
&\qquad
+\epsilon
\Biggr].
\end{aligned}
\label{eq:appendix_short_term_loudness}
\end{equation}

The implemented loudness-range proxy is
\begin{equation}
\mathrm{LR}_i(z)
=
\operatorname{Pct}_{95}
\{L_i(m;z)\}
-
\operatorname{Pct}_{10}
\{L_i(m;z)\},
\label{eq:appendix_loudness_range}
\end{equation}
with error
\begin{equation}
D_{\mathrm{LR},i}
=
|\mathrm{LR}_i(\hat y)-\mathrm{LR}_i(y)|.
\label{eq:appendix_loudness_range_error}
\end{equation}
The short-term loudness-correlation loss and standard-deviation error are
\begin{align}
D_{\rho L,i}
&=
1-\max\left\{
\operatorname{corr}
\left(
L_i(\cdot;y),L_i(\cdot;\hat y)
\right),0
\right\},
\label{eq:appendix_loudness_corr_loss}\\
D_{\sigma L,i}
&=
\left|
\operatorname{std}_m L_i(m;y)
-
\operatorname{std}_m L_i(m;\hat y)
\right|.
\label{eq:appendix_loudness_std_error}
\end{align}

\paragraph{P.56-Inspired Active-Level, Energy-Ratio, Crest, and Duty Metrics.}
The implemented active RMS level is
\begin{equation}
L_{A,i}(z)
=
20\log_{10}
\left[
\sqrt{
\frac{1}{T_a}
\sum_{n=1}^{T_a}z_i^a[n]^2+\epsilon
}
+\epsilon
\right],
\label{eq:appendix_active_level}
\end{equation}
and its error is
\begin{equation}
D_{A,i}
=
|L_{A,i}(\hat y)-L_{A,i}(y)|.
\label{eq:appendix_active_level_error}
\end{equation}

The active sub-band energy ratio is

\begin{equation}
\begin{aligned}
r_{A,i}(z)
&=
\frac{
\sum_n z_i^a[n]^2
}{
\sum_n z^a[n]^2+\epsilon
},\\
D_{A\mathrm{E},i}
&=
|r_{A,i}(\hat y)-r_{A,i}(y)|.
\end{aligned}
\label{eq:appendix_active_energy_ratio}
\end{equation}

For a total HF region, $\mathcal B_i$ in
Eq.~\eqref{eq:appendix_active_energy_ratio} is replaced by the complete
declared reconstructed band.

The active crest factor is expressed in decibels:
\begin{equation}
\mathrm{CF}_{A,i}(z)
=
20\log_{10}
\frac{
\max_n|z_i^a[n]|+\epsilon
}{
\sqrt{T_a^{-1}\sum_nz_i^a[n]^2+\epsilon}+\epsilon
},
\label{eq:appendix_active_crest_factor}
\end{equation}
with
\begin{equation}
D_{\mathrm{ACF},i}
=
|\mathrm{CF}_{A,i}(\hat y)-\mathrm{CF}_{A,i}(y)|.
\label{eq:appendix_active_crest_error}
\end{equation}

For the band-specific duty cycle, let
\begin{equation}
E_i(m;z)
=
\frac{1}{N_m}
\sum_{n\in\mathcal W_m}z_i^a[n]^2.
\label{eq:appendix_band_frame_energy}
\end{equation}
Then

\begin{equation}
\begin{aligned}
d_i(z)
&=
\frac{1}{M}
\sum_{m=1}^{M}
\mathbbm{1}
\Big[
E_i(m;z)
\\
&\qquad\ge
10^{-4.5}\max_{\ell}E_i(\ell;z)
\Big],\\
D_{d,i}
&=
|d_i(\hat y)-d_i(y)|.
\end{aligned}
\label{eq:appendix_duty_cycle}
\end{equation}

\paragraph{P.862-Inspired Spectral Disturbances.}
The disturbance metrics use the BS.1387-weighted STFT levels of the
active signals:

\begin{equation}
\begin{aligned}
\Delta_i(k,m)
&=
20\log_{10}
\left[
\max\{
\widetilde A_{\hat y}(k,m),\epsilon
\}
\right]\\
&\quad -
20\log_{10}
\left[
\max\{
\widetilde A_y(k,m),\epsilon
\}
\right].
\end{aligned}
\label{eq:appendix_log_spectral_difference}
\end{equation}

Let $[u]_+=\max(u,0)$. The added and missing components are
\begin{align}
D_{\mathrm{add},i}
&=
\frac{1}{M|\Omega_i|}
\sum_{m=1}^{M}
\sum_{k\in\Omega_i}
[\Delta_i(k,m)]_+,
\label{eq:appendix_added_disturbance}\\
D_{\mathrm{miss},i}
&=
\frac{1}{M|\Omega_i|}
\sum_{m=1}^{M}
\sum_{k\in\Omega_i}
[-\Delta_i(k,m)]_+.
\label{eq:appendix_missing_disturbance}
\end{align}
The symmetric and asymmetric disturbances are
\begin{align}
D_{\mathrm{sym},i}
&=
\frac{1}{M|\Omega_i|}
\sum_{m,k}
|\Delta_i(k,m)|
\notag\\
&=
D_{\mathrm{add},i}+D_{\mathrm{miss},i},
\label{eq:appendix_symmetric_disturbance}\\
D_{\mathrm{asym},i}
&=
D_{\mathrm{miss},i}
+
1.5D_{\mathrm{add},i}.
\label{eq:appendix_asymmetric_disturbance}
\end{align}
The remaining distributional diagnostics are

\begin{align}
D_{\mathrm{skew},i}
&=
D_{\mathrm{add},i}
-D_{\mathrm{miss},i},
\label{eq:appendix_disturbance_skew}\\
D_{\mathrm{rms},i}
&=
\Bigg[
\frac{1}{M|\Omega_i|}
\sum_{m,k}
\notag\\
&\qquad
\Delta_i(k,m)^2
\Bigg]^{1/2},
\label{eq:appendix_disturbance_rms}\\
D_{95,i}
&=
\operatorname{Pct}_{95}
\Big\{
|\Delta_i(k,m)|:
\notag\\
&\qquad
k\in\Omega_i,\ 1\le m\le M
\Big\}.
\label{eq:appendix_disturbance_p95}
\end{align}

\paragraph{P.863-Inspired Temporal-Envelope Metrics.}
The implementation uses the analytic-signal envelope rather than a
framewise RMS envelope:
\begin{equation}
e_i[n;z]
=
\left|
z_i^a[n]
+
j\,\mathscr H\{z_i^a\}[n]
\right|,
\label{eq:appendix_hilbert_envelope}
\end{equation}
where $\mathscr H$ denotes the Hilbert transform. The temporal-envelope
error is the mean absolute log-envelope difference

\begin{equation}
\begin{aligned}
D_{\mathrm{env},i}
&=
\frac{1}{T_a}
\sum_{n=1}^{T_a}
\Big|
20\log_{10}
\left(e_i[n;\hat y]+\epsilon\right)\\
&\quad -
20\log_{10}
\left(e_i[n;y]+\epsilon\right)
\Big|.
\end{aligned}
\label{eq:appendix_envelope_error}
\end{equation}

The envelope-correlation loss is
\begin{equation}
D_{\rho e,i}
=
1-
\max\left\{
\operatorname{corr}
\left(
e_i[\cdot;y],e_i[\cdot;\hat y]
\right),0
\right\}.
\label{eq:appendix_envelope_corr_loss}
\end{equation}

Let
\begin{equation}
\delta_i[n;z]
=
\frac{
e_i[n+1;z]-e_i[n;z]
}{
T_a^{-1}\sum_{\ell=1}^{T_a}e_i[\ell;z]+\epsilon
}.
\label{eq:appendix_normalized_envelope_difference}
\end{equation}
The normalized attack and decay rates are

\begin{equation}
\begin{aligned}
a_i(z)
&=
\frac{1}{T_a-1}
\sum_n[\delta_i[n;z]]_+,\\
c_i(z)
&=
\frac{1}{T_a-1}
\sum_n[-\delta_i[n;z]]_+.
\end{aligned}
\label{eq:appendix_attack_decay_rates}
\end{equation}

with errors

\begin{equation}
\begin{aligned}
D_{\mathrm{att},i}
&=
|a_i(\hat y)-a_i(y)|,\\
D_{\mathrm{dec},i}
&=
|c_i(\hat y)-c_i(y)|.
\end{aligned}
\label{eq:appendix_attack_decay_errors}
\end{equation}

\paragraph{P.863-Inspired Modulation Metrics.}
Divide $e_i[n;z]$ into 20-ms frames with a 10-ms hop and define the
framewise envelope RMS
\begin{equation}
\mu_i(m;z)
=
\left[
\frac{1}{N_m}
\sum_{n\in\mathcal W_m}
e_i[n;z]^2+\epsilon
\right]^{1/2}.
\label{eq:appendix_envelope_modulation_series}
\end{equation}
The modulation-correlation loss and modulation-depth error are

\begin{align}
D_{\mathrm{modcorr},i}
&=
1-
\max\Big\{
\operatorname{corr}
\Bigl(
\mu_i(\cdot;y),
\notag\\[-0.2em]
&\qquad
\mu_i(\cdot;\hat y)
\Bigr),
0
\Big\},
\label{eq:appendix_modulation_corr_loss}\\
D_{\mathrm{moddepth},i}
&=
\Bigg|
\frac{
\operatorname{std}_m\mu_i(m;\hat y)
}{
\operatorname{mean}_m\mu_i(m;\hat y)+\epsilon
}
\notag\\
&\quad -
\frac{
\operatorname{std}_m\mu_i(m;y)
}{
\operatorname{mean}_m\mu_i(m;y)+\epsilon
}
\Bigg|.
\label{eq:appendix_modulation_depth_error}
\end{align}

For the modulation centroid, subtract the mean envelope and define

\begin{equation}
\begin{aligned}
E_i^{\circ}[n;z]
&=
e_i[n;z]
-
\frac{1}{T_a}
\sum_{\ell=1}^{T_a}e_i[\ell;z],\\
S_i^{\mathrm{mod}}(\nu;z)
&=
\left|
\operatorname{DFT}
\{E_i^{\circ}[\cdot;z]\}(\nu)
\right|^2.
\end{aligned}
\label{eq:appendix_modulation_spectrum}
\end{equation}

After setting the DC component to zero,

\begin{equation}
\begin{aligned}
c_i^{\mathrm{mod}}(z)
&=
\frac{
\sum_{\nu>0}
\nu S_i^{\mathrm{mod}}(\nu;z)
}{
\sum_{\nu>0}
S_i^{\mathrm{mod}}(\nu;z)+\epsilon
},\\
D_{\mathrm{modcent},i}
&=
\left|
c_i^{\mathrm{mod}}(\hat y)
-
c_i^{\mathrm{mod}}(y)
\right|.
\end{aligned}
\label{eq:appendix_modulation_centroid}
\end{equation}

\paragraph{Custom Composite Quality Index.}
The evaluator's custom quality index combines the K-weighted
loudness error, active-energy-ratio error, asymmetric disturbance,
temporal-envelope error, and modulation-correlation loss:

\begin{equation}
\begin{aligned}
\mathrm{QI}_i
&=100\Bigg[
1-\frac{1}{5}\Bigg(\\
&\quad
\operatorname{clip}_{[0,1]}
\frac{D_{K,i}}{12}\\
&\quad+
\operatorname{clip}_{[0,1]}
\frac{D_{A\mathrm{E},i}}{0.02}\\
&\quad+
\operatorname{clip}_{[0,1]}
\frac{D_{\mathrm{asym},i}}{12}\\
&\quad+
\operatorname{clip}_{[0,1]}
\frac{D_{\mathrm{env},i}}{12}\\
&\quad+
\operatorname{clip}_{[0,1]}
D_{\mathrm{modcorr},i}
\Bigg)\Bigg].
\end{aligned}
\label{eq:appendix_custom_quality_index}
\end{equation}

Thus, $\mathrm{QI}_i\in[0,100]$, and higher values are preferred.
The score is unavailable when any of its five components is
unavailable.

\paragraph{Sub-Band and Utterance Aggregation.}
At 48~kHz, total-HF versions of the ITU-inspired metrics are formed
from the 8--12, 12--16, 16--20, and 20--24~kHz sub-bands using

\begin{equation}
\begin{aligned}
\mathbf w
&= (0.40,0.30,0.20,0.10),\\
D_{\mathrm{HF}}
&=
\frac{
\sum_{i=1}^{4}w_iD_i
}{
\sum_{i=1}^{4}w_i
}.
\end{aligned}
\label{eq:appendix_hf_weighted_aggregation}
\end{equation}

The total active-HF energy ratio is instead recomputed directly over
8~kHz--Nyquist, and total asymmetric disturbance is recomputed from
all STFT bins in that region. At 16~kHz, where an 8--24~kHz region
does not exist, the evaluator reports the available 0--4 and
4--8~kHz sub-band diagnostics without constructing a fictitious
8--24~kHz total.

Finally, every metric is first evaluated per utterance. For $N_u$
test utterances, the reported corpus mean is
\begin{equation}
\overline D
=
\frac{1}{N_u}
\sum_{u=1}^{N_u}D^{(u)}.
\label{eq:appendix_metric_aggregation}
\end{equation}
Lower values are preferred for all quantities explicitly named
``error,'' ``distance,'' ``loss,'' ``disturbance,'' or NMR. Higher
values are preferred for HF SI-SDR, spectral correlation, the custom
quality index, and the BS.1387-inspired $Q^\star$ score, for which
values closer to zero are better. Reference and candidate energy
ratios, levels, loudness ranges, and duty cycles are descriptive
values; their absolute discrepancies are minimized. None of the
ITU-inspired quantities should be interpreted as a certified score
under the corresponding ITU recommendation.

\subsection{Prior model/sub-band and NISQA comparisons}

\noindent\textbf{Explanation of Table~\ref{tab:ap_udss_v18_selected_itu_metrics_}.}
Rows are grouped in pairs for AP-BWE and Q-BWE within each sub-band.
Q-BWE is strongest in the observed 0--4~kHz band
(Rows~\textcircled{\textbf{\scriptsize 1}}--\textcircled{\textbf{\scriptsize 2}})
but is not uniformly best above 4~kHz.
This supports reporting source preservation and generated-band fidelity separately.
\par\smallskip

\begin{table*}[!tbp]
\centering
\tablefontsize
\setlength{\tabcolsep}{4pt}
\resetapprow
\begin{appfit}
\begin{tabular}{llrrr}
\toprule
Band (kHz) & Model & $D_{SF,i}$ & $D_{\sigma L,i}$ & $D_{miss,i}$ \\
\midrule
\appnextrow{} 0--4 & AP-BWE & $1.04{\times}10^{-4}$ & 0.0521 & 0.227 \\
\appnextrow{} 0--4 & Q-BWE & \textbf{$2.30{\times}10^{-5}$} & 0.00467 & 0.0290 \\
\appnextrow{} 4--8 & AP-BWE & \textbf{$1.57{\times}10^{-4}$} & 0.375 & 0.704 \\
\appnextrow{} 4--8 & Q-BWE & $3.56{\times}10^{-4}$ & 0.215 & 1.610 \\
\appnextrow{} 8--12 & AP-BWE & 0.00114 & 0.831 & 3.979 \\
\appnextrow{} 8--12 & Q-BWE & 0.000968 & 0.726 & 4.832 \\
\appnextrow{} 12--16 & AP-BWE & 0.00124 & 0.689 & 3.214 \\
\appnextrow{} 12--16 & Q-BWE & 0.00113 & 0.992 & 3.259 \\
\appnextrow{} 16--20 & AP-BWE & 0.00113 & 0.775 & 3.616 \\
\appnextrow{} 16--20 & Q-BWE & 0.000978 & 0.830 & 2.549 \\
\appnextrow{} 20--24 & AP-BWE & 0.00133 & 0.942 & 3.342 \\
\appnextrow{} 20--24 & Q-BWE & 0.000833 & 1.079 & 2.981 \\
\bottomrule
\end{tabular}
\end{appfit}
\caption{Sub-band comparison of AP-BWE and Q-BWE using evaluator-defined spectral-flatness, loudness-variation, and missing-disturbance proxies. Lower is better.} 
\label{tab:ap_udss_v18_selected_itu_metrics_}
\end{table*}

\noindent\textbf{Explanation of Table~\ref{tab:nisqa_mos_stats_udss_1000_}.}
Rows~\textcircled{\textbf{\scriptsize 1}}--\textcircled{\textbf{\scriptsize 4}}
report utterance-level and speaker-level MOS descriptive statistics
for AP-BWE and Q-BWE. Q-BWE achieves higher mean predicted MOS at both aggregation levels. 
\par\smallskip

\begin{table*}[!tbp]
\centering
\tablefontsize
\setlength{\tabcolsep}{3pt}
\resetapprow
\begin{appfit}
\begin{tabular}{llccccc}
\hline
Agg. & Method & $N$ & Mean $\uparrow$ & SD & SEM & 95\% CI $\uparrow$ \\
\hline
\appnextrow{} Utt. & AP-BWE & 1000 & 4.458 & 0.472 & 0.015 & [4.429, 4.488] \\
\appnextrow{} Utt. & Q-BWE  & 1000 & 4.498 & 0.453 & 0.014 & [4.470, 4.526] \\
\appnextrow{} Spk. & AP-BWE & 7 & 4.448 & 0.196 & 0.074 & [4.267, 4.630] \\
\appnextrow{} Spk. & Q-BWE  & 7 & 4.491 & 0.151 & 0.057 & [4.351, 4.630] \\
\hline
\end{tabular}
\end{appfit}
\caption{NISQA-MOS descriptive statistics for AP-BWE and Q-BWE across 1,000 utterances and seven speakers. Utt.: utterance-level aggregation; Spk.: speaker-level aggregation. Mean, standard deviation (SD), standard error of the mean (SEM), and 95\% confidence intervals (CI) are reported. Higher predicted MOS indicates better perceptual speech quality.}
\label{tab:nisqa_mos_stats_udss_1000_}
\end{table*}

Q-BWE obtains a higher utterance-level mean NISQA-MOS (4.498 vs.\ 4.458) and speaker-level mean (4.491 vs.\ 4.448) compared to AP-BWE. 

\subsection{Perceptual-score and subjective-evaluation boundary}
Table~\ref{tab:nisqa_mos_stats_udss_1000_} remain diagnostic evidence for MOS statistical analysis. 

\subsection{Cross-band diagnostic proxies}
All three diagnostic tables below use custom ITU-inspired proxies, not certified PEAQ/POLQA scores. The target-band aggregates span 0--8~kHz for 16~kHz targets and 0--24~kHz for 48~kHz targets. Each table retains the original task-specific values and separates two measurement panels for readability.

\noindent\textbf{Explanation of Table~\ref{tab:qcis-itu-spectral-level-12}.} Rows~\textcircled{\textbf{\scriptsize 1}}--\textcircled{\textbf{\scriptsize 10}} in each panel share the same bandwidth tasks. Errors generally shrink as the source rate approaches the target, especially within the 48-kHz group. This pattern is consistent with less missing spectrum, but the custom proxies are not certified ITU scores.\par\smallskip
\begin{table*}[!tbp]
\centering

\tablefontsize
\setlength{\tabcolsep}{4pt}
\resetapprow
\begin{appfit}
\begin{tabular}{lrrrrrr}
\toprule
Task & \shortstack{Flux err.\\$10^{-3}$} & Flat. err. & \shortstack{Cent. err.\\Hz} & \shortstack{Spread err.\\Hz} & \shortstack{NMR\\dB} & \shortstack{BW err.\\Hz} \\
\midrule
\appnextrow{} 2--16 & 1.336 & 0.032 & 189.5 & 109.8 & 22.12 & 352.4 \\
\appnextrow{} 4--16 & 1.458 & 0.043 & 178.1 & 133.1 & 18.39 & 417.7 \\
\appnextrow{} 8--16 & 0.876 & 0.024 & 153.6 & 85.0 & 11.87 & 180.1 \\
\appnextrow{} 2--48 & 1.422 & 0.039 & 158.8 & 91.4 & 27.97 & 412.7 \\
\appnextrow{} 4--48 & 1.319 & 0.035 & 181.3 & 98.9 & 22.94 & 392.2 \\
\appnextrow{} 8--48 & 1.315 & 0.028 & 148.9 & 75.7 & 19.57 & 290.2 \\
\appnextrow{} 12--48 & 0.857 & 0.026 & 105.7 & 61.7 & 16.46 & 146.8 \\
\appnextrow{} 16--48 & 0.714 & 0.030 & 118.9 & 59.6 & 13.57 & 114.4 \\
\appnextrow{} 24--48 & 0.610 & 0.035 & 125.8 & 66.5 & 8.50 & 85.5 \\
\appnextrow{} 32--48 & 0.541 & 0.034 & 103.8 & 51.8 & 3.73 & 62.9 \\
\bottomrule\end{tabular}
\end{appfit}
\par\medskip
\resetapprow
\begin{appfit}
\begin{tabular}{lrrrrrr}
\toprule
Task & \shortstack{$\Delta K$\\LU} & \shortstack{$\Delta\sigma_L$\\LU} & \shortstack{$\Delta R_L$\\LU} & $1-\rho_L$ & \shortstack{$\Delta L_A$\\dB} & \shortstack{$\Delta C_A$\\dB} \\
\midrule
\appnextrow{} 2--16 & 3.116 & 0.989 & 3.167 & 0.091 & 3.112 & 1.716 \\
\appnextrow{} 4--16 & 3.845 & 1.177 & 3.660 & 0.073 & 3.844 & 1.336 \\
\appnextrow{} 8--16 & 2.946 & 0.853 & 2.798 & 0.029 & 2.946 & 0.984 \\
\appnextrow{} 2--48 & 5.050 & 1.889 & 6.284 & 0.389 & 5.018 & 2.609 \\
\appnextrow{} 4--48 & 6.600 & 2.430 & 7.898 & 0.311 & 6.578 & 2.571 \\
\appnextrow{} 8--48 & 3.934 & 1.316 & 4.456 & 0.182 & 3.930 & 1.961 \\
\appnextrow{} 12--48 & 2.094 & 0.677 & 2.450 & 0.108 & 2.093 & 1.780 \\
\appnextrow{} 16--48 & 1.700 & 0.641 & 2.104 & 0.097 & 1.700 & 1.688 \\
\appnextrow{} 24--48 & 1.352 & 0.544 & 1.672 & 0.080 & 1.352 & 1.542 \\
\appnextrow{} 32--48 & 1.318 & 0.433 & 1.343 & 0.046 & 1.318 & 1.095 \\
\bottomrule\end{tabular}
\end{appfit}
\par\medskip
\caption{Spectral, loudness, and active-level anlysis of \method{}, in two panels. Tasks are input--output sampling rates (kHz); values are means over 2,937 VCTK utterances. Errors are lower-is-better; $Q_{HF}$ is higher-is-better; reference/test values describe activity.}
\label{tab:qcis-itu-spectral-level-12}

\end{table*}

\noindent\textbf{Explanation of Table~\ref{tab:qcis-itu-activity-disturbance-12}.} Rows~\textcircled{\textbf{\scriptsize 1}}--\textcircled{\textbf{\scriptsize 10}} align across the activity and disturbance panels. The narrowest missing-band cases reduce duty and energy-ratio errors, while missing/added disturbance remains asymmetric. The analysis assumes fixed evaluator definitions across rates.\par\smallskip
\begin{table*}[!tbp]
\centering

\tablefontsize
\setlength{\tabcolsep}{4pt}
\resetapprow
\begin{appfit}
\begin{tabular}{lrrrrrr}
\toprule
Task & Duty err. & Duty ref. & Duty test & E-ratio err. & E-ratio ref. & E-ratio test \\
\midrule
\appnextrow{} 2--16 & 0.051 & 0.891 & 0.884 & 0.0122 & 0.997 & 0.998 \\
\appnextrow{} 4--16 & 0.052 & 0.891 & 0.915 & 0.0133 & 0.997 & 0.999 \\
\appnextrow{} 8--16 & 0.036 & 0.891 & 0.905 & 0.0113 & 0.997 & 0.999 \\
\appnextrow{} 2--48 & 0.065 & 0.923 & 0.987 & 0.0063 & 0.996 & 0.999 \\
\appnextrow{} 4--48 & 0.071 & 0.923 & 0.993 & 0.0071 & 0.996 & 1.000 \\
\appnextrow{} 8--48 & 0.045 & 0.923 & 0.952 & 0.0054 & 0.996 & 0.998 \\
\appnextrow{} 12--48 & 0.025 & 0.923 & 0.936 & 0.0023 & 0.996 & 0.996 \\
\appnextrow{} 16--48 & 0.017 & 0.923 & 0.923 & 0.0013 & 0.996 & 0.996 \\
\appnextrow{} 24--48 & 0.008 & 0.923 & 0.928 & 0.0006 & 0.996 & 0.996 \\
\appnextrow{} 32--48 & 0.004 & 0.923 & 0.925 & 0.0002 & 0.996 & 0.996 \\
\bottomrule\end{tabular}
\end{appfit}
\par\medskip
\resetapprow
\begin{appfit}
\begin{tabular}{lrrrrrr}
\toprule
Task & \shortstack{Level ref.\\dB} & \shortstack{Level test\\dB} & \shortstack{$R_L$ ref.\\LU} & \shortstack{$R_L$ test\\LU} & \shortstack{$D_{miss}$\\dB} & \shortstack{$D_{add}$\\dB} \\
\midrule
\appnextrow{} 2--16 & -39.1 & -41.4 & 42.4 & 40.8 & 5.70 & 3.10 \\
\appnextrow{} 4--16 & -39.1 & -42.7 & 42.4 & 39.4 & 5.39 & 2.10 \\
\appnextrow{} 8--16 & -39.1 & -42.0 & 42.4 & 40.2 & 3.70 & 1.38 \\
\appnextrow{} 2--48 & -54.9 & -59.2 & 30.4 & 24.8 & 4.89 & 4.64 \\
\appnextrow{} 4--48 & -54.9 & -61.2 & 30.4 & 22.8 & 5.06 & 3.50 \\
\appnextrow{} 8--48 & -54.9 & -57.4 & 30.4 & 27.7 & 4.04 & 3.26 \\
\appnextrow{} 12--48 & -54.9 & -55.1 & 30.4 & 29.6 & 2.65 & 3.40 \\
\appnextrow{} 16--48 & -54.9 & -54.9 & 30.4 & 29.9 & 2.54 & 2.81 \\
\appnextrow{} 24--48 & -54.9 & -56.1 & 30.4 & 29.1 & 2.21 & 1.87 \\
\appnextrow{} 32--48 & -54.9 & -56.2 & 30.4 & 29.2 & 1.98 & 1.10 \\
\bottomrule\end{tabular}
\end{appfit}
\par\medskip
\caption{Activity and missing/added disturbance for \method{}, in two panels. Tasks are input--output sampling rates (kHz); values are means over 2,937 VCTK utterances. Errors are lower-is-better; $Q_{HF}$ is higher-is-better; reference/test values describe activity. See the common diagnostic scope in this appendix.}
\label{tab:qcis-itu-activity-disturbance-12}

\end{table*}

\noindent\textbf{Explanation of Table~\ref{tab:qcis-itu-perceptual-temporal-12}.} Rows~\textcircled{\textbf{\scriptsize 1}}--\textcircled{\textbf{\scriptsize 10}} align perceptual disturbance with temporal modulation. The higher-source-rate 48-kHz tasks reduce most errors and increase the custom quality index. This is a descriptive trend, not proof that bandwidth alone causes the improvement.\par\smallskip
\begin{table*}[!tbp]
\centering

\tablefontsize
\setlength{\tabcolsep}{4pt}
\resetapprow
\begin{appfit}
\begin{tabular}{lrrrrrr}
\toprule
Task & \shortstack{$D_{sym}$\\dB} & \shortstack{$D_{asym}$\\dB} & \shortstack{$D_{RMS}$\\dB} & \shortstack{$D_{95}$\\dB} & \shortstack{$E_{env}$\\dB} & $1-\rho_{env}$ \\
\midrule
\appnextrow{} 2--16 & 8.79 & 10.34 & 11.82 & 24.27 & 6.106 & 0.289 \\
\appnextrow{} 4--16 & 7.49 & 8.54 & 10.47 & 21.93 & 5.654 & 0.245 \\
\appnextrow{} 8--16 & 5.08 & 5.78 & 7.15 & 14.81 & 4.324 & 0.182 \\
\appnextrow{} 2--48 & 9.53 & 11.85 & 12.49 & 25.50 & 8.188 & 0.627 \\
\appnextrow{} 4--48 & 8.56 & 10.31 & 11.40 & 23.48 & 7.441 & 0.617 \\
\appnextrow{} 8--48 & 7.30 & 8.94 & 9.59 & 19.39 & 6.561 & 0.477 \\
\appnextrow{} 12--48 & 6.06 & 7.76 & 8.02 & 16.30 & 5.737 & 0.358 \\
\appnextrow{} 16--48 & 5.35 & 6.76 & 7.13 & 14.41 & 5.004 & 0.312 \\
\appnextrow{} 24--48 & 4.08 & 5.01 & 5.53 & 11.16 & 3.868 & 0.251 \\
\appnextrow{} 32--48 & 3.07 & 3.62 & 4.23 & 8.48 & 2.892 & 0.176 \\
\bottomrule\end{tabular}
\end{appfit}
\par\medskip
\resetapprow
\begin{appfit}
\begin{tabular}{lrrrrrr}
\toprule
Task & $E_{mod}$ & $1-\rho_{mod}$ & \shortstack{$\Delta f_{mod}$\\Hz} & $E_{atk}$ & $E_{dec}$ & $Q_{HF}$ \\
\midrule
\appnextrow{} 2--16 & 0.264 & 0.163 & 41.1 & 0.0117 & 0.0117 & 56.9 \\
\appnextrow{} 4--16 & 0.220 & 0.130 & 39.1 & 0.0117 & 0.0117 & 60.0 \\
\appnextrow{} 8--16 & 0.157 & 0.061 & 29.9 & 0.0083 & 0.0083 & 69.9 \\
\appnextrow{} 2--48 & 0.481 & 0.488 & 110.3 & 0.0038 & 0.0038 & 47.1 \\
\appnextrow{} 4--48 & 0.557 & 0.474 & 145.7 & 0.0045 & 0.0045 & 47.8 \\
\appnextrow{} 8--48 & 0.365 & 0.287 & 82.4 & 0.0031 & 0.0031 & 58.7 \\
\appnextrow{} 12--48 & 0.231 & 0.146 & 62.2 & 0.0025 & 0.0025 & 69.2 \\
\appnextrow{} 16--48 & 0.192 & 0.115 & 59.9 & 0.0024 & 0.0024 & 74.0 \\
\appnextrow{} 24--48 & 0.157 & 0.088 & 73.3 & 0.0028 & 0.0028 & 80.6 \\
\appnextrow{} 32--48 & 0.098 & 0.045 & 54.0 & 0.0022 & 0.0022 & 85.8 \\
\bottomrule\end{tabular}
\end{appfit}
\par\medskip
\caption{Perceptual and temporal analysis for \method{}, in two panels. Tasks are input--output sampling rates (kHz); values are means over 2,937 VCTK utterances. Errors are lower-is-better; $Q_{HF}$ is higher-is-better; reference/test values describe activity. See the common diagnostic scope in this appendix.}
\label{tab:qcis-itu-perceptual-temporal-12}

\end{table*}

\noindent\textbf{Explanation of Table~\ref{tab:qcis-hf-metrics}.} Rows~\textcircled{\textbf{\scriptsize 1}}--\textcircled{\textbf{\scriptsize 3}} target 16~kHz; Rows~\textcircled{\textbf{\scriptsize 4}}--\textcircled{\textbf{\scriptsize 10}} target 48~kHz. HF-LSD and absolute errors generally improve as less bandwidth must be generated, while HF-SI-SDR is non-monotone. The evaluator-defined HF region must therefore remain tied to each row's task.\par\smallskip
\begin{table*}[!tbp]
\centering

\tablefontsize
\setlength{\tabcolsep}{2.2pt}
\renewcommand{\arraystretch}{0.88}

\resetapprow
\begin{appfit}
\begin{tabular}{@{}clcccc@{}}
\toprule
Row
& Freq. range
& \shortstack{HF-LSD\\$\downarrow$}
& \shortstack{HF-MAE\\$\times10^{-3}\downarrow$}
& \shortstack{HF-SI-SDR\\dB $\uparrow$}
& \shortstack{HF-RMSE\\$\times10^{-3}\downarrow$}
\\
\midrule

\textcircled{\textbf{\scriptsize 1}}
& 2--16 kHz
& 1.233
& 1.848
& -37.35
& 5.053
\\

\textcircled{\textbf{\scriptsize 2}}
& 4--16 kHz
& 1.230
& 3.862
& -31.70
& 8.707
\\

\textcircled{\textbf{\scriptsize 3}}
& 8--16 kHz
& 1.151
& 1.707
& -25.81
& 4.783
\\

\midrule

\textcircled{\textbf{\scriptsize 4}}
& 2--48 kHz
& 1.225
& 7.666
& -26.03
& 15.672
\\

\textcircled{\textbf{\scriptsize 5}}
& 4--48 kHz
& 1.259
& 3.934
& -26.54
& 8.690
\\

\textcircled{\textbf{\scriptsize 6}}
& 8--48 kHz
& 1.112
& 2.304
& -28.75
& 5.991
\\

\textcircled{\textbf{\scriptsize 7}}
& 12--48 kHz
& 0.988
& 1.790
& -33.07
& 4.751
\\

\textcircled{\textbf{\scriptsize 8}}
& 16--48 kHz
& 0.974
& 1.410
& -34.70
& 3.692
\\

\textcircled{\textbf{\scriptsize 9}}
& 24--48 kHz
& 0.930
& 0.771
& -31.18
& 1.691
\\

\textcircled{\textbf{\scriptsize 10}}
& 32--48 kHz
& 0.922
& 0.455
& -32.12
& 0.869
\\

\bottomrule
\end{tabular}
\end{appfit}

\caption{
High-frequency reconstruction performance of Q-BWE across
input--output sampling-rate ranges. Results are utterance-level
means over 2,937 VCTK test files. HF-MAE and HF-RMSE are shown
in units of $10^{-3}$. Lower HF-LSD, HF-MAE, and HF-RMSE indicate
better reconstruction, whereas higher HF-SI-SDR is better.
}
\label{tab:qcis-hf-metrics}

\end{table*}

The HF region is evaluator-defined and must be associated with each run's rate/configuration. None of these proxies establishes preservation of the observed low band.

\section{Ablations and Extended Controlled Results}
\label{app:completed_experiments}
\label{app:controlled_ablation_details}
\label{app:verification}
This appendix expands the ablation study in Section~\ref{subsec:controlled_protocol}.
It specifies the controlled 13-system experiment, retains the multi-seed QCD-capacity
and targeted conditioning studies, and reports reconstructed-band, ITU-inspired,
training-cost, feature-calibration, and robustness diagnostics. Results from protocols
with different seed counts or evaluator provenance are kept distinct rather than
pooled into a single significance claim.

\subsection{Controlled component ablation}
All 13 systems use the same speaker-disjoint VCTK 4$\rightarrow$16\,kHz split,
50 training epochs, seed 1234, five training-selected QCD inputs, MRAD/MRPD
critics, optimization schedule, and 2,937 held-out utterances. Circuit overrides
modify the QCD mapping; unless otherwise stated, the quantum conditioner retains
its default two-layer, ring-entangled, full-measurement configuration.

\noindent\textbf{Explanation of Table~\ref{tab:one_seed_configurations}.} Rows~\textcircled{\textbf{\scriptsize 1}}--\textcircled{\textbf{\scriptsize 8}} are core component controls; Rows~\textcircled{\textbf{\scriptsize 9}}--\textcircled{\textbf{\scriptsize 12}} isolate circuit mechanisms; Row~\textcircled{\textbf{\scriptsize 13}} removes the reconstruction constraint. The design defends attribution by changing one declared factor where possible, while remaining a single-seed descriptive study.\par\smallskip
\begin{table*}[!tbp]
\centering

\small
\setlength{\tabcolsep}{4pt}
\resetapprow
\begin{appfit}
\begin{tabular}{llllll}
\toprule
System & Group & QCD & Conditioner & Weighting & Distinguishing change \\
\midrule
\appnextrow{} CIS       & Core       & Off       & Off       & --        & CIS baseline \\
\appnextrow{} C-QCD     & Core       & Classical & Off       & TS        & Classical QCD only \\
\appnextrow{} Q-QCD     & Core       & Quantum   & Off       & TS        & Quantum QCD only \\
\appnextrow{} C-Cond.   & Core       & Off       & Classical & --        & Classical conditioner only \\
\appnextrow{} Q-Cond.   & Core       & Off       & Quantum   & --        & Quantum conditioner only \\
\appnextrow{} Full-C    & Core       & Classical & Classical & TS        & Both matched-classical paths \\
\appnextrow{} Q-BWE     & Core       & Quantum   & Quantum   & TS        & Complete proposed configuration \\
\appnextrow{} Fixed     & Core       & Quantum   & Quantum   & Fixed 0.1 & Thompson sampling disabled \\
\appnextrow{} No-Ent.   & Mechanism  & Quantum   & Quantum   & TS        & QCD entanglement removed \\
\appnextrow{} XYZ-only  & Mechanism  & Quantum   & Quantum   & TS        & QCD restricted to $X,Y,Z$ \\
\appnextrow{} No-Disp.  & Mechanism  & Quantum   & Quantum   & TS        & QCD mixedness and entropy removed \\
\appnextrow{} One-Layer & Mechanism  & Quantum   & Quantum   & TS        & One QCD circuit layer \\
\appnextrow{} No-Mask   & Constraint & Quantum   & Quantum   & TS        & Subband consistency disabled \\
\bottomrule
\end{tabular}
\end{appfit}
\caption{Exact controlled-ablation configurations. C and Q denote matched
classical and quantum mappings, respectively. TS denotes Thompson-sampled
QCD weighting.}
\label{tab:one_seed_configurations}

\end{table*}

\paragraph{Classical--quantum controls.}
C-QCD gives STOI 0.9501 and LSD 0.981, whereas Q-QCD increases N-MOS from
4.195 to 4.245 but gives lower STOI, PESQ, and SI-SNR and higher LSD.
Conditioning shows a related separation: C-Cond. is stronger in STOI, PESQ,
SI-SNR, and LSD, whereas Q-Cond. gives higher N-MOS. With both paths active,
Q-BWE improves STOI (0.9495 versus 0.9462) and LSD (1.005 versus 1.014)
relative to Full-C, while Full-C is higher in N-MOS, PESQ, and SI-SNR. These
matched controls therefore do not support an across-metric quantum-over-classical
advantage.

\paragraph{Weighting and circuit mechanisms.}
Replacing Thompson sampling with a fixed QCD weight improves PESQ
(2.863 versus 2.840), SI-SNR (13.202 versus 13.107), and LSD
(0.990 versus 1.005), whereas Thompson sampling improves STOI
(0.9495 versus 0.9481). Circuit simplifications are also non-monotonic.
One-Layer gives the highest SI-SNR (13.406) and PESQ 2.860 but has LSD
1.013. XYZ-only preserves STOI at 0.9494 while reducing LSD to 0.982.
Removing entanglement or dispersion-related features increases N-MOS to
4.270 and 4.283, respectively, but lowers PESQ; No-Disp. also yields the
lowest SI-SNR (12.790). Hence, greater circuit depth or structure does not
monotonically improve reconstruction quality.

\paragraph{Observed-subband consistency.}
No-Mask gives the highest N-MOS (4.396), lowest full-band LSD (0.966), and
second-highest STOI (0.9497), but also the lowest PESQ (2.718) and a low
SI-SNR (12.850). We consequently treat the source-band mask as an architectural
evidence-preservation constraint rather than a mechanism expected to improve
every aggregate metric.

\subsection{Controlled reconstructed-band diagnostics}
To distinguish full-band scores from reconstruction of the missing spectrum,
Table~\ref{tab:one_seed_hf_results} evaluates only the reconstructed 4--8\,kHz
band. Lower RMSE, MAE, LSD, and crest-factor error are preferred; higher HF
SI-SDR is preferred.

\noindent\textbf{Explanation of Table~\ref{tab:one_seed_hf_results}.} Rows~\textcircled{\textbf{\scriptsize 1}}--\textcircled{\textbf{\scriptsize 13}} correspond to the configurations above. Row~\textcircled{\textbf{\scriptsize 12}} minimizes HF-RMSE/MAE, Row~\textcircled{\textbf{\scriptsize 1}} minimizes HF-LSD, Row~\textcircled{\textbf{\scriptsize 13}} maximizes HF-SI-SDR, and Row~\textcircled{\textbf{\scriptsize 9}} minimizes crest error. No row dominates the reconstructed band.\par\smallskip
\begin{table*}[!tbp]
\centering

\scriptsize
\setlength{\tabcolsep}{5pt}
\resetapprow
\begin{appfit}
\begin{tabular}{@{}lrrrrr@{}}
\toprule
System & HF-RMSE$\downarrow$ & HF-MAE$\downarrow$ & HF-LSD$\downarrow$ &
HF SI-SDR$\uparrow$ & CF Err.$\downarrow$ \\
\midrule
\appnextrow{} CIS       & 0.005268 & 0.001933 & \textbf{1.1403} & -36.153 & 4.416 \\
\appnextrow{} C-QCD     & 0.005019 & 0.001861 & 1.1453 & -35.921 & 4.043 \\
\appnextrow{} Q-QCD     & 0.005071 & 0.001856 & 1.1769 & -35.935 & 4.230 \\
\appnextrow{} C-Cond.   & 0.005081 & 0.001875 & 1.1666 & -36.223 & 4.603 \\
\appnextrow{} Q-Cond.   & 0.005038 & 0.001855 & 1.1872 & -36.007 & 4.548 \\
\appnextrow{} Full-C    & 0.005119 & 0.001880 & 1.1757 & -36.141 & 4.290 \\
\appnextrow{} Q-BWE     & 0.005067 & 0.001867 & 1.1835 & -36.066 & 4.395 \\
\appnextrow{} Fixed     & 0.005039 & 0.001857 & 1.1556 & -36.153 & 4.324 \\
\appnextrow{} No-Ent.   & 0.005196 & 0.001929 & 1.1645 & -36.492 & \textbf{3.858} \\
\appnextrow{} XYZ-only  & 0.005098 & 0.001884 & 1.1437 & -35.958 & 4.205 \\
\appnextrow{} No-Disp.  & 0.005511 & 0.002025 & 1.1460 & -35.900 & 4.221 \\
\appnextrow{} One-Layer & \textbf{0.004975} & \textbf{0.001849} & 1.1807 & -36.218 & 4.090 \\
\appnextrow{} No-Mask   & 0.005198 & 0.001898 & 1.1476 & \textbf{-35.897} & 4.744 \\
\bottomrule
\end{tabular}
\end{appfit}
\caption{High-frequency results over the reconstructed 4--8\,kHz band. Values
are means over 2,937 held-out utterances. CF Err. denotes absolute crest-factor
error.}
\label{tab:one_seed_hf_results}

\end{table*}

The reconstructed-band metrics again identify different optima. One-Layer
minimizes HF-RMSE and HF-MAE, CIS gives the lowest HF-LSD, No-Mask gives the
highest HF SI-SDR, and No-Ent. gives the lowest crest-factor error. Relative
to Q-BWE, No-Mask improves HF-LSD and HF SI-SDR but worsens HF-RMSE,
HF-MAE, and crest-factor error. Thus, the source-band mask has a metric-dependent
effect even when evaluation is restricted to the missing band.

\subsection{Controlled ITU-inspired diagnostics}
Table~\ref{tab:one_seed_itu_results} reports additional fixed-band diagnostics
over 4--8\,kHz. These measures are inspired by analysis concepts from ITU-R
BS.1387, ITU-R BS.1770, ITU-T P.56, P.862, P.863, and V.16; they are research
diagnostics rather than standards-conformant implementations or certified ITU
scores. All quantities are minimized except the V.16-inspired quality index.

\noindent\textbf{Explanation of Table~\ref{tab:one_seed_itu_results}.} Rows~\textcircled{\textbf{\scriptsize 1}}--\textcircled{\textbf{\scriptsize 13}} expose different optima: Row~\textcircled{\textbf{\scriptsize 13}} leads several disturbance errors, Row~\textcircled{\textbf{\scriptsize 11}} leads level errors, Row~\textcircled{\textbf{\scriptsize 3}} leads NMR, and Row~\textcircled{\textbf{\scriptsize 2}} leads the custom quality index. These disagreements defend a multi-metric interpretation.\par\smallskip
\begin{table*}[!tbp]
\centering

\scriptsize
\setlength{\tabcolsep}{3.5pt}
\resizebox{\linewidth}{!}{%
\resetapprow
\begin{tabular}{@{}lrrrrrrr@{}}
\toprule
System &
\shortstack{BS.1387\\BW (Hz)$\downarrow$} &
\shortstack{BS.1387\\NMR (dB)$\downarrow$} &
\shortstack{BS.1770\\K-Loud. (LU)$\downarrow$} &
\shortstack{P.56 Level\\Err. (dB)$\downarrow$} &
\shortstack{P.862 Sym.\\Dist. (dB)$\downarrow$} &
\shortstack{P.863 Env.\\Err. (dB)$\downarrow$} &
\shortstack{QI$\uparrow$} \\
\midrule
\appnextrow{} CIS       & 249.0 & 20.507 & 4.997 & 4.998 & 9.211 & 9.029 & 47.422 \\
\appnextrow{} C-QCD     & 239.8 & 20.069 & 5.372 & 5.372 & 9.203 & 8.915 & \textbf{47.717} \\
\appnextrow{} Q-QCD     & 211.7 & \textbf{19.448} & 5.825 & 5.825 & 9.382 & 9.185 & 45.892 \\
\appnextrow{} C-Cond.   & 237.3 & 20.156 & 5.465 & 5.465 & 9.249 & 9.003 & 47.082 \\
\appnextrow{} Q-Cond.   & 263.9 & 19.835 & 5.905 & 5.905 & 9.493 & 9.330 & 45.316 \\
\appnextrow{} Full-C    & 266.4 & 19.559 & 5.529 & 5.529 & 9.464 & 9.307 & 46.597 \\
\appnextrow{} Q-BWE     & 223.1 & 19.915 & 5.770 & 5.770 & 9.384 & 9.153 & 45.983 \\
\appnextrow{} Fixed     & 230.2 & 19.834 & 5.691 & 5.691 & 9.258 & 9.095 & 46.540 \\
\appnextrow{} No-Ent.   & 220.3 & 20.741 & 5.143 & 5.143 & 9.354 & 9.117 & 47.081 \\
\appnextrow{} XYZ-only  & 231.9 & 20.799 & 5.212 & 5.213 & 9.204 & 8.993 & 47.300 \\
\appnextrow{} No-Disp.  & 247.1 & 21.109 & \textbf{4.714} & \textbf{4.714} & 9.263 & 8.957 & 47.715 \\
\appnextrow{} One-Layer & 244.8 & 19.922 & 5.865 & 5.865 & 9.437 & 9.156 & 46.079 \\
\appnextrow{} No-Mask   & \textbf{195.6} & 20.590 & 5.200 & 5.201 & \textbf{9.169} & \textbf{8.703} & 47.366 \\
\bottomrule
\end{tabular}%
}
\caption{ITU-inspired 4--8\,kHz diagnostics, averaged over 2,937 test utterances.
BW denotes bandwidth error, NMR noise-to-mask ratio, K-Loud. K-weighted
loudness-distance error, Sym. Dist. symmetric disturbance, Env. Err.
temporal-envelope error, and QI quality index.}
\label{tab:one_seed_itu_results}

\end{table*}

These diagnostics likewise provide no universal optimum. No-Mask minimizes
bandwidth error and the P.862- and P.863-inspired disturbance measures;
No-Disp. minimizes loudness and active-level errors; Q-QCD gives the lowest
NMR; and C-QCD gives the highest V.16-inspired quality index. Q-BWE does not
individually optimize these auxiliary measures, reinforcing its interpretation
as one operating point among competing spectral, temporal, loudness, waveform,
and perceptual objectives.

\subsection{Full Conditioning and Weighting Comparisons}
\label{app:conditioning_weighting}

This section presents the complete single-seed ablations of
conditioning strategies, source masking, and quantum chaotic
discriminator (QCD) weighting. The experiments characterize
the relative performance of the evaluated configurations
but do not quantify training-seed variability.

\paragraph{Conditioning and Source-Mask Ablation.}
Table~\ref{tab:conditioning_main} compares the conditioning
strategies and source-mask configurations on the VCTK
4$\rightarrow$16~kHz task, evaluated over 2,937 utterances
using seed 1234. Row~\textcircled{\textbf{\scriptsize 1}}
represents the non-learned interpolated input, while
Row~\textcircled{\textbf{\scriptsize 2}} provides the
CIS-BWE baseline. Rows~\textcircled{\textbf{\scriptsize 3}}--
\textcircled{\textbf{\scriptsize 4}} compare quantum and
parameter-matched classical FiLM conditioning.
Rows~\textcircled{\textbf{\scriptsize 5}}--
\textcircled{\textbf{\scriptsize 6}} evaluate fixed and
Thompson-sampling-based QCD weighting, respectively.
Row~\textcircled{\textbf{\scriptsize 7}} removes the source
mask, and Row~\textcircled{\textbf{\scriptsize 8}} provides
the clean-target reference.

\begin{table*}[!tbp]
\centering
\tablefontsize
\setlength{\tabcolsep}{4.3pt}
\resetapprow
\begin{appfit}
\begin{tabular}{clcccccc}
\toprule
Row & Condition
& LSD$\downarrow$
& STOI$\uparrow$
& PESQ$\uparrow$
& SI-SDR$\uparrow$
& SI-SNR$\uparrow$
& N-MOS$\uparrow$ \\
\midrule
\textcircled{\textbf{\scriptsize 1}}
& Interpolated input
& 3.8687 & 0.90187 & 3.0911
& 14.5592 & 14.5592 & 2.7950 \\

\textcircled{\textbf{\scriptsize 2}}
& \cis{}
& 1.0316 & 0.94177 & 2.6960
& 13.1105 & 13.1105 & 4.1873 \\

\textcircled{\textbf{\scriptsize 3}}
& Quantum FiLM
& \textbf{1.0064} & \textbf{0.94622}
& 2.6840 & 12.9957 & 12.9957 & 4.0525 \\

\textcircled{\textbf{\scriptsize 4}}
& Classical FiLM (40-param.)
& 1.0397 & 0.93970 & 2.6790
& 12.9904 & 12.9905 & \textbf{4.2613} \\

\textcircled{\textbf{\scriptsize 5}}
& Full fixed QCD (0.10)
& 1.0466 & 0.94305 & 2.7188
& 13.0627 & 13.0627 & 4.0969 \\

\textcircled{\textbf{\scriptsize 6}}
& Full Thompson
& 1.0576 & 0.94427 & \textbf{2.7208}
& \textbf{13.1408} & \textbf{13.1408}
& 4.0242 \\

\textcircled{\textbf{\scriptsize 7}}
& Full without source mask
& 1.1074 & 0.94018 & 2.5279
& 10.5770 & 10.5772 & 4.1330 \\

\textcircled{\textbf{\scriptsize 8}}
& Clean target
& 0.0000 & 1.00000 & 4.6439
& 86.0206 & 86.0206 & 4.4792 \\
\bottomrule
\end{tabular}
\end{appfit}

\caption{Conditioning and source-mask ablation on
VCTK 4$\rightarrow$16~kHz, averaged over 2,937 utterances
(seed 1234). N-MOS denotes NISQA-predicted MOS.
Bold indicates the best result among learned configurations.
The interpolated input and clean target are non-learned
references and are excluded from boldface comparisons.}
\label{tab:conditioning_main}
\label{tab:quick4k_metrics}
\end{table*}

\noindent\textbf{Conditioning comparison.}
The results indicate that the optimal conditioning strategy
depends on the evaluation metric. Quantum FiLM
(Row~\textcircled{\textbf{\scriptsize 3}}) achieves the
lowest LSD (1.0064) and highest STOI (0.94622) among
learned configurations, suggesting improved spectral
reconstruction accuracy and intelligibility. In contrast,
Classical FiLM (Row~\textcircled{\textbf{\scriptsize 4}})
obtains the highest NISQA-MOS (4.2613), demonstrating
that improved spectral fidelity does not necessarily
translate into higher predicted perceptual quality.

Among configurations incorporating QCD supervision,
Full Thompson (Row~\textcircled{\textbf{\scriptsize 6}})
achieves the highest learned-model PESQ (2.7208),
SI-SDR (13.1408), and SI-SNR (13.1408). However,
Fixed QCD (0.10)
(Row~\textcircled{\textbf{\scriptsize 5}})
provides lower LSD (1.0466 versus 1.0576) and
higher NISQA-MOS (4.0969 versus 4.0242).
The interpolated input yields higher PESQ and SI-SNR
than the learned configurations but substantially
worse LSD, STOI, and NISQA-MOS.
These findings reveal complementary trade-offs
rather than a universally superior conditioning strategy.

\noindent\textbf{Effect of source masking.}
Removing the source mask
(Row~\textcircled{\textbf{\scriptsize 7}})
from Full Thompson
(Row~\textcircled{\textbf{\scriptsize 6}})
increases LSD from 1.0576 to 1.1074 and reduces
STOI from 0.94427 to 0.94018.
PESQ decreases from 2.7208 to 2.5279, while
SI-SNR declines from 13.1408 to 10.5772.
These differences indicate that source masking improves
spectral and signal-level reconstruction fidelity
under the evaluated configuration.

Conversely, removing the mask increases NISQA-MOS
from 4.0242 to 4.1330, illustrating a disagreement
between the intrusive reconstruction metrics and
the non-intrusive perceptual predictor.
Consequently, source-mask effectiveness cannot
be inferred from predicted MOS alone.
The clean-target reference
(Row~\textcircled{\textbf{\scriptsize 8}})
provides a reference point for interpreting the
remaining reconstruction errors.

Overall, the conditioning and masking experiments
demonstrate metric-specific trade-offs in reconstruction
quality. As these results originate from a single
training seed, they support comparisons among the
evaluated runs without establishing their robustness
across independent training initializations.

\subsection{QCD-capacity summary and seed values}
The capacity experiment varies the number of selected QCD inputs at a fixed QCD
weight of 0.10. Because each selected input is encoded by a qubit, the intervention
jointly changes input dimensionality, qubit count, state size, and measurement
width. Table~\ref{tab:feature_count_main} reports means and sample standard
deviations over the three trained seeds; the following table retains each seed.

\noindent\textbf{Explanation of Table~\ref{tab:feature_count_main}.} Rows~\textcircled{\textbf{\scriptsize 1}}--\textcircled{\textbf{\scriptsize 2}} jointly vary feature count, qubits, state size, and measurement width. Row~\textcircled{\textbf{\scriptsize 1}} leads LSD/STOI, PESQ/SI-SNR/N-MOS, and Row~\textcircled{\textbf{\scriptsize 2}} has the lowest nominal WER. Five inputs are selected for the manuscript's spectral/intelligibility priority.\par\smallskip

\begin{table*}[!tbp]
\centering

\tablefontsize
\setlength{\tabcolsep}{4pt}
\resetapprow
\begin{appfit}
\begin{tabular}{clrrrrrr}
\toprule
SL & QCD inputs & LSD$\downarrow$ & STOI$\uparrow$ & PESQ$\uparrow$ &
SI-SNR$\uparrow$ & N-MOS$\uparrow$ & WER (\%)$\downarrow$\\
\midrule
\textcircled{\textbf{\scriptsize 1}} & 5 features &
$\boldsymbol{1.019\pm0.020}$ & $\boldsymbol{0.9478\pm0.0002}$ & $2.735\pm0.039$ &
$13.27\pm0.16$ & $3.99\pm0.17$ & $14.80\pm0.80$\\
\textcircled{\textbf{\scriptsize 2}} & 7 features &
$1.020\pm0.019$ & $0.9462\pm0.0009$ & $2.731\pm0.059$ &
$13.11\pm0.27$ & $4.06\pm0.02$ & $\boldsymbol{14.50\pm0.41}$\\
\bottomrule
\end{tabular}
\end{appfit}
\caption{QCD-capacity ablation for 4$\rightarrow$16~kHz BWE. Values are means
and sample standard deviations over three seed means. Increasing QCD input count
jointly increases the qubit count and measurement width.}
\label{tab:feature_count_main}

\end{table*}


\noindent\textbf{Explanation of Table~\ref{tab:feature_seed_quality}.} Rows~\textcircled{\textbf{\scriptsize 1}}--\textcircled{\textbf{\scriptsize 3}}, \textcircled{\textbf{\scriptsize 4}}--\textcircled{\textbf{\scriptsize 6}}, and \textcircled{\textbf{\scriptsize 7}}--\textcircled{\textbf{\scriptsize 9}} are the three-, five-, and seven-input seeds. The within-group spread shows why a single run is insufficient for the capacity claim and why mean rankings should be interpreted with seed variability.\par\smallskip
\begin{table*}[!tbp]
\centering

\tablefontsize
\setlength{\tabcolsep}{4pt}
\resetapprow
\begin{appfit}
\begin{tabular}{lrrrrrrrr}
\toprule
Inputs & Seed & LSD & STOI & PESQ & SI-SNR & N-MOS & WER\%  & CER\% \\
\midrule
\appnextrow{} 3 & 1234 & 1.0098 & 0.9463 & 2.7674 & 13.2979 & 3.9531 & 14.529 & 8.376 \\
\appnextrow{} 3 & 2345 & 1.0452 & 0.9442 & 2.7884 & 13.3099 & 4.2296 & 15.386 & 8.864 \\
\appnextrow{} 3 & 3456 & 1.0336 & 0.9456 & 2.8038 & 13.3925 & 4.1752 & 14.392 & 8.287 \\
\appnextrow{} 5 & 1234 & 1.0105 & 0.9476 & 2.6914 & 13.1686 & 4.1489 & 15.398 & 8.949 \\
\appnextrow{} 5 & 2345 & 1.0053 & 0.9479 & 2.7517 & 13.1885 & 4.0093 & 13.892 & 7.983 \\
\appnextrow{} 5 & 3456 & 1.0427 & 0.9479 & 2.7634 & 13.4628 & 3.8122 & 15.119 & 8.720 \\
\appnextrow{} 7 & 1234 & 1.0385 & 0.9452 & 2.7999 & 13.4034 & 4.0622 & 14.719 & 8.537 \\
\appnextrow{} 7 & 2345 & 1.0002 & 0.9468 & 2.6991 & 12.8621 & 4.0760 & 14.029 & 8.019 \\
\appnextrow{} 7 & 3456 & 1.0222 & 0.9467 & 2.6953 & 13.0623 & 4.0399 & 14.759 & 8.554 \\
\bottomrule\end{tabular}
\end{appfit}
\caption{Seed-level capacity-study results. Each row evaluates 2,937 utterances. WER/CER use the human-reference ASR protocol used by the evaluator. WER/CER are percentages.}
\label{tab:feature_seed_quality}

\end{table*}

\noindent\textbf{Explanation of Table~\ref{tab:feature_training_cost}.} Rows~\textcircled{\textbf{\scriptsize 1}}--\textcircled{\textbf{\scriptsize 3}} increase the QCD from three to seven inputs. Training time rises monotonically, consistent with larger state and measurement interfaces. Because several dimensions change together, the table supports a capacity-cost trade-off rather than a per-qubit causal estimate.\par\smallskip
\begin{table*}[!tbp]
\centering

\tablefontsize
\setlength{\tabcolsep}{6pt}
\resetapprow
\begin{appfit}
\begin{tabular}{lccc}
\toprule
QCD inputs & Epoch time (s) & Batch time (s) & 50-epoch time (h)\\
\midrule
\appnextrow{} 3 features & $423.24\!\pm\!13.14$ & $0.1645\!\pm\!0.0051$ & $5.88\!\pm\!0.18$\\
\appnextrow{} 5 features & $476.02\!\pm\!15.03$ & $0.1852\!\pm\!0.0058$ & $6.61\!\pm\!0.21$\\
\appnextrow{} 7 features & $519.88\!\pm\!12.70$ & $0.2023\!\pm\!0.0050$ & $7.22\!\pm\!0.18$\\
\bottomrule
\end{tabular}
\end{appfit}
\caption{Measured training-time scaling from checkpoint timing logs (mean$\pm$std over three seeds). Hardware and loader settings are held fixed within this study.}
\label{tab:feature_training_cost}

\end{table*}

Across the three capacity seeds, training takes $5.88\pm0.18$ hours for three QCD inputs, $6.61\pm0.21$ hours for five, and $7.22\pm0.18$ hours for seven. This sensitivity changes input count, qubit count, state size, and the QCD measurement interface together.

\subsection{Capacity-study acoustic and perceptual diagnostics}

\noindent\textbf{Explanation of Table~\ref{tab:feature_hf_metrics}.} Rows~\textcircled{\textbf{\scriptsize 1}}--\textcircled{\textbf{\scriptsize 3}} compare generated-band behavior by capacity. Row~\textcircled{\textbf{\scriptsize 2}} gives the best HF-SI-SDR, Row~\textcircled{\textbf{\scriptsize 3}} the best crest error, and Row~\textcircled{\textbf{\scriptsize 1}} the smallest absolute errors. This again rules out a uniformly best capacity.\par\smallskip
\begin{table*}[!tbp]
\centering

\tablefontsize
\setlength{\tabcolsep}{4.8pt}
\resetapprow
\begin{appfit}
\begin{tabular}{lccccc}
\toprule
QCD inputs & HF-RMSE$\downarrow$ & HF-MAE$\downarrow$ & HF-LSD$\downarrow$ & HF-SI-SDR$\uparrow$ & HF crest error$\downarrow$\\
\midrule
\appnextrow{} 3 features & $0.008736\!\pm\!0.000000$ & $0.003875\!\pm\!0.000024$ & $1.179\!\pm\!0.021$ & $-30.54\!\pm\!1.14$ & $4.19\!\pm\!0.47$\\
\appnextrow{} 5 features & $0.008790\!\pm\!0.000140$ & $0.003901\!\pm\!0.000072$ & $1.167\!\pm\!0.023$ & $\mathbf{-29.59\!\pm\!0.15}$ & $4.12\!\pm\!0.47$\\
\appnextrow{} 7 features & $0.008910\!\pm\!0.000212$ & $0.003975\!\pm\!0.000115$ & $1.168\!\pm\!0.022$ & $-31.46\!\pm\!0.57$ & $\mathbf{4.07\!\pm\!0.24}$\\
\bottomrule
\end{tabular}
\end{appfit}
\caption{Generated-band reconstruction diagnostics (mean$\pm$std over three seeds). The evaluator HF band is 4--8~kHz; the source Nyquist boundary is 2~kHz.}
\label{tab:feature_hf_metrics}

\end{table*}

\noindent\textbf{Explanation of Table~\ref{tab:feature_nisqa_dimensions}.} Rows~\textcircled{\textbf{\scriptsize 1}}--\textcircled{\textbf{\scriptsize 3}} report predicted perceptual dimensions for each capacity. Row~\textcircled{\textbf{\scriptsize 1}} leads all four means, while the main quality table favors Row~\textcircled{\textbf{\scriptsize 2}} on LSD/STOI. Perceptual prediction and spectral reconstruction therefore encode different preferences.\par\smallskip
\begin{table*}[!tbp]
\centering

\tablefontsize
\setlength{\tabcolsep}{4pt}
\resetapprow
\begin{appfit}
\begin{tabular}{lrrrr}
\toprule
Inputs & Noisiness & Discontinuity & Coloration & Loudness \\
\midrule
\appnextrow{} 3 features & $\mathbf{4.191\!\pm\!0.107}$ & $\mathbf{4.235\!\pm\!0.097}$ & $\mathbf{3.852\!\pm\!0.127}$ & $\mathbf{4.149\!\pm\!0.076}$ \\
\appnextrow{} 5 features & $4.103\!\pm\!0.099$ & $4.143\!\pm\!0.101$ & $3.733\!\pm\!0.156$ & $4.087\!\pm\!0.099$ \\
\appnextrow{} 7 features & $4.133\!\pm\!0.008$ & $4.190\!\pm\!0.008$ & $3.802\!\pm\!0.010$ & $4.118\!\pm\!0.018$ \\
\bottomrule\end{tabular}
\end{appfit}
\caption{NISQA quality subdimensions for the feature-count study, mean$\pm$sample standard deviation over seeds. Overall MOS is reported in Table~\ref{tab:feature_count_main}.}
\label{tab:feature_nisqa_dimensions}

\end{table*}

\noindent\textbf{Explanation of Table~\ref{tab:feature_proxies}.} Rows~\textcircled{\textbf{\scriptsize 1}}--\textcircled{\textbf{\scriptsize 10}} compare error, correlation, level, envelope, and custom-quality diagnostics across capacities. Winners alternate across rows and bands. This heterogeneity supports the claim that added QCD inputs reshape trade-offs rather than monotonically improving fidelity.\par\smallskip
\begin{table*}[!tbp]
\centering

\tablefontsize
\setlength{\tabcolsep}{4pt}
\resetapprow
\begin{appfit}
\begin{tabular}{lrrr}
\toprule
Diagnostic & 3 inputs & 5 inputs & 7 inputs \\
\midrule
\appnextrow{} NMR 0--4 kHz (dB) & $\mathbf{16.07\!\pm\!0.15}$ & $16.34\!\pm\!0.16$ & $16.47\!\pm\!0.73$ \\
\appnextrow{} NMR 4--8 kHz (dB) & $\mathbf{21.44\!\pm\!1.10}$ & $22.72\!\pm\!0.16$ & $23.23\!\pm\!1.86$ \\
\appnextrow{} Spectral corr. 0--4 kHz & $0.9316\!\pm\!0.0007$ & $\mathbf{0.9346\!\pm\!0.0004}$ & $0.9333\!\pm\!0.0015$ \\
\appnextrow{} Spectral corr. 4--8 kHz & $\mathbf{0.7400\!\pm\!0.0033}$ & $0.7328\!\pm\!0.0043$ & $0.7289\!\pm\!0.0129$ \\
\appnextrow{} Active-level error 0--4 kHz (dB) & $0.105\!\pm\!0.002$ & $0.099\!\pm\!0.006$ & $\mathbf{0.099\!\pm\!0.012}$ \\
\appnextrow{} Active-level error 4--8 kHz (dB) & $6.23\!\pm\!0.63$ & $6.40\!\pm\!0.92$ & $\mathbf{5.64\!\pm\!0.58}$ \\
\appnextrow{} Envelope error 0--4 kHz (dB) & $\mathbf{1.153\!\pm\!0.011}$ & $1.156\!\pm\!0.012$ & $1.173\!\pm\!0.026$ \\
\appnextrow{} Envelope error 4--8 kHz (dB) & $\mathbf{9.63\!\pm\!0.07}$ & $9.68\!\pm\!0.13$ & $9.73\!\pm\!0.23$ \\
\appnextrow{} Custom QI 0--4 kHz & $80.51\!\pm\!0.20$ & $80.60\!\pm\!0.36$ & $\mathbf{80.82\!\pm\!0.27}$ \\
\appnextrow{} Custom QI 4--8 kHz & $43.72\!\pm\!1.25$ & $43.05\!\pm\!1.92$ & $\mathbf{44.23\!\pm\!0.61}$ \\
\bottomrule\end{tabular}
\end{appfit}
\caption{Consolidated ITU-inspired sub-band diagnostics for the capacity study, mean$\pm$sample standard deviation over three seeds. These are ITU-inspired scores, not certified standards scores. Errors/NMR are lower-is-better; correlation/QI are higher-is-better.}
\label{tab:feature_proxies}

\end{table*}

\subsection{Paired capacity-study differences}
\noindent\textbf{Explanation of Table~\ref{tab:feature_bootstrap}.} Scores are averaged across the three trained seeds before 2,000 shared-identifier bootstrap draws. Rows~\textcircled{\textbf{\scriptsize 1}}--\textcircled{\textbf{\scriptsize 5}} show conditional acoustic/perceptual differences; Row~\textcircled{\textbf{\scriptsize 6}} shows WER intervals crossing zero. Positive differences are larger raw scores, so direction depends on the metric. These intervals quantify utterance variation and do not replace uncertainty over new training runs.\par\smallskip
\begin{table*}[!tbp]
\centering

\tablefontsize
\setlength{\tabcolsep}{6pt}
\resetapprow
\begin{appfit}
\begin{tabular}{lcc}
\toprule
Metric & $\Delta$(5 features $-$ 3 features) [95\% CI] & $\Delta$(7 features $-$ 3 features) [95\% CI]\\
\midrule
\appnextrow{} LSD$\downarrow$ & $-0.01003\;[-0.01120,-0.00886]$ & $-0.00923\;[-0.01048,-0.00794]$\\
\appnextrow{} STOI$\uparrow$ & $+0.002446\;[+0.002264,+0.002641]$ & $+0.000889\;[+0.000725,+0.001046]$\\
\appnextrow{} PESQ$\uparrow$ & $-0.05106\;[-0.05521,-0.04688]$ & $-0.05513\;[-0.05902,-0.05137]$\\
\appnextrow{} SI-SNR$\uparrow$ & $-0.0601\;[-0.0700,-0.0508]$ & $-0.2242\;[-0.2332,-0.2150]$\\
\appnextrow{} N-MOS$\uparrow$ & $-0.1292\;[-0.1375,-0.1211]$ & $-0.0600\;[-0.0671,-0.0531]$\\
\appnextrow{} WER$\downarrow$ & $+0.00034\;[-0.00293,+0.00381]$ & $-0.00267\;[-0.00629,+0.00101]$\\
\bottomrule
\end{tabular}
\end{appfit}
\caption{Paired utterance-bootstrap differences for the QCD feature-count study. WER is expressed as a fraction in this table.}
\label{tab:feature_bootstrap}

\end{table*}

\subsection{Earlier conditioning dimensions}
\noindent\textbf{Explanation of Table~\ref{tab:quick4k_nisqa_dims}.} Rows~\textcircled{\textbf{\scriptsize 1}}--\textcircled{\textbf{\scriptsize 7}} compare inputs and learned systems; Row~\textcircled{\textbf{\scriptsize 8}} is the clean target. Row~\textcircled{\textbf{\scriptsize 4}} is strongest among learned rows on these predicted dimensions, which differs from the LSD/STOI ranking and motivates the multi-objective discussion.\par\smallskip
\begin{table*}[!tbp]
\centering

\tablefontsize
\setlength{\tabcolsep}{4pt}
\resetapprow
\begin{appfit}
\begin{tabular}{lrrrr}
\toprule
Condition & Noisiness & Discontinuity & Coloration & Loudness \\
\midrule
\appnextrow{} Interpolated input & 4.1672 & 4.1583 & 2.2001 & 2.9538 \\
\appnextrow{} \cis{} & 4.2490 & 4.2962 & 3.9155 & 4.1857 \\
\appnextrow{} Quantum FiLM only & 4.1409 & 4.1931 & 3.7891 & 4.1277 \\
\appnextrow{} Matched FiLM & 4.3091 & 4.3514 & 3.9971 & 4.2270 \\
\appnextrow{} Full fixed 0.10 & 4.2109 & 4.2273 & 3.8464 & 4.1516 \\
\appnextrow{} Full Thompson & 4.1390 & 4.1790 & 3.7758 & 4.1081 \\
\appnextrow{} Full without source mask & 4.2250 & 4.2222 & 3.8364 & 4.1636 \\
\appnextrow{} Clean target & 4.3526 & 4.4658 & 4.1796 & 4.3889 \\
\bottomrule\end{tabular}
\end{appfit}
\caption{Conditioning evaluation for seed-1234 : NISQA subdimensions over 2,937 utterances. Overall MOS is reported once in Table~\ref{tab:quick4k_metrics}.}
\label{tab:quick4k_nisqa_dims}

\end{table*}

\subsection{Subsequent suite diagnostic families}
\noindent\textbf{Explanation of Table~\ref{tab:summary_diagnostics}.} Rows~\textcircled{\textbf{\scriptsize 1}}--\textcircled{\textbf{\scriptsize 5}} assess HF reconstruction, Rows~\textcircled{\textbf{\scriptsize 6}}--\textcircled{\textbf{\scriptsize 12}} custom spectral/level proxies, and Rows~\textcircled{\textbf{\scriptsize 13}}--\textcircled{\textbf{\scriptsize 16}} NISQA dimensions. Matched training improves many noisy/codec averages but does not restore clean performance. Codec and packet columns are equal-weight condition averages, not clean controls; no separate numerical low-band preservation score is available.\par\smallskip
\begin{table*}[!tbp]
\centering

\tablefontsize
\setlength{\tabcolsep}{3pt}
\resetapprow
\begin{appfit}
\begin{tabular}{lrrrrrrr}
\toprule
Diagnostic & Clean & No mask & ZS noise & M noise & ZS codec & M codec & Packet \\
\midrule
\appnextrow{} HF LSD & 1.155 & 1.170 & 1.649 & 1.401 & 1.302 & 1.186 & 1.170 \\
\appnextrow{} HF RMSE ($10^{-3}$) & 5.005 & 5.189 & 4.819 & 4.670 & 4.745 & 4.946 & 4.965 \\
\appnextrow{} HF MAE ($10^{-3}$) & 1.854 & 1.886 & 2.224 & 1.887 & 1.765 & 1.852 & 1.844 \\
\appnextrow{} HF SI-SDR (dB) & -35.734 & -35.741 & -44.190 & -40.224 & -36.798 & -36.301 & -36.012 \\
\appnextrow{} HF crest error & 4.185 & 4.534 & 6.360 & 5.635 & 4.426 & 4.221 & 4.344 \\
\appnextrow{} Spectral correlation & 0.7566 & 0.7587 & 0.4270 & 0.5181 & 0.7232 & 0.7453 & 0.7493 \\
\appnextrow{} Noise-mask ratio (dB) & 20.504 & 20.531 & 40.080 & 35.346 & 20.454 & 21.138 & 21.019 \\
\appnextrow{} Loudness error (LU) & 5.291 & 5.274 & 9.237 & 8.006 & 8.157 & 5.519 & 5.513 \\
\appnextrow{} Active-level error (dB) & 5.291 & 5.274 & 9.239 & 8.006 & 8.157 & 5.519 & 5.513 \\
\appnextrow{} Disturbance proxy (dB) & 9.342 & 9.446 & 14.737 & 13.017 & 10.126 & 9.669 & 9.451 \\
\appnextrow{} Envelope error (dB) & 9.055 & 8.757 & 16.518 & 13.493 & 9.836 & 9.207 & 9.348 \\
\appnextrow{} Custom quality index & 46.992 & 46.957 & 21.327 & 27.855 & 39.276 & 45.989 & 45.232 \\
\appnextrow{} NISQA noisiness & 4.271 & 4.358 & 1.581 & 2.000 & 4.029 & 4.170 & 4.048 \\
\appnextrow{} NISQA discontinuity & 4.333 & 4.422 & 3.411 & 3.544 & 4.003 & 4.162 & 2.908 \\
\appnextrow{} NISQA coloration & 3.942 & 4.083 & 2.452 & 3.095 & 3.516 & 3.731 & 3.689 \\
\appnextrow{} NISQA loudness & 4.224 & 4.330 & 2.384 & 2.828 & 3.954 & 4.088 & 3.996 \\
\bottomrule\end{tabular}
\end{appfit}
\caption{Evaluation of \method{} on waveform, spectral, BS1387, BS1770, P.56, P.862, P.863, custom quality-index, and NISQA metrics. HF is 4--8~kHz at 16~kHz evaluation. ZS/M: zero-shot/matched training. Codec and packet columns equally average their four conditions. ITU-inspired quantities are not certified scores. Errors are lower-is-better; correlations, quality indices, and NISQA are higher-is-better.}
\label{tab:summary_diagnostics}

\end{table*}

\subsection{Measured training profiles for the subsequent suite}
\noindent\textbf{Explanation of Table~\ref{tab:efficiency}.} Rows~\textcircled{\textbf{\scriptsize 1}}--\textcircled{\textbf{\scriptsize 13}} compare measured 50-epoch profiles. Similar parameter counts do not imply similar hours: data regime and active paths matter. The missing deployment RTF is left unfilled, defending the boundary between measured training cost and unmeasured inference latency.\par\smallskip
\begin{table*}[!t]
\centering

\tablefontsize
\setlength{\tabcolsep}{3pt}
\resetapprow
\begin{appfit}
\begin{tabular}{lrrrr}
\toprule
Run & Deploy M & Train M & Peak GB & Train hours \\
\midrule
\appnextrow{} canonical & 18.207 & 19.412 & 1.683 & 8.13 \\
\appnextrow{} cis & 18.141 & 19.342 & 1.672 & 3.97 \\
\appnextrow{} fixed\_005 & 18.207 & 19.412 & 1.682 & 8.19 \\
\appnextrow{} fixed\_010 & 18.207 & 19.412 & 1.682 & 8.48 \\
\appnextrow{} fixed\_025 & 18.207 & 19.412 & 1.682 & 8.31 \\
\appnextrow{} fixed\_050 & 18.207 & 19.412 & 1.682 & 8.15 \\
\appnextrow{} fixed\_100 & 18.207 & 19.412 & 1.682 & 8.04 \\
\appnextrow{} full & 18.207 & 19.412 & 1.918 & 9.29 \\
\appnextrow{} matched\_codec & 18.207 & 19.412 & 1.683 & 13.27 \\
\appnextrow{} matched\_noise & 18.207 & 19.412 & 1.683 & 8.35 \\
\appnextrow{} matched & 18.207 & 19.407 & 1.683 & 4.01 \\
\appnextrow{} no\_mask & 18.207 & 19.412 & 1.683 & 8.28 \\
\appnextrow{} qcd\_only & 18.141 & 19.346 & 1.673 & 6.83 \\
\bottomrule\end{tabular}
\end{appfit}
\caption{Analysis of 50-epoch training profiles on an RTX 4090. Parameter counts are in millions; peak CUDA allocated training memory is reported in decimal GB and includes optimizer memory but excludes validation. Training hours sum epoch times.} 
\label{tab:efficiency}

\end{table*}

\subsection{Ablation scope of inference}
The controlled component, targeted conditioning, controlled high-frequency,
and controlled ITU-inspired results are utterance-level means from one trained
seed. Averaging 2,937 test utterances characterizes each trained model but does
not quantify variation across stochastic training runs, so these differences are
reported descriptively. The QCD-capacity study is separately repeated over three
seeds and reports variation across seed means. Paired utterance-bootstrap intervals
condition on those trained runs and therefore do not substitute for additional
training-seed replication. Taken together, the ablations establish component-dependent
trade-offs rather than a uniformly superior QCD mapping, conditioner, weighting
policy, circuit structure, or reconstruction constraint.

\end{document}